\documentclass[eqsecnum,
 reprint,
 amsmath,amssymb,
 aps,
]{revtex4-2}

\usepackage{graphicx}% Include figure files
\usepackage{dcolumn}% Align table columns on decimal point
\usepackage{bm}% bold math

\usepackage{url}

\DeclareMathAlphabet{\pazocal}{OMS}{zplm}{m}{n}
\usepackage{braket}
\usepackage{amsmath}
\usepackage{amssymb}
\usepackage{setspace}
\usepackage{leftidx}
\usepackage{hyperref}
\usepackage[colorinlistoftodos]{todonotes}
\usepackage{bm}
\usepackage{slashed}
\usepackage{calrsfs}
\usepackage{cancel}
\usepackage{dirtytalk}
\begin{document}

\title{The eigenmodes for spinor quantum field theory in global de Sitter space-time}% Force line breaks with \\

\author{Vasileios A. Letsios}%
 \email{vl624@york.ac.uk}
\affiliation{%
Department of Mathematics, University of York\\ Heslington, York, YO10 5DD, United Kingdom 
}%

\date{\today}% It is always \today, today,

\begin{abstract}
The mode solutions of the Dirac equation on $N$-dimensional de~Sitter space-time ($dS_{N}$) with $(N-1)$-sphere spatial sections are obtained by analytically continuing the spinor eigenfunctions of the Dirac operator on the $N$-sphere ($S^{N}$). The analogs of flat space-time positive frequency modes are identified and a vacuum is defined. The transformation properties of the mode solutions under the de~Sitter group double cover (Spin($N$,1)) are studied. We reproduce the expression for the massless spinor Wightman two-point function in closed form using the mode-sum method. By using this closed-form expression and taking advantage of the maximal symmetry of $dS_{N}$ we find an analytic expression for the spinor parallel propagator. The latter is used to construct the massive Wightman two-point function in closed form. 

\end{abstract}

\maketitle
%%%%%%%%%%%%%%%%%%%%%%%%%%%%%%%%%%%%%%%%%%%%%%%%%%%%%%%%%%%%%%%%%%%%%%%%%%%%%%%%%%%%%%%%%%%%%%%%%%%%%%%%%%%%%%%%%%%%%%%%%%%%%%%%%%%%%%%%%%%%%%%%%%%%%%%%%%%%%%%%%%%%%%%%%%%%%%%%%%%%%%%%%%%%%%%%%%%%%%%%%%%%%%%%%%%%%%%%%%
\section{Introduction} \label{Introduction}
 The spinor functions that satisfy the eigenvalue equation of the Dirac operator on $S^{N}$
 \begin{equation}\label{eigenvalue_equation_Dirac}
     \slashed{\nabla} \psi=i \lambda \psi
 \end{equation}
 have been studied by Camporesi and Higuchi~\cite{Camporesi}. More specifically, the eigenspinors on $S^{N}$ have been recursively constructed in terms of eigenspinors on $S^{N-1}$ using separation of variables in geodesic polar coordinates and their eigenvalues have been calculated. The line element for $S^{N}$ may be written as 
\begin{equation} \label{sphere_metric}
    ds^{2}_{N}=d\theta^{2}_{N} + \sin^{2}{\theta_{N}} ds_{N-1}^{2},
\end{equation}
where $\theta_{N}$ is the geodesic distance from the North Pole and $ds^{2}_{N-1}$ is the line element of $S^{N-1}$. Similarly, the line element of $S^{n}$ ($n=2,3,...,N-1$) can be expressed as  
\begin{align}
    ds^{2}_{n}=d\theta^{2}_{n} + \sin^{2}{\theta_{n}} ds_{n-1}^{2},
\end{align} 
 while $ds_{1}^{2}=d \theta_{1}^{2}$.

The $N$-dimensional de~Sitter space-time is the maximally symmetric solution of the vacuum Einstein field equations with positive cosmological constant $\Lambda$~\cite{hawking_ellis_1973}
\begin{equation}
    R_{\mu \nu}-\frac{1}{2} g_{\mu \nu} R + \Lambda  g_{\mu \nu}=0.
\end{equation}
The cosmological constant is given by 
\begin{equation}
    \Lambda=\frac{(N-2)(N-1)}{2 \hspace{1mm} \mathcal{R}^{2}   },
\end{equation}
where $\mathcal{R}$ is the de~Sitter radius. Throughout this paper we use units in which $\mathcal{R}=1$.

The $N$-dimensional de~Sitter space-time can also be obtained by an ``analytic continuation" of $S^{N}$. More specifically, by replacing
\begin{equation} \label{analytic_con}
\theta_{N}\rightarrow x\equiv \pi/2 - it 
\end{equation}
in the $S^{N}$ metric (\ref{sphere_metric}) we find the line element for $dS_{N}$ with $S^{N-1}$ spatial sections (see Eq.~(\ref{global_coordinates}))
\begin{equation} \label{dS_metric}
    ds^{2}=-dt^{2}+\cosh^{2}{t} ds^{2}_{N-1}.
\end{equation}
Motivated by the above, one can obtain the mode solutions to the Dirac equation on $dS_{N}$
\begin{equation}\label{Dirac_equation}
    \slashed{\nabla} \psi- M \psi=0
\end{equation}
just by analytically continuing the eigenmodes of (\ref{eigenvalue_equation_Dirac}). The Dirac spinors obtained by analytic continuation can be used to describe spin-1/2 particles in de~Sitter space-time and they form a representation of Spin($N$,1). The latter has to be unitary to ensure that negative probabilities will not arise. In order to study the unitarity of the representation we are going to introduce a de~Sitter invariant inner product among the analytically continued eigenspinors (see Sec.\ \ref{Section_Group_Theory}). Note that this approach has been previously applied for the divergence-free and traceless tensor eigenfunctions of the Laplace-Beltrami operator on $S^{N}$~\cite{STSHS}, where the restriction of unitarity gave rise to the forbidden mass range for the spin-2 field on $dS_{N}$.

In this paper our main aim is the identification of the mode functions for the free Dirac field on global $dS_{N}$ with $S^{N-1}$ spatial sections. As a consistency check, we reproduce the expected form for the massless spinor Wightman function~\cite{Muck:1999mh} using the mode-sum method. We also use this Wightman function to find an analytic expression for the spinor parallel propagator. To our knowledge, such an expression is absent from the literature. Solutions of the free Dirac equation on de~Sitter space-time with static charts may be found in Ref.~\cite{Otchik_1985}, with moving charts in Refs.~\cite{Barut,Shishkin:1991ma,movingcharts} and with open charts in Ref.~\cite{hyperbolic_dS_Kanno}.     

The rest of this paper is organized as follows. In Sec.~\ref{Geometry_of_de_Sitter} we discuss the global coordinate system that is relevant to the analytic continuation of $S^{N}$ and we review the geodesic structure of $dS_{N}$. In Sec.~\ref{Section_spinors_and_clifford} we present the basics about  Dirac spinors and Clifford algebras on $dS_{N}$. In Sec.~\ref{solutions_section} we begin by reviewing the eigenspinors of the Dirac operator on $S^{N}$ following Ref.~\cite{Camporesi}. Then we obtain the mode solutions of the Dirac equation on $dS_{N}$ by analytically continuing the eigenmodes on $S^{N}$ and we give a criterion for generalized positive frequency modes. We also construct spinors satisfying the Dirac equation with the sign of the mass term changed. These spinors are used in Appendix~\ref{appendix_charge_conjugation} for an alternative construction of the negative frequency modes via charge conjugation. In Sec.~\ref{Section_Group_Theory} we define a de~Sitter invariant inner product among the analytically continued eigenmodes and we show that the associated norm is positive-definite (i.e. the representation is unitary). Using this norm we normalize the analytically continued eigenspinors. Then the transformation properties of the positive frequency solutions under Spin($N$,1) are studied using the spinorial Lie derivative~\cite{Kosmann}. It is shown that the positive frequency solution subspace is Spin($N$,1) invariant (hence, so is the corresponding vacuum). In Sec.~\ref{Section_2_point_function}, after presenting the negative frequency solutions of the Dirac equation, we perform the canonical quantization procedure for the free Dirac quantum field. Then we review the coordinate independent construction of Dirac spinor Green's functions on $dS_{N}$ following Ref.~\cite{Muck:1999mh}. We present a closed-form expression for the massless spinor Wightman two-point function obtained by the mode-sum method.  This closed-form expression is in agreement with the construction given in Ref.~\cite{Muck:1999mh}. Then we find an analytic expression for the spinor parallel propagator and we use it to obtain a closed-form expression for the massive Wightman two-point function in terms of intrinsic geometric objects. Our summary and concluding remarks are given in Sec.~\ref{summary and conclusions}. 

There are six appendices. In Appendix~\ref{appendix_charge_conjugation} we construct the negative frequency solutions of the Dirac equation on $dS_{N}$ by charge conjugating our analytically continued eigenspinors. In Appendix~\ref{appendix_conjecture} we compare the mode-sum method for the massive spinor Wightman function with the construction presented in Ref.~\cite{Muck:1999mh} and we arrive at a closed-form conjecture for a series containing the Gauss hypergeometric function. The rest of the appendices concern technical details. Some minor details omitted in the main text are presented in Appendices~\ref{appendix_raising_lowering_hypergeom} and \ref{Appendix_Spinor_Lie}. In Appendix~\ref{appendix_2pt_function_calculations} we present details about the mode-sum construction of the massless spinor Wightman function. In Appendix~\ref{appendix_defining_test} we demonstrate that our analytic expression for the spinor parallel propagator satisfies the defining properties given in Ref.~\cite{Muck:1999mh}.

We use the mostly plus convention for the metric signature. When it comes to tensors, lower case Greek indices refer to components with respect to the ``coordinate basis" while Latin ones refer to components with respect to the vielbein (i.e. orthonormal frame) basis. Spinor indices (when not suppressed) are denoted with capital Latin letters. For bitensors (or bispinors) that depend on two space-time points $x,x'$, unprimed indices refer to the tangent space at $x$ while primed ones refer to the tangent space at $x'$. Summation over repeated indices is understood throughout this paper.
%%%%%%%%%%%%%%%%%%%%%%%%%%%%%%%%%%%%%%%%%%%%%%%%%%%%%%%%%%%%%%%%%%%
\section{Geometry of \texorpdfstring{$N$}{N}-dimensional de Sitter space-time}\label{Geometry_of_de_Sitter}
\subsection{Coordinate system, Christoffel symbols and spin connection}

The $N$-dimensional de~Sitter space-time can be represented as a hyperboloid embedded in $(N+1)$-dimensional Minkowski space.  The de~Sitter hyperboloid is described by 
\begin{align}\label{hyperboloid}
    \eta_{ab}X^{a} X^{b}=1,
\end{align}
where 
$ \eta_{ab}=\text{diag}(-1,1,1,...,1) \, (a,b=0,1,...,N)$
is the flat metric for the embedding space and $X^{0},X^{1},...,X^{N}$ are the standard Minkowski coordinates. The global coordinates used in this paper are given by 
\begin{align}\label{global_coordinates}
    &X^{0}=X^{0}{(t,\bm{\theta})} =\sinh{t} \nonumber \\
   & X^{i}=X^{i}{(t,\bm{\theta})}=\cosh{t} \,\, Z^{i}, \hspace{4mm} i=1,...,N,
\end{align}
where $t\in \mathbb{R},\, \bm{\theta}=(\theta_{N-1},\theta_{N-2},...,\theta_{1})$ and the $Z^{i}$'s are the spherical coordinates for $S^{N-1}$ in $N$-dimensional Euclidean space
\begin{align}\label{spherical_coordinates}
    &Z^{1}= \sin{\theta_{N-1}} \, \sin{\theta_{N-2}}  \, ... \sin{\theta_{2}}\, \sin{\theta_{1}} \nonumber \\
     &Z^{2}= \sin{\theta_{N-1}} \, \sin{\theta_{N-2}}  \, ... \sin{\theta_{2}}\,\cos{\theta_{1}} \, \nonumber \\
    &  \vdots \nonumber\\
     &Z^{N-1}= \sin{\theta_{N-1}} \, \cos{\theta_{N-2}} \nonumber\\
     &Z^{N}=  \cos{\theta_{N-1}},
    \end{align}
    where $0 \leq \theta_{1} <2 \pi$ and $0\leq \theta_{i} \leq  \pi$ ($i \neq 1$).
Using the coordinates (\ref{global_coordinates}) we obtain the line element (\ref{dS_metric}) for $dS_{N}$.

The non-zero Christoffel symbols for the coordinates (\ref{global_coordinates}) are 
\begin{align}\label{Christoffels_dS}
    &\Gamma^{t}_{\hspace{0.2mm}\theta_{i} \theta_{j}}=\cosh{t} \sinh{t} \hspace{1mm}\tilde{g}_{\theta_{i} \theta_{j}}, \hspace{2mm} \Gamma^{\theta_{i}}_{\hspace{0.2mm}\theta_{j} t} =\tanh{t}  \hspace{1mm}\tilde{g}^{\theta_{i}}_{\theta_{j}}, \nonumber \\ 
& \Gamma^{\theta_{k}}_{\hspace{0.2mm}\theta_{i} \theta_{j}}=\tilde{\Gamma}^{\theta_{k}}_{\hspace{0.2mm}\theta_{i} \theta_{j}},
\end{align}
where $\tilde{g}_{\theta_{i} \theta_{j} },\tilde{\Gamma}^{\theta_{k}}_{\hspace{0.2mm}\theta_{i} \theta_{j}}$ are the metric tensor and the Christoffel symbols, respectively, on $S^{N-1}$. The vielbein fields are given by
\begin{equation}\label{vielbeins}
    e^{t}{\hspace{0.2mm}}_{0}=1, \hspace{5mm}  e^{\theta_{i}}{\hspace{0.2mm}}_{i}=\frac{1}{\cosh{t}} \tilde{e}^{\theta_{i}}{\hspace{0.2mm}}_{i} , \hspace{5mm}i=1,...,N-1,
\end{equation}
 where $\tilde{e}^{\theta_{i}}{\hspace{0.2mm}}_{i}$ are the vielbein fields on $S^{N-1}$. The latter are given by
 \begin{align}\label{vielbeins_S_(N-1)}
    & \tilde{e}^{\theta_{N-1}}{\hspace{0.2mm}}_{N-1}=1, \nonumber \\
    & \tilde{e}^{\theta_{j}} {\hspace{0.2mm}}_{j} = \frac{1}{\sin{\theta_{N-1}}\, \sin{\theta_{N-2}}\,...  \,\sin{\theta_{j+1}}}, \hspace{2mm} j=1,...,N-2.
 \end{align}
  The spin connection ${\omega}_{abc}={\omega}_{a[bc]} \equiv ({\omega}_{abc}-{\omega}_{acb})/2 $ is given by
 \begin{equation}
           {\omega}_{abc} =  e^{\mu}\hspace{0.2mm}_{a} \Big( \partial_{\mu}  e^{\lambda}\hspace{0.2mm}_{b} + {\Gamma}^{\lambda}_{\mu \nu} e^{\nu}\hspace{0.2mm}_{b} \Big) e_{\lambda c}      
       \end{equation}
       and its only non-zero components are
\begin{equation}\label{spin_connection_dS}
   \omega_{ijk} = \frac{  \tilde{\omega}_{ijk}}{ \cosh{t}  } , \hspace{4mm}  \omega_{i0k} =  \tanh{t}\hspace{1mm} \delta_{ik} \hspace{3mm} i,j,k=1,...,N-1,
        \end{equation} 
where $\tilde{\omega}_{ijk}$ are the spin connection components on $S^{N-1}$ and $\delta_{ij}$ is the Kronecker delta symbol. (Note that the sign convention we use for the spin connection is the opposite of the one used in most supersymmetry texts.)
%%%%%%%%%%%%%%%%%%%%%%%%%%%%%%%%%%%%%%%%%%%%%%%%%%%%%%%%%%%%%%%%%%%%%%%%%%%%%%%%%%%%%%%%%%%%%%%%%%%%%%%%%%%%%%%%%%%%%%%%%%%%%%%%%%%
\subsection{Geodesics on \texorpdfstring{$dS_{N}$}{dSN}}
Geodesics on $dS_{N}$ are obtained by intersecting the hyperboloid (\ref{hyperboloid}) with two-planes passing through the origin~\cite{Two-pointFunctionsAndQFieldsIndS}. Note that, contrary to the case of maximally symmetric Euclidean spaces ($\mathbb{R}^{N},S^{N}, H^{N}$), on pseudo-Riemannian spaces two points cannot always be connected by a geodesic.

Let $x,x'$ be two points on the de~Sitter hyperboloid (\ref{hyperboloid}) and $\mu(x,x')$ the geodesic distance between them. Using the scalar product of the ambient space
\begin{align}
    \mathcal{Z}(x,x')=\eta_{a b } X^{a}(x) X^{b}(x')
\end{align}
one can define the useful quantity
\begin{align}\label{globally_defined_geodesic_distance}
    z{(x,x')}=\frac{1}{2}\Big( 1+ \eta_{ab}X^{a}{(x)} X^{b}{(x')}    \Big).
\end{align}
If $-1\leq \mathcal{Z}(x,x')< 1$ (i.e. $z \in [0,1)$) the points $x,x'$ are spacelike separated ($\mu \in \mathbb{R}$) and they can be connected by a spacelike geodesic. (The equality sign corresponds to antipodal points.) The geodesic distance is then defined by $\mathcal{Z}(x,x')=\cos{(\mu(x,x'))}$ or equivalently
\begin{align}\label{spacelike_sep}
   z =\cos^{2}{\frac{\mu}{2}}.
\end{align}
If $\mathcal{Z}(x,x')<- 1$ (i.e. $z<0$) the points are spacelike separated but there is no geodesic connecting them. However, the function $\mu{(x,x')}$ can still be defined by Eq.~(\ref{globally_defined_geodesic_distance}) via analytic continuation~\cite{allen1986}. (Let $\bar{x}$ be the antipodal point of $x$ and let $x'$ be any point in the interior of the past or future light cone of $\bar{x}$. Then there is no geodesic connecting $x$ and $x'$~\cite{allen1986}.)
If $\mathcal{Z}(x,x')=1$ (i.e. $z=1$) the geodesic distance is zero and the two points can be connected by a null geodesic (or they coincide).
If $\mathcal{Z}(x,x')> 1$ (i.e. $z>1$) the two points are timelike separated ($\mu = i \kappa,\, \kappa \in \mathbb{R}$) and they can be connected by a timelike geodesic. The geodesic distance for timelike separation is given by 
\begin{align}\label{timelike_sep}
   z =\cos^{2}{\frac{\mu}{2}}=\cosh^{2}{\frac{\kappa}{2}}.
\end{align}
In the rest of this paper we suppose that the points under consideration can be connected by a spacelike geodesic (unless otherwise stated). The corresponding results for the timelike case can be obtained just by replacing $\mu \rightarrow i \kappa$.

The unit tangent vectors at $x$ and $x'$ to the geodesic connecting the two points are defined by 
\begin{align}\label{tangent_vectors}
  & n_{\kappa}{(x,x')} = \nabla_{\kappa} \mu{(x,x')}, \hspace{4mm}n_{\kappa'}{(x,x')} = \nabla_{\kappa'} \mu{(x,x')},
\end{align}
respectively. Since $dS_{N}$ is a maximally symmetric space-time, the unit tangents satisfy~\cite{allen1986}
\begin{align}
   & \nabla_{\mu} n_{\nu}= \cot{\mu} ( g_{\mu \nu} -n_{\mu} n_{\nu}),\label{cov_deriv_of_tangent_vector} \\ 
   &\nabla_{\mu'} n_{\nu}= -\frac{1}{ \sin{\mu}} ( g_{\mu' \nu} +n_{\mu'} \eta_{\nu}), \label{calculate_vec_parallel_propagator}
    \\& \nabla_{\kappa}g_{\mu \nu'}=\tan\frac{\mu}{2}(g_{\kappa \mu}n_{\nu'} + g_{\kappa \nu'}  n_{\mu} ),
\end{align}
where $g_{\mu \nu}(x)$ is the metric tensor and $g_{\mu \nu'}(x,x')$ is the bivector of parallel transport. The latter is also known as the vector parallel propagator and it performs the parallel transport of a vector field $V^{\nu'}(x')$ from $x'$ to $x$ along the geodesic connecting these points~\cite{allen1986}
\begin{align}
    V^{\mu}_{||}(x)=g^{\mu}_{\hspace{2mm}\nu'}V^{\nu'}(x'),
\end{align}
where $ V^{\mu}_{||}(x)$ is the parallelly transported vector at $x$. (In this paper by geodesic we mean the shortest geodesic connecting the two points.)
It is worth noting the relations~\cite{allen1986}
\begin{align}
  & n_{\mu}=-g_{\mu}^{\hspace{2mm}\nu'}\,n_{\nu'},  \hspace{4mm}  n_{\mu'}=-g_{\mu'}^{\hspace{2mm}\nu}\,n_{\nu},\label{parallel_transport_of_tangent_vector}\\
  & g^{\mu}_{\hspace{2mm}\nu'}\, g^{\nu'}_{\hspace{2mm}\lambda}=\delta^{\mu}_{\hspace{2mm}\lambda},  \hspace{4mm} g^{\mu'}_{\hspace{2mm}\kappa}\, g^{\kappa}_{\hspace{2mm}\nu'}=\delta^{\mu'}_{\hspace{2mm}\nu'}.
\end{align}

Using the coordinates (\ref{global_coordinates}) we obtain the following expression for the geodesic distance:
\begin{align}\label{geodesic_distance_dS}
    \cos \big( \,\mu{(x,x')} \, \big)=-\sinh{t}\sinh{t'}+\cosh{t}\cosh{t'} \cos{\Omega_{N-1}},
\end{align}
where 
\begin{align}\label{geodesic_distacne_sphere}
    \cos{\Omega_{n}}=&\cos{\theta_{n}}\cos{\theta_{n}'}+\sin{\theta_{n}}\sin{\theta_{n}'}\cos{\Omega_{n-1}}, 
\end{align}
for $n=2,...,N-1$ and
\begin{align}\label{geodesic_distance_S1}
    \cos{\Omega_{1}}=\cos{(\theta_{1}-\theta_{1}')}.
\end{align}
Then the coordinate basis components of the tangent vector $n_{\mu}(x,x')=(n_{t}(x,x'),n_{\theta_{i}}(x,x'))$ ($i=1,...,N-1$) are given by
\begin{align}\label{tangent_vectors_general}
   &n_{t}=\frac{1}{\sin{\mu}}(  \cosh{t} \sinh{t'} -
    \sinh{t} \cosh{t'} \cos{\Omega_{N-1}}   ), \\
    &n_{\theta_{i}}=-\frac{1}{\sin{\mu}}\cosh{t} \cosh{t'} \frac{\partial}{\partial {\theta_{i}}}(\cos{\Omega_{N-1}}) ,
\end{align}
where 
\begin{align}
  \frac{\partial}{\partial {\theta_{i}}}&(\cos{\Omega_{N-1}})    =\,\Big(\,\prod_{r=1}^{N-(i+1)}\sin{\theta_{N-r}}\sin{\theta_{N-r}'} \Big) \nonumber \\&\times (-\sin{\theta_{i}}\cos{\theta_{i}'}+\cos{\theta_{i}}\sin{\theta_{i}'} \cos{\Omega_{i-1}}).
\end{align}
The components of $n_{\mu'}(x,x')$ are given by analogous expressions with $t \leftrightarrow t', \theta_{i} \leftrightarrow \theta_{i}'$. The vielbein basis components of the tangent vector at $x$, $n_{a}{(x,x')}=e^{\mu}_{\hspace{2mm} a}(x)\, n_{\mu}(x,x')$ ($a=0,1,...,N-1$), are given by
\begin{align}
    n_{0}=&n_{t} ,\label{orthonormal_basis_components_tangent_vectors_n0}\\
    n_{N-1}=&-\frac{\cosh{t'}}{\sin{\mu}}(-\sin{\theta_{N-1}}\cos{\theta_{N-1}'}\nonumber \\
   & +\cos{\theta_{N-1}}\sin{\theta_{N-1}'} \cos{\Omega_{N-2}}),\label{orthonormal_basis_components_tangent_vectors_n_N-1}\\
    n_{b}=&-\frac{\cosh{t'}}{\sin{\mu}}\,\Big( \,\prod_{r=1}^{N-(b+1)}\sin{\theta_{N-r}'} \Big) \nonumber \\
   & \times(-\sin{\theta_{b}}\cos{\theta_{b}'}+\cos{\theta_{b}}\sin{\theta_{b}'} \cos{\Omega_{b-1}}),\label{orthonormal_basis_components_tangent_vectors_spatial_comps}
\end{align}
($b=1,...,N-2$) while the components of $n_{a'}{(x,x')}=e^{\mu'}_{\hspace{2mm} a'}(x')\, n_{\mu'}(x,x')$ ($a'=0',1',...,(N-1)'$) can be obtained from Eqs.~(\ref{orthonormal_basis_components_tangent_vectors_n0})-(\ref{orthonormal_basis_components_tangent_vectors_spatial_comps}) with $t \leftrightarrow t', \, \theta_{a} \leftrightarrow \theta_{a}'$. (Note that we define $\cos \Omega_{0} \equiv 1$.) 
%%%%%%%%%%%%%%%%%%%%%%%%%%%%%%%%%%%%%%%%%%%%%%%%%%%%%%%%%%%%%%%%%%%%
\section{Dirac Spinors and Clifford Algebra on \texorpdfstring{$N$}{N}-dimensional de Sitter space-time}\label{Section_spinors_and_clifford}

Dirac spinors are $2^{[N/2]}$-dimensional column vectors that appear naturally in Clifford algebra representations, where $[N/2]=N/2$ if $N$ is even and $[N/2]=(N-1)/2$ if $N$ is odd. A Clifford algebra representation in $(N-1)+1$ dimensions is generated by $N$ gamma matrices satisfying the anti-commutation relations
\begin{equation}
   \{\gamma^{a}, \gamma^{b}\}  = 2 \eta^{ab} \bm{1}, \hspace{10mm} a,b=0,1,...,N-1,
\end{equation}
where $\bm{1}$ is the identity matrix and $\eta^{ab}$ is the inverse of the $N$-dimensional Minkowski metric 
$\eta_{ab}=\text{diag}{(-1,+1,...,+1)}$.
We follow the inductive construction of Ref.~\cite{Camporesi} where gamma matrices in $(N-1)+1$ dimensions are expressed in terms of spacelike gamma matrices in $(N-1)$ dimensions ($\widetilde{\gamma}^{i}$) as follows:

\begin{itemize}
    \item For $N$ even

\begin{equation}\label{even_gammas}
 \gamma^{0}= i\begin{pmatrix}  
   0 & \bm{1} \\
   \bm{1} & 0
    \end{pmatrix} , \hspace{5mm}
    \gamma^{i}=\begin{pmatrix}  
   0 & i\widetilde{\gamma}^{i} \\
   -i\widetilde{ \gamma}^{i} & 0
    \end{pmatrix} ,
    \hspace{5mm} i=1,..., N-1,
     \end{equation} 
where the lower-dimensional gamma matrices satisfy the Euclidean Clifford algebra anti-commutation relations
\begin{equation} \label{Euclidean_Clifford_relns}
    \{ \widetilde{\gamma}^{i}, \widetilde{\gamma}^{j}\} = 2 \delta^{ij} \bm{1}, \hspace{5mm}i,j=1,...,N-1.
\end{equation}

\item For $N$ odd

\begin{equation*}
   \gamma^{0}= i\begin{pmatrix}  
    \bm{1} & 0 \\
 0& -\bm{1} 
    \end{pmatrix} ,\hspace{5mm}\gamma^{N-1}= \begin{pmatrix}  
   0 &  \bm{1} \\
   \bm{1} & 0
    \end{pmatrix} , 
    \end{equation*}

\begin{equation}\label{odd_gammas}
    \gamma^{j}= \widetilde{\gamma}^{j}=\begin{pmatrix}  
   0 & i\widetilde{\widetilde{ \gamma}}^{j} \\
   -i\widetilde{\widetilde{ \gamma}}^{j} & 0
    \end{pmatrix},\hspace{5mm}j=1,...,N-2.
\end{equation}
The double-tilde is used to denote gamma matrices in $N-2$ dimensions. For $N=1$ the only (one-dimensional) gamma matrix is equal to $1$. 
\end{itemize}
Note that the gamma matrices we use here for $dS_{N}$ can be obtained by the Euclidean gamma matrices on $S^{N}$ used in Ref.~\cite{Camporesi} via the coordinate change (\ref{analytic_con}). (Gamma matrices transform as vectors under coordinate transformations and it can be checked that all Euclidean $\gamma^{a}$'s remain the same under (\ref{analytic_con}) apart from $\gamma^{N}$; the latter transforms into the timelike gamma matrix: $\gamma^{N} \rightarrow\ i \gamma^{N} = \gamma^{0}$.) 

 Spinors transform under $2^{[N/2]}$-dimensional spinor representations of Spin($N-1$,1) (double cover of SO($N-1,$1)) as
\begin{equation}
    \psi{(x)} \rightarrow S(\Lambda{(x)}) \hspace{1mm}\psi{(x)},
\end{equation}
where $S(\Lambda{(x)}) \in $ Spin($N-1$,1) is a spinorial matrix. The $N(N-1)/2$ generators of Spin$(N-1,1)$ are given by the commutators
\begin{align}\label{Spin(N-1,1)_generators}
    \Sigma^{ab}&=\frac{1}{4}[\gamma^{a},\gamma^{b}] \\
    &=\frac{1}{2}\gamma^{a}\, \gamma^{b} -\frac{1}{2}\eta^{ab},\hspace{5mm} a,b=0,...,N-1
\end{align}
and they satisfy the Spin$(N-1,1)$ algebra commutation relations
\begin{equation}
       [\Sigma^{ab},\Sigma^{cd}]= \eta^{bc} \Sigma^{ad}-\eta^{ac} \Sigma^{bd} + \eta^{ad} \Sigma^{bc} - \eta^{bd} \Sigma^{ac}.
   \end{equation}
 The covariant derivative for a spinor along the vielbein is
  \begin{equation}\label{covariant_deriv_spinor}
      \nabla_{a} \psi = \bm e_{a} \psi  - \frac{1}{2} \omega_{abc} \Sigma^{bc} \psi ,
  \end{equation}
where $\bm{e}_{a}=e^{\mu}_{\hspace{2mm}a} \partial_{\mu}$.
 The Dirac adjoint of a spinor is defined as 
 \begin{equation*}
     \bar{\psi}\equiv i \psi^{\dagger} \gamma^{0}
 \end{equation*}
 with covariant derivative given by 
 \begin{equation}
     \nabla_{a} \bar{\psi}=\bm{e}_{a}\bar{\psi} + \frac{1}{2} \bar{\psi}\, \omega_{abc} \Sigma^{bc} .
 \end{equation}
The covariant derivative of the gamma matrices is 
 \begin{align}
      \nabla_{a}\gamma^{k} &= \bm{e}_{a} \gamma^{k}  - \omega_{a}\hspace{0.1mm}^{k}\hspace{0.1mm}_{c} \gamma^{c} -\frac{1}{2} \omega_{abc} [\Sigma^{bc}, \gamma^{k}] \nonumber \\
      &=0.\label{gamma_constant}
 \end{align}

One can show the following properties of the gamma matrices given by Eqs.~(\ref{even_gammas}) and (\ref{odd_gammas}):
\begin{equation} \label{transpose}
     (\gamma^{0})^{T}=\gamma^{0},\hspace{2mm} (\gamma^{r})^{T}=(-1)^{[\frac{N}{2}]}(-1)^{[\frac{r}{2}]} \gamma^{r}, 
 \end{equation}
  \begin{equation} \label{c.c._of_gamma}
     (\gamma^{0})^{*}=-\gamma^{0},\hspace{2mm} (\gamma^{r})^{*}=(-1)^{[\frac{N}{2}]}(-1)^{[\frac{r}{2}]} \gamma^{r}, 
 \end{equation}
 ($r=1,...,N-1$) and 
 \begin{equation}\label{dagger_of_gammas}
     (\gamma^{a})^{\dagger}=\gamma^{0}\gamma^{a}\gamma^{0},\hspace{5mm} \hspace{5mm} a=0,...,N-1,
 \end{equation}
where the star symbol denotes complex conjugation. Note that the timelike gamma matrix is anti-hermitian while the spacelike ones are hermitian. 
%%%%%%%%%%%%%%%%%%%%%%%%%%%%%%%%%%%%%%%%%%%%%%%%%%%%%%%%%%%%%%%%%%%%%%%%%%%%%%%%%%%%%%%%%

%%%%%%%%%%%%%%%%%%%%%%%%%%%%%%%%%%%%%%%%%%%%%%%%%%%%%%%%%%%%%%%%%%%%%%%%%%%%%%%%%%%%%%%%%%%%%%%%%%%%%%%%%%%%%%%%%%%%%%%%%%%%%%%%%%%%%%%
%%%%%%%%%%%%%%%%%%%%%%%%%%%%%%%%%%%%%%%%%%%%%%%%%%%%%%%%%%%%%%%%%%%%%%%%%%%%%%%%%%%%%%%%%%%%%%%%%%%%%%%%%%%%%%%%%%%%%%%%%%%%%%%%%%%%%%%%

 %%%%%%%%%%%%%%%%%%%%%%%%%%%%%%%%%%%%%%%%%%%%%%%%%%%%%%%%%%%%%%%%%%%%%%%%%%%%%%%%%%%%%%%%%%%%%%%%%%%%%%%%%%%%%%%%%%%%%%%%%%%%%%%%%%%%%%%%%%%%%%%%%%%%%%%%%%%%%%%%%%%%%%%%%%%%%%%%%%%%%%%%%%%%%%%%%%%%%%%%%
  \section{Solutions of the Dirac equation on \texorpdfstring{$N$}{N}-dimensional de Sitter space-time}\label{solutions_section}
 We first present the basic results from Ref.~\cite{Camporesi} regarding the eigenmodes of the Dirac operator on $S^{N}$ and then we perform analytic continuation for the two cases with $N$~even and $N$ odd. 
 
\noindent \textbf{Case 1:} $\bm{N}$ \textbf{even}. The eigenvalue equation for the Dirac operator on $S^{N}$ is 
  \begin{equation}\label{eigenvalue_Equation_for_Dirac_op_eigenvalues_explicitly_written}
      \slashed{\nabla} \psi^{(s,\tilde{s})}_{\pm n \ell \sigma}=\pm i (n+\frac{N}{2})\psi^{(s,\tilde{s})}_{\pm n \ell \sigma},
  \end{equation}
 where $n=0,1,...$ and $\ell=0,...,n$ are the angular momentum quantum numbers on $S^{N}$ and $S^{N-1}$ respectively. The index $s$ indicates the two different spin projections ($s= \pm$). The symbol $\sigma$ stands for the angular momentum quantum numbers $\ell_{N-2} \geq \ell_{N-3} \geq ... \geq \ell_{2} \geq \ell_{1} \geq 0$ on the lower-dimensional spheres while $\tilde{s}$ stands for the $(N/2-1)$ spin projection indices $s_{N-2},s_{N-4},...,s_{2}$ on the lower-dimensional spheres $S^{N-2}, S^{N-4},...,S^{2}$ respectively. (Note that there exists one spin projection index for each lower-dimensional sphere of even dimension.) For each value of $n$ we have a representation of Spin$(N+1)$ on the space of the eigenspinors $\psi^{(s,\tilde{s})}_{+n \ell \sigma}$ (or $\psi^{(s,\tilde{s})}_{-n \ell \sigma}$) with dimension~\cite{Camporesi}
 \begin{align}\label{Spin(N+1)_degeneracy}
     d_{n}=\frac{2^{[N/2]} (N+n-1)!}{n! (N-1)!}.
 \end{align} 
 
 The solutions of the eigenvalue equation for the Dirac operator on $S^{N}$~(\ref{eigenvalue_equation_Dirac}) are found by writing the spinor $\psi$ in terms of ``upper'' ($\varphi_{+}$) and ``lower'' ($\varphi_{-}$) components as follows:
 \begin{align}\label{split_spinor_in_upper_and_lower}
     \psi \equiv \begin{pmatrix}  \varphi_{+} \\ \varphi_{-}
     \end{pmatrix}.
 \end{align}
 By substituting Eq.~(\ref{split_spinor_in_upper_and_lower}) into Eq.~(\ref{eigenvalue_equation_Dirac}) one obtains two coupled differential equations for $\varphi_{+}, \varphi_{-}$. By eliminating $\varphi_{+}$ (or $\varphi_{-}$) one finds~\cite{Camporesi} 
 \begin{align}\label{itterated_Dirac_op_on_S_N_even}
  \Big[ \big( \frac{\partial}{\partial \theta_{N}}&+ \frac{N-1}{2}\cot{\theta_{N}} \big)^{2}+\frac{1}{\sin^{2}{\theta_{N}}}\tilde{\slashed{\nabla}}^{2}\pm \frac{\cos{\theta_{N}}}{\sin^{2}{\theta_{N}}}i\tilde{\slashed{\nabla}}\Big]\varphi_{\pm} \nonumber \\
  &=-\lambda^{2}\varphi_{\pm},
 \end{align}
 where $\tilde{\slashed{\nabla}}$ is the Dirac operator on $S^{N-1}$. (Equations~(\ref{itterated_Dirac_op_on_S_N_even}) are equivalent to $\slashed{\nabla}^{2} \psi =-\lambda^{2} \psi$.) Then, by separating variables, the normalized eigenspinors of $\slashed{\nabla}|_{S^{N}}$ are found to be~\cite{Camporesi}
 \begin{equation}\label{psi_minus_even_S_N}
     \psi^{(-,\tilde{s})}_{ \pm n \ell \sigma}(\theta_{N},\Omega_{{N-1}})=\frac{c_{N}(n\ell)} {\sqrt{2}} \begin{pmatrix} \phi_{n\ell}(\theta_{N})\chi^{(\tilde{s})}_{-\ell \sigma}(\Omega_{{N-1}})
 \\  \pm i \psi_{n\ell} (\theta_{N})\chi^{(\tilde{s})}_{-\ell \sigma}(\Omega_{{N-1}})\end{pmatrix}
 \end{equation}
and
\begin{equation}\label{psi_plus_even_S_N}
\psi^{(+,\tilde{s})}_{ \pm n \ell \sigma}(\theta_{N},\Omega_{{N-1}})=\frac{c_{N}(n\ell )} {\sqrt{2}} \begin{pmatrix} i\psi_{M\ell}(\theta_{N})\chi^{(\tilde{s})}_{+\ell \sigma}(\Omega_{{N-1}})
 \\  \pm \phi_{M\ell}(\theta_{N})\chi^{(\tilde{s})}_{+\ell \sigma}(\Omega_{{N-1}})\end{pmatrix},
 \end{equation}
 where $\Omega_{N-1} \in S^{N-1}$ and the normalization factor is given by
 \begin{equation}\label{S_N_normlzn_fac}
     |c_{N}(n \ell)|^{2}=\frac{\Gamma(n-\ell +1) \Gamma(n+N+\ell)}{2^{N-2}|\Gamma{(N/2+n)}|^{2}}.
 \end{equation}
The eigenspinors on $S^{N-1}$, $\chi^{(\tilde{s})}_{\pm \ell \sigma}(\Omega_{{N-1}})$, satisfy the eigenvalue equation
\begin{equation}\label{S_n_eigenvalue_equation}
\tilde{\slashed{\nabla}} \chi^{(\tilde{s})}_{\pm \ell \sigma}= \pm i ( \ell + \frac{N-1}{2}) \chi^{(\tilde{s})}_{\pm \ell \sigma}.
\end{equation}
 They are normalized by
 \begin{align} \label{S_N-1_normalzn}
     \int_{S^{N-1}} &d\Omega_{N-1} \chi^{(\tilde{s})}_{s \ell \sigma}(\Omega_{{N-1}})^{\dagger} \chi^{(\tilde{s}')}_{s' \ell' \sigma'}(\Omega_{{N-1}}) \nonumber \\
     &=\delta_{s s'}\delta_{\ell \ell'}  \delta_{\sigma \sigma'}\delta_{\tilde{s} \tilde{s}'},
 \end{align}
 while the eigenspinors on $S^{N}$ are normalized by
 \begin{align}\label{S_N_normlzn}
     \int_{S^{N}} & d\Omega_{N} \psi^{(s,\tilde{s})}_{
     \pm n \ell \sigma}(\theta_{N},\Omega_{{N-1}})^{\dagger} \psi^{(s',\tilde{s}')}_{\pm n' \ell' \sigma'}(\theta_{N},\Omega_{{N-1}}) \nonumber \\ & =\delta_{s s'}\delta_{n n'}\delta_{\ell \ell'}\delta_{\sigma \sigma'} \delta_{\tilde{s}\tilde{s}' } ,
 \end{align}
where all the $\psi_{+}$ eigenspinors are orthogonal to all the $\psi_{-}$ eigenspinors. The functions $\phi_{n\ell}(\theta_{N}),\psi_{n\ell}(\theta_{N})$ are given in terms of the Gauss hypergeometric function by
  \begin{align}\label{phi_nl}
    \phi_{n\ell}(\theta_{N})=\, &\kappa^{(N)}_{\phi}(n \ell)\, (\cos{\frac{\theta_{N}}{2}})^{\ell+1}( \sin{\frac{\theta_{N}}{2}})^{\ell} \nonumber \\ &\times  F(n+N+\ell,-n+\ell; {N}/{2}+\ell; \sin^{2}{\frac{\theta_{N}}{2}} )
\end{align}
 and
  \begin{align}\label{psi_nl}
    \psi_{n\ell}(&\theta_{N}) \nonumber \\ 
    =\,& \frac{{\kappa^{(N)}_{\phi}(n \ell)\,(n+{N}/{2})}}{ {  {N}/{2}+\ell }}  (\cos{\frac{\theta_{N}}{2}})^{\ell}( \sin{\frac{\theta_{N}}{2}})^{\ell+1} \nonumber \\ &\times  F(n+N+\ell,-n+\ell; {N}/{2}+\ell+1; \sin^{2}{\frac{\theta_{N}}{2}} ),
\end{align}
where
\begin{align}\label{scalar_functions_normalization_factors}
    \kappa^{(N)}_{\phi}(n \ell)=\frac{\Gamma{(n+{N}/{2})}}{ {\Gamma{(n-\ell+1)}   \Gamma{({N}/{2}+\ell)} }}.
\end{align}
The condition $n \geq \ell$ as well as the quantization of the eigenvalue of the Dirac operator $\lambda^{2}=(n+N/2)^{2}$ ($n=0,1,...$) arise by requiring that the mode functions are not singular \cite{Camporesi}. The functions $\phi_{n\ell},\psi_{n\ell}$ are related to each other by
  \begin{align}\label{phi->psi}
      \Big[ \frac{d}{d\theta_{N}}&+\frac{N-1}{2} \cot{\theta_{N}} -\frac{1}{\sin{\theta_{N}}}(\ell+\frac{N-1}{2}) \Big] \phi_{n \ell}(\theta_{N})\nonumber \\ &= -(n+\frac{N}{2}) \psi_{n \ell}(\theta_{N}),
 \end{align}
 \begin{align}\label{psi->phi}
      \Big[ \frac{d}{d\theta_{N}}&+\frac{N-1}{2} \cot{\theta_{N}} +\frac{1}{\sin{\theta_{N}}}(\ell+\frac{N-1}{2}) \Big] \psi_{n\ell}(\theta_{N})\nonumber \\ &= +(n+\frac{N}{2}) \phi_{n \ell}(\theta_{N}).
 \end{align}
 
 As mentioned in the Introduction, we can obtain the Dirac spinors which solve the Dirac equation $( \gamma^{a} \nabla_{a} - M) \psi = 0$
on $dS_{N}$ by analytically continuing the eigenmodes of the Dirac operator on $S^{N}$. The eigenvalues on $S^{N}$ will be replaced by the spinor's mass $M$. It is easy to check that under the replacement $\theta_{N} \rightarrow \pi/2 - it$ one finds $\slashed{\nabla}|_{S^{N}}\rightarrow \slashed{\nabla}|_{dS_{N}}$.
Without loss of generality, we choose to analytically continue the eigenspinors $\psi_{+}$ with the positive sign for the eigenvalue (see Eqs.~(\ref{psi_minus_even_S_N})-(\ref{psi_plus_even_S_N})) by making the replacements 
 \begin{equation}\label{replacements}
        \theta_{N} \rightarrow x\equiv\pi /2 -it , \hspace{10mm} n \rightarrow -i M - \frac{N}{2}.
    \end{equation}
 The solutions of the Dirac equation on $dS_{N}$ are then
 \begin{equation}\label{negative_spin_even_dS}
     \psi^{(-,\tilde{s})}_{ M\ell \sigma}(t,\Omega_{N-1})=\frac{c_{N}(M\ell)} {\sqrt{2}} \begin{pmatrix} \phi_{M\ell}(t)\chi^{(\tilde{s})}_{-\ell \sigma}(\Omega_{{N-1}})
 \\  i \psi_{M\ell} (t)\chi^{(\tilde{s})}_{-\ell \sigma}(\Omega_{{N-1}})\end{pmatrix}
 \end{equation}
 and
\begin{equation}\label{positive_spin_even_dS}
     \psi^{(+,\tilde{s})}_{ M\ell \sigma}(t,\Omega_{{N-1}})=\frac{c_{N}(M\ell )} {\sqrt{2}} \begin{pmatrix} i\psi_{M\ell}(t)\chi^{(\tilde{s})}_{+\ell \sigma}(\Omega_{{N-1}})
 \\  \phi_{M\ell}(t)\chi^{(\tilde{s})}_{+\ell \sigma}(\Omega_{{N-1}})\end{pmatrix},
 \end{equation}
where $c_{N}(M \ell)$ is a normalization factor that will be determined later ($\ell=0,1,...$). The un-normalized functions that describe the time dependence are
    \begin{align} \label{phiM_a}
\phi_{M\hspace{0.2mm}\ell}(t)=& \big(\cos\frac{x}{2}\big)^{\ell+1} \big(\sin \frac{x}{2}\big)^{\ell} \nonumber \\& \times F( \frac{N}{2}+ \ell + iM , \frac{N}{2}+ \ell - iM ; \frac{N}{2}+\ell ; \sin^{2}\frac{x}{2} ) 
  \end{align}
and
  \begin{align}\label{psiM_a}
 \psi_{M\hspace{0.2mm}\ell}(&t)=    \frac{-iM}{N/2+\ell} \big(\cos\frac{x}{2}\big)^{\ell} \big(\sin \frac{x}{2}\big)^{\ell+1}  \nonumber \\ &\times F( \frac{N}{2}+ \ell + iM , \frac{N}{2}+ \ell - iM ; \frac{N}{2}+\ell +1; \sin^{2}\frac{x}{2} ),
  \end{align}
  where 
  \begin{align}
   &\cos{\frac{x}{2}}=\frac{\sqrt{2}}{2}(\cosh{\frac{t}{2}} + i \sinh{\frac{t}{2}} )  ,  \label{cosx/2} \\
   &\sin{\frac{x}{2}}=\frac{\sqrt{2}}{2}(\cosh{\frac{t}{2}} - i \sinh{\frac{t}{2}} )  , \label{sinx/2} \\
   &\sin^{2}\frac{x}{2}=\frac{1- i \sinh{t}}{2}.
       \end{align}
It is clear from Eq.~(\ref{psiM_a}) that $\psi_{M\ell}(t)$ vanishes in the massless limit.
Note the analytically continued version of Eqs.~(\ref{phi->psi}) and (\ref{psi->phi})
    \begin{align}\label{psiM_in_terms_of_phiM_and_derivs}
         \Big( \frac{d}{dt}+& \frac{N-1}{2} \tanh{t} +\frac{i}{\cosh{t}}(\ell+\frac{N-1}{2}) \Big) \phi_{M\ell}(t)\nonumber \\&= +M\psi_{M\ell}(t) ,
     \end{align}
\begin{align}\label{phiM_in_terms_of_psiM_and_derivs}
    \Big( \frac{d}{dt}+& \frac{N-1}{2} \tanh{t} -\frac{i}{\cosh{t}}(\ell+\frac{N-1}{2}) \Big) \psi_{M\ell}(t)\nonumber \\&= -M\phi_{M\ell}(t).
\end{align}
 Using the following relation~\cite{NIST:DLMF}:
\begin{align}
    F(a,b;c;z)=(1-z)^{c-a-b}  F(c-a,c-b;c;z)
\end{align}
we can rewrite the functions $\phi_{M \ell},\psi_{M \ell}$ as
 \begin{align}\label{phiM_b}
     \phi_{M \ell}(t) = &\big(\cos\frac{x}{2}\big)^{-N-\ell+1} \big(\sin \frac{x}{2}\big)^{\ell}\nonumber \\& \times  F(  iM , - iM ; \frac{N}{2}+\ell ; \sin^{2}\frac{x}{2}  ) 
\end{align}
and
 \begin{align}\label{psiM_b}
     \psi_{M\ell}(t) =&\frac{-iM}{N/2+\ell} \big(\cos\frac{x}{2}\big)^{-N-\ell+2} \big(\sin \frac{x}{2}\big)^{\ell+1}\nonumber\\ &\times F(  iM +1, - iM +1; \frac{N}{2}+\ell+1 ; \sin^{2}\frac{x}{2} ).
 \end{align}
The short wavelength limit ($\ell \gg 1$) of these functions can be found, by noting that the hypergeometric functions here tend to $1$ in this limit, as
\begin{align}
    &\frac{d}{dt}\phi_{M\ell}(t) \sim -i \label{pos_freq_flat_spacetime_time_der_one} \frac{\ell}{\cosh{t}}\phi_{M\ell}(t), \\  &\frac{d}{dt}\psi_{M\ell}(t) \sim -i \label{pos_freq_flat_spacetime_time_der_two} \frac{\ell}{\cosh{t}}\psi_{M\ell}(t).
\end{align}
We see that the time derivative of our mode solutions (\ref{negative_spin_even_dS}) and (\ref{positive_spin_even_dS}) reproduces locally the positive frequency behaviour of flat space-time. Thus, our modes can serve as the analogs of the positive frequency modes and we can use this criterion as well as de~Sitter invariance (see Sec.~\ref{Section_Group_Theory}) in order to define a vacuum.

Note that by making the replacements (\ref{replacements}) in the expressions for the spinors $\psi_{-}$ with the negative sign for the eigenvalue on $S^{N}$ (see Eqs.~(\ref{psi_minus_even_S_N})-(\ref{psi_plus_even_S_N})), we obtain the spinors
\begin{align}\label{neg_mass_negative_spin_even_dS}
     \psi^{(-,\tilde{s})}_{ -M\ell \sigma}(t,\Omega_{N-1})= \begin{pmatrix} \phi_{M\ell}(t)\chi^{(\tilde{s})}_{-\ell \sigma}(\Omega_{{N-1}})
 \\ - i \psi_{M\ell} (t)\chi^{(\tilde{s})}_{-\ell \sigma}(\Omega_{{N-1}})\end{pmatrix}
 \end{align}and
\begin{equation}\label{neg_mass_positive_spin_even_dS}
     \psi^{(+,\tilde{s})}_{ -M\ell \sigma}(t,\Omega_{{N-1}})= \begin{pmatrix} i\psi_{M\ell}(t)\chi^{(\tilde{s})}_{+\ell \sigma}(\Omega_{{N-1}})
 \\  -\phi_{M\ell}(t)\chi^{(\tilde{s})}_{+\ell \sigma}(\Omega_{{N-1}})\end{pmatrix}.
 \end{equation}
These spinors satisfy the equation $\slashed{\nabla} \psi_{-M }=-M \psi_{-M }$ and they serve as a tool in the construction of the negative frequency solutions using charge conjugation (see Appendix~\ref{appendix_charge_conjugation}). 

%%%%%%%%%%%%%%%%%%%%%%%%%%%%%%%%%%%%%%%%%%%%%%%%%%%%%%%%%%%%%%%%%%%%%%%%%%%%%%%%%%%%%%%%%%%%%%%%%%%%%%%%%%%%%%%%%%%%%%%%%%%%%%%%%%%%%%%   

\noindent \textbf{Case 2:} $\bm{N}$ \textbf{odd}. For the construction of the eigenmodes of Eq.~(\ref{eigenvalue_equation_Dirac}) it is convenient to consider the eigenvalue equation for the iterated Dirac operator $\slashed{\nabla}^{2}\psi=-\lambda^{2}\psi$. The latter may be written as follows~\cite{Camporesi}:
\begin{align}\label{iterated_Dirac_operator_N=odd_sphere}
\Big[ \big( \frac{\partial}{\partial \theta_{N}}&+ \frac{N-1}{2}\cot{\theta_{N}} \big)^{2}+\frac{1}{\sin^{2}{\theta_{N}}}\tilde{\slashed{\nabla}}^{2}- \frac{\cos{\theta_{N}}}{\sin^{2}{\theta_{N}}}\gamma^{N}\tilde{\slashed{\nabla}}\Big]\psi \nonumber \\
  &=-\lambda^{2}\psi.
\end{align}
By separating variables, the spinor eigenfunctions of the Dirac operator on $S^{N}$ are found to be~\cite{Camporesi}
 \begin{align} \label{odd_eigenspinor}
      \psi^{(s,\tilde{s})}_{
     \pm n \ell \sigma}&(\theta_{N},\Omega_{{N-1}})= \frac{c_{N}(n \ell)}{\sqrt{2}} \nonumber \\ &\times (  \phi_{n\ell}(\theta_{N}) \hat{\chi}^{(s,\tilde{s})}_{-\ell \sigma}(\Omega_{N-1})
      \pm i  \psi_{n\ell}(\theta_{N})\hat{\chi}^{(s,\tilde{s})}_{+\ell \sigma}(\Omega_{{N-1}}) ),
 \end{align}
where 
 \begin{equation}\label{chi_hat_in_terms_of_chi}
     \hat{\chi}^{(s,\tilde{s})}_{-\ell \sigma}=\frac{1}{\sqrt{2}}(\bm{1}+i \gamma^{N})\chi^{(s,\tilde{s})}_{-\ell \sigma}
 \end{equation}
 and the eigenvalues are the same as in Eq.~(\ref{eigenvalue_Equation_for_Dirac_op_eigenvalues_explicitly_written}) (i.e. $\lambda=~ \pm (n+N/2)$ with $n=0,1,...$).
 The spinors $\chi^{(s,\tilde{s})}_{+ \ell \sigma}$ and $\hat{\chi}^{(s,\tilde{s})}_{+ \ell \sigma}$ are given by
 \begin{equation}\label{gamma_changes_sign_for_chi_minus_to_chi_plus}
   \gamma^{N}\chi^{(s,\tilde{s})}_{-\ell \sigma} = \chi^{(s,\tilde{s})}_{+\ell \sigma}
 \end{equation}
and
\begin{equation}\label{gamma_changes_sign_for_chi_hat}
    \hat{\chi}^{(s,\tilde{s})}_{+\ell \sigma}=\gamma^{N}\hat{\chi}^{(s,\tilde{s})}_{-\ell \sigma}.
\end{equation}
 Here $s$ is the spin projection index on $S^{N-1}$ and $\tilde{s}$ stands for the rest of the spin projection indices on the lower-dimensional spheres of even dimensions.
The functions $\phi_{n \ell}, \psi_{n \ell}$ are given by Eqs.~(\ref{phi_nl}) and (\ref{psi_nl}), while the spinors $\hat{\chi}^{(s,\tilde{s})}_{\pm \ell \sigma}(\Omega_{N-1})$ are eigenfunctions of the hermitian operator $\gamma^{N}\tilde{\slashed{\nabla}}$ (that commutes with the iterated Dirac operator ${\slashed{\nabla}}^{2}$) satisfying~\cite{Camporesi}
\begin{equation}\label{chi_hat_equation}
     \gamma^{N}\tilde{\slashed{\nabla}} \hat{\chi}^{(s,\tilde{s})}_{\pm \ell \sigma} = \pm (\ell+\frac{N-1}{2})   \hat{\chi}^{(s,\tilde{s})}_{\pm\ell \sigma}.
 \end{equation}

 As in the even-dimensional case, for each value of $n$  the eigenspinors $\psi^{(s,\tilde{s})}_{+n \ell \sigma}$ (or $\psi^{(s,\tilde{s})}_{-n \ell \sigma}$) form a representation of Spin$(N+1)$ with dimension $d_{n}$ given by Eq.~(\ref{Spin(N+1)_degeneracy}). (The dimension is half the dimension for the case with $N$ even because there is no contribution from spin projections on $S^{N}$.) Notice that on $S^{1}$ the Dirac operator is just $\partial / \partial \theta_{1}$ and the eigenspinors are $\chi_{\pm \ell_{1}}(\theta_{1})=\exp{(\pm i\,(\ell_{1} + 1/2)\theta_{1})}$ (the normalization constant is $(2 \pi)^{-1/2}$).
 The eigenspinors (\ref{odd_eigenspinor}) are normalized as in the case with $N$ even and the normalization factors are given again by Eq.~(\ref{S_N_normlzn_fac}).

We choose to analytically continue the $\psi_{+}$ eigenmodes. By making the replacements (\ref{replacements}) in the expression for the eigenspinors $\psi^{(s,\tilde{s})}_{+n\ell \sigma}(\theta_{N},\Omega_{N-1})$ (Eq.~(\ref{odd_eigenspinor})) we obtain the solutions of the Dirac equation on odd-dimensional $dS_{N}$
\begin{align} \label{odd_eigenspinor_dS}
      \psi_{ M \ell \sigma}^{(s,\tilde{s})}&(t,\Omega_{{N-1}})= \frac{c_{N}(M \ell)}{\sqrt{2}} \nonumber \\ &\times (  \phi_{M\ell}(t) \hat{\chi}^{(s,\tilde{s})}_{-\ell \sigma}(\Omega_{N-1})
      + i  \psi_{M\ell}(t)\hat{\chi}^{(s,\tilde{s})}_{+\ell \sigma}(\Omega_{{N-1}}) ),
 \end{align}
where the normalization factor will be determined later. The functions $\phi_{M \ell}(t), \psi_{M \ell}(t)$ are given again by Eqs.~(\ref{phiM_a}) and (\ref{psiM_a}). Hence, the solutions (\ref{odd_eigenspinor_dS}) can be used as positive frequency modes. 

As in the even-dimensional case, we can analytically continue the eigenspinors $\psi_{-}$ to obtain 
\begin{align} \label{neg_mass_odd_N_eigenspinor_dS}
      \psi^{(s,\tilde{s})}_{- M \ell \sigma}&(t,\Omega_{{N-1}}) \nonumber \\ &= (  \phi_{M\ell}(t) \hat{\chi}^{(s,\tilde{s})}_{-\ell \sigma}(\Omega_{N-1})
      - i  \psi_{M\ell}(t)\hat{\chi}^{(s,\tilde{s})}_{+\ell \sigma}(\Omega_{{N-1}}) ),
 \end{align}
 which satisfy the Dirac equation~(\ref{Dirac_equation}) with $M \rightarrow -M$.
%%%%%%%%%%%%%%%%%%%%%%%%%%%%%%%%%%%%%%%%%%%%%%%%%%%%%%%%%%%%%%%%%%%%%%%%%%%%%%%%%%%%%%%%%%%%%%%%%%%%%%%%%%%%%%%%%%%%%%%%%%%%%%%%%%%%%%%
\section{Normalization factors and Transformation properties under Spin(\texorpdfstring{$N$}{N},1) of the analytically continued eigenspinors of the Dirac Operator on the \texorpdfstring{$N$}{N}-sphere}\label{Section_Group_Theory}
For each value of $M$ the set of the analytically continued eigenspinors of the Dirac operator $\slashed{\nabla}|_{S^{N}}$ forms a representation of the Lie algebra of Spin($N$,1) (which is also a representation of the group Spin($N$,1)). If we want to use these mode functions to describe spin-1/2 particles on $N$-dimensional de~Sitter space-time, the corresponding representation has to be unitary. Unitarity ensures that no negative probabilities will arise. A representation is unitary if there is a positive definite inner product that is preserved under the action of the group.
In this section we show that the representation formed by our analytically continued eigenspinors is unitary by introducing a Spin($N$,1) invariant inner product among the solutions of the Dirac equation and by verifying the positive-definiteness of the norm associated with this inner product. In addition, we calculate the normalization factors $c_{N}({M \ell})$ and we show that our positive frequency modes transform among themselves under infinitesimal Spin($N$,1) transformations. In view of a mode expansion of the quantum Dirac field using our analytically continued modes, the transformation properties thus obtained imply that the corresponding vacuum is de~Sitter invariant.  

%%%%%%%%%%%%%%%%%%%%%%%%%%%%%%%%%%%%%%%%%%%%%%%%%%%%%%%%%%%%%%%%%%%%%%%%%%%%%%%%%%%%%%%%%%%%%%%%%%%%%%%%%%%%%%%%%%%%%%%%%%%%%%%%%%%%%%%

  \subsection{Unitarity of the Spin(\texorpdfstring{$N$}{N},1) representation and normalization factors}\label{Unitarity_of_rep_and_norm_factors}
     We define the following inner product for spinors with the same mass $M$:
\begin{align}\label{inner_prod}
        (\psi^{(s,\tilde{s})}_{ M\ell \sigma}, \psi^{(s',\tilde{s}')}_{ M\ell'\sigma'}) &= i\int d \bm{\theta} \sqrt{-g} \hspace{2mm} \overline{\psi}_{ M\ell \sigma}^{(s,\tilde{s})} \gamma^{0} \psi_{ M\ell' \sigma'}^{(s',\tilde{s}')} \\
        &= \int d \bm{\theta} \sqrt{-g} \hspace{2mm} {\psi}_{ M\ell \sigma}^{(s,\tilde{s})\dagger}  \psi_{ M\ell' \sigma'}^{(s',\tilde{s}')},
    \end{align}
    where $d\bm{\theta}$ stands for $d\theta_{1} d\theta_{2}... d\theta_{N-1}$ . The square root of the determinant of the de~Sitter metric is 
    \begin{align}\label{determinant_sqrt_dS}
        \sqrt{-g}= \cosh^{N-1}{t} \hspace{0.5mm} \sin^{N-2}{\theta_{{N-1}}} \dots \sin{\theta_{2}}=\cosh^{N-1}{t}\sqrt{\tilde{g}},
    \end{align}
    where $\tilde{g}$ is the determinant of the $S^{N-1}$ metric. First, we show that the inner product (\ref{inner_prod}) is both time independent and Spin($N$,1) invariant. Let $\psi^{(1)},\psi^{(2)}$ be two analytically continued eigenspinors which satisfy the Dirac equation (\ref{Dirac_equation}). The Dirac equation and Eq.~(\ref{gamma_constant}) imply that the vector current
    \begin{equation}\label{current}
     J^{\mu}= i\overline{\psi}^{(1)} \gamma^{\mu} \psi^{(2)}
    \end{equation}
is covariantly conserved. Hence, the inner product~(\ref{inner_prod}) is time independent. As for the invariance under Spin($N$,1), we can show that the change in the inner product due to infinitesimal Spin($N$,1) transformations vanishes (as in Ref.~\cite{STSHS}). 
Let $\xi^{\mu}$ be a Killing vector of $dS_{N}$ satisfying
\begin{equation} \label{Killing_equation}
    \nabla_{\mu}\xi_{\nu}+ \nabla_{\nu}\xi_{\mu}=0.
\end{equation}
The Lie derivative of $J^{\mu}$ with respect to the Killing vector $\xi^{\mu}$ ($\mathcal{L}_{\xi} J^{\mu}$) gives the change in $J^{\mu}$ under the corresponding transformation; that is
\begin{align}
    \delta J ^{\mu}=\mathcal{L}_{\xi} J^{\mu}&= \xi^{\nu} \nabla_{\nu} J^{\mu}- J^{\nu}  \nabla_{\nu} \xi^{\mu} \nonumber \\&=\nabla_{\nu} (  \xi^{\nu} J^{\mu}- J^{\nu}   \xi^{\mu}),
\end{align}
where we used the fact that both $J^{\mu},\xi^{\mu}$ are divergence free.
Then we find
\begin{equation}\label{delta_J_0}
      \delta J ^{0}=  \nabla_{\nu} (  \xi^{\nu}  J^{0}- J^{\nu}  \xi^{0}) = \frac{1}{\sqrt{-g}}\partial_{\theta_{\kappa}}\big[ \sqrt{-g} (  \xi^{\theta_{\kappa}}  J^{0}- J^{\theta_{\kappa}}   \xi^{0}) \big],
\end{equation}
where $\kappa={1},...,{N-1}$. By integrating Eq.~(\ref{delta_J_0}) over $S^{N-1}$ we find
\begin{equation}
  \delta(\psi^{(1)}, \psi^{(2)})=  \int d \bm{\theta} \sqrt{-g} \hspace{1mm} \delta J^{0}= 0.
\end{equation}
 Below we study the positive-definiteness of the norm associated with the inner product~(\ref{inner_prod}).

\noindent \textbf{Case 1:} $\bm{N}$ \textbf{even}. Substituting the analytically continued eigenspinors (\ref{negative_spin_even_dS}) (or (\ref{positive_spin_even_dS})) into the inner product (\ref{inner_prod}) we find 
   \begin{align}\label{inner_product_breaked_down}
        (\psi^{(s,\tilde{s})}_{ M\ell \sigma}, \psi^{(s',\tilde{s}')}_{ M\ell'\sigma'}) =&\frac{|c_{N}(M\ell)|^{2}}{2}  \cosh^{N-1}t \Big( \phi_{M\ell}^{*}(t)   \phi_{M\ell}(t)   \nonumber\\ & +  \psi_{M\ell}^{*}(t)   \psi_{M\ell}(t)      \Big) \delta_{ss'}\delta_{\tilde{s} \tilde{s}'} \delta_{\ell \ell'} \delta_{\sigma \sigma'},
    \end{align}
   where the positive-definiteness is obvious (i.e. the representation is unitary). 
    
    Using Eqs.~(\ref{psiM_in_terms_of_phiM_and_derivs}) and (\ref{phiM_in_terms_of_psiM_and_derivs}) one finds
\begin{align}
    \frac{d}{dt} \big [   \cosh^{(N-1)/2}t \hspace{1mm} \phi_{M\ell}    \big] =& -i \cosh^{({N-3})/{2}}t (\ell+\frac{N-1}{2})\nonumber \\ &\times   \phi_{M\ell}  + M \cosh^{(N-1)/2}t\hspace{1mm} \psi_{M\ell},
\end{align}
and
\begin{align}
      \frac{d}{dt} \big [   \cosh^{(N-1)/2}t  \psi_{M \ell}    \big]=& +i\cosh^{({N-3})/{2}}t(\ell+\frac{N-1}{2}) \nonumber \\ &\times \psi_{M\ell}  - M \cosh^{(N-1)/2}t\hspace{1mm} \phi_{M \ell}
\end{align}
respectively. Consequently
\begin{align}\label{C}
 \cosh^{N-1}t  \Big( \phi_{M\ell}^{*}(t)   \phi_{M\ell}(t)  +  \psi_{M\ell}^{*}(t)   \psi_{M\ell}(t)\Big)= K,
\end{align}
where $K$ is a positive real constant (since the time derivative of the left-hand side vanishes). We can determine the value of $K$ just by letting $t=0$ in Eq.~(\ref{C}). The functions (\ref{phiM_a}) and (\ref{psiM_a}) for $t=0$ are 
 \begin{equation} \label{phiM_zero}
     \phi_{M\ell}(t=0)   =\frac{\sqrt{2}}{2} \big(  \frac{1}{2} \big)^{\ell}  \hspace{1mm} F(\delta, \delta^{*},  \frac{\delta+ \delta^{*}}{2} ; \frac{1}{2}  )
    \end{equation}
and
\begin{equation}\label{psiMzero}
   \psi_{M\ell}(t=0)   =\frac{-i\sqrt{2}M}{N+2\ell} \big(  \frac{1}{2} \big)^{\ell}  \hspace{1mm} F(\delta, \delta^{*},  \frac{\delta+ \delta^{*}}{2}+1 ; \frac{1}{2}  ) 
 \end{equation}
 respectively, where $$\delta=\frac{N}{2}+\ell+ i M .$$ Using the following two formulas~\cite{wolf1}, \cite{wolf2}:
 \begin{align}\label{Fab_(a+b)/2}
   F(a, b,  \frac{a+b}{2} ; \frac{1}{2})=&\sqrt{\pi}\Gamma(\frac{a+b}{2})\big[ \frac{1}{\Gamma((a+1)/2)   \Gamma(b/2)}  \nonumber \\  &+ \frac{1}{\Gamma((b+1)/2)   \Gamma(a/2)}   \big],
\end{align}
\begin{align}\label{Fab_(a+b)/2+1}
    F(a, b,  \frac{a+b}{2}+1 ; \frac{1}{2})=&\frac{2\sqrt{\pi}}{a-b} \Gamma(\frac{a+b}{2}+1)\big[ \frac{1}{\Gamma((b+1)/2)}\nonumber \\ &\times \frac{1}{   \Gamma(a/2)}- \frac{1}{\Gamma((a+1)/2)   \Gamma(b/2)} \big]
\end{align}
we find 
\begin{align}\label{C_final}
K&=\phi_{M\ell}^{*}(0)   \phi_{M\ell}(0)  +  \psi_{M\ell}^{*}(0)   \psi_{M\ell}(0) \nonumber\\
    &=2^{N-1}   \frac{|\Gamma(\frac{N}{2}+\ell)  |^{2}}{|\Gamma(\frac{N}{2}+\ell+iM)|^{2}},
\end{align}
where we also used the Legendre duplication formula
\begin{equation}\label{Legendre_duplication_formula}
    \Gamma(2z)=\frac{1}{\sqrt{\pi}} 2^{2z-1} \Gamma(z) \Gamma(z+1/2).
\end{equation}
Since $(\psi^{(s,\tilde{s})}_{ M\ell \sigma}, \psi^{(s,\tilde{s})}_{ M\ell \sigma}) ={|c_{N}(M\ell)|^{2}}K/{2} $, it is straightforward to calculate the normalization factor as
 \begin{align}\label{dS_normalization_fac}
        |c_{N}(M\ell)|^{2} =2 ^{(2-N)}  
    \frac{|\Gamma(\frac{N}{2}+\ell+iM)|^{2}}{|\Gamma(\frac{N}{2}+\ell)  |^{2}}.
    \end{align}
Our analytically continued eigenspinors are now normalized by
\begin{equation}\label{dS_normalization}
  (\psi^{(s,\tilde{s})}_{ M\ell \sigma}, \psi^{(s',\tilde{s}')}_{ M\ell'\sigma'})=\delta_{s s'} \delta_{\tilde{s} \tilde{s}'}\delta_{\ell \ell '} \delta_{\sigma \sigma'}  .
\end{equation} 

\noindent \textbf{Case 2:} $\bm{N}$ \textbf{odd}. Substituting the analytically continued eigenspinors (\ref{odd_eigenspinor_dS}) into the inner product (\ref{inner_prod}) we obtain again Eq.~({\ref{inner_product_breaked_down}}). Thus, the Spin($N$,1) representation is unitary (due to the positive-definiteness of the norm) and the normalization is again given by Eqs.~(\ref{dS_normalization_fac}) and (\ref{dS_normalization}).
%%%%%%%%%%%%%%%%%%%%%%%%%%%%%%%%%%%%%%%%%%%%%%%%%%%%%%%%%%%%%%%%%%%%%%%%%%%%%%%%%%%%%%%%%%%%%%%%%%%%%%%%%%%%%%%%%%%%%%%%%%%%%%%%%%%%%%%%

\subsection{Transformation properties of the positive frequency solutions under Spin(\texorpdfstring{$N$}{N},1)}\label{transformation_properties_section}
In this section we use the {spinorial Lie derivative}~\cite{Kosmann} with respect to the Killing vector field $\xi$ in order to study the Spin($N$,1) transformations of the analytically continued modes of $\slashed{\nabla}|_{S^{N}}$ generated by $\xi$. More specifically, we show that our positive frequency modes transform among themselves under the action of an infinitesimal boost in the $\theta_{N-1}$ direction.

The coordinate expression for the spinorial Lie derivative of a spinor field $\psi$ with respect to the Killing vector $\xi$ is~\cite{Kosmann}
\begin{equation}\label{spinorial_Lie}
   \mathcal{L}^{s}_{\xi} \psi  =   \xi^{\mu} \nabla_{\mu} \psi + \frac{1}{4}  \nabla_{\kappa} \xi_{\lambda}  \gamma^{\kappa} \gamma^{\lambda }    \psi.
\end{equation}
 (We use the superscript $s$ to distinguish the spinorial Lie derivative from the usual Lie derivative.)
We are interested in the transformation generated by the boost Killing vector
\begin{equation}\label{Killing_vector}
    \xi=\cos{\theta_{N-1}} \frac{\partial}{\partial {t}} - \tanh{t} \hspace{0.5mm} \sin{\theta_{N-1}} \frac{\partial}{\partial\theta_{N-1}}.
\end{equation}
After a straightforward calculation we find
\begin{align}
  \mathcal{L}^{s}_{\xi} \psi  =& \xi^{\mu} \partial_{\mu} \psi  + \frac{\sin{\theta_{N-1}}}{2\cosh{t}} \gamma^{N-1} \gamma^{0} \psi\\
    = &\cos{\theta_{N-1}} \partial_{t} \psi- \tanh{t} \sin{\theta_{N-1}} \partial_{\theta_{N-1}} \psi \nonumber \\ &+ \frac{\sin{\theta_{N-1}}}{2\cosh{t}} \gamma^{N-1} \gamma^{0} \psi.\label{spinorial_Lie_final}
\end{align}
The spinorial Lie derivative with respect to Killing vectors commutes with the Dirac operator~\cite{Kosmann}. Hence if $\psi$ is an analytically continued eigenspinor of $\slashed{\nabla}|_{S^{N}}$ we can express Eq.~(\ref{spinorial_Lie_final}) as a linear combination of other such eigenspinors. In order to proceed, it is useful to introduce the ladder operators for the functions ${\phi}_{M\ell }(t),{\psi}_{M\ell }(t),\tilde{\phi}_{\ell \, \ell_{N-2}}(\theta_{N-1}),\tilde{\psi}_{\ell \, \ell_{N-2}}(\theta_{N-1})$ sending the angular momentum quantum number $\ell$ to $\ell \pm 1$. (The functions $\tilde{\phi}_{\ell\, \ell_{N-2}},\tilde{\psi}_{\ell\, \ell_{N-2}} $ are given by Eqs.~(\ref{phi_nl}) and (\ref{psi_nl}) respectively, with $N \rightarrow N-1,\, n \rightarrow \ell$ and $\ell \rightarrow \ell_{N-2}$.) The ladder operators are given by the following expressions:
\begin{align}
  T^{(+)}_{\phi }=&  \frac{d}{dt}-(\ell+\frac{1}{2}) \tanh{t} - \frac{i}{2 \cosh{t}},  \\
  T^{(+)}_{\psi }=&  \frac{d}{dt}-(\ell+\frac{1}{2}) \tanh{t} + \frac{i}{2 \cosh{t}},\\
   T^{(-)}_{\phi} =& \frac{d}{dt}+(\ell+N-\frac{3}{2}) \tanh{t} + \frac{i}{2 \cosh{t}}, \\
    T^{(-)}_{\psi} =& \frac{d}{dt}+(\ell+N-\frac{3}{2}) \tanh{t} - \frac{i}{2 \cosh{t}},
   \end{align} 
   
   \begin{align}
  \tilde{T}^{(+)}_{\tilde{\phi}} = &  \sin{\theta_{N-1}} \frac{d}{d\theta_{N-1}}+(\ell+N-\frac{3}{2})\cos{\theta_{N-1}} \nonumber \\ &- \frac{\ell_{N-2}+\frac{N-2}{2}}{2(\ell+\frac{N}{2})},   \\
   \tilde{T}^{(+)}_{\tilde{\psi}} =&  \sin{\theta_{N-1}} \frac{d}{d\theta_{N-1}}+(\ell+N-\frac{3}{2})\cos{\theta_{N-1}} \nonumber \\ &+ \frac{\ell_{N-2}+\frac{N-2}{2}}{2(\ell+\frac{N}{2})}
   \end{align}
   
   \begin{align}
    \tilde{T}^{(-)}_{\tilde{\phi}}=&  \sin{\theta_{N-1}} \frac{d}{d\theta_{N-1}}-\cos{\theta_{N-1}}(\ell+\frac{1}{2})\nonumber \\ &+\frac{\ell_{N-2}+\frac{N-2}{2}}{2(\ell+\frac{N-2}{2})} ,\\
     \tilde{T}^{(-)}_{\tilde{\psi}}=&  \sin{\theta_{N-1}} \frac{d}{d\theta_{N-1}}-\cos{\theta_{N-1}}(\ell+\frac{1}{2})\nonumber \\&-\frac{\ell_{N-2}+\frac{N-2}{2}}{2(\ell+\frac{N-2}{2})}.
  \end{align}
 The corresponding ladder relations are
  \begin{align}
      &T^{(+)}_{f} f_{M \ell}(t)= k^{(+)}f_{M \ell+1}(t), \label{raising_phi_psi}   \\
       &T^{(-)}_{f} f_{M \ell}(t)= k^{(-)}f_{M \ell-1}(t), \label{lowering_phi_psi} \\ 
        &\tilde{T}^{(+)}_{\tilde{f}} \tilde{f}_{ \ell \, \ell_{N-2}}(\theta_{N-1})= \tilde{k}^{(+)}\tilde{f}_{ \ell+1 \, \ell_{N-2}}(\theta_{N-1}), \label{raising_tilde} \\
         &\tilde{T}^{(-)}_{\tilde{f}} \tilde{f}_{ \ell \, \ell_{N-2}}(\theta_{N-1})= \tilde{k}^{(-)}\tilde{f}_{ \ell-1 \, \ell_{N-2}}(\theta_{N-1}),\label{lowering_tilde}
     \end{align}
     where $f_{M \ell}(t) \in \set{ \phi_{M \ell}(t),\psi_{M \ell}(t)}$, $\tilde{f}_{ \ell \, \ell_{N-2}}(\theta_{N-1}) \in \set { \tilde{\phi}_{\ell \,\ell_{N-2}}(\theta_{N-1}),\tilde{\psi}_{\ell \, \ell_{N-2}}(\theta_{N-1})}$ and 
     \begin{align}
        & k^{(+)}=-i \frac{(N/2+\ell)^{2}+M^{2}}{N/2+\ell}, \label{kappa_plus}\\
         & k^{(-)}=-i (N/2+\ell-1), \label{kappa_minus}\\
         & \tilde{k}^{(+)}=\frac{(\ell+N-1+\ell_{N-2})(\ell-\ell_{N-2}+1)}{(\ell+{N}/{2})  },\label{tilde_kappa_plus} \\
        &\tilde{k}^{(-)}= -\frac{(({N-1})/{2}+\ell-1)(({N-1})/{2}+\ell)  }{ ({N-2})/{2}+\ell}.\label{tilde_kappa_minus}
     \end{align}
The ladder relations (\ref{raising_phi_psi})-(\ref{lowering_tilde}) can be proved using the raising and lowering operators for the parameters of the Gauss hypergeometric function given in Appendix~\ref{appendix_raising_lowering_hypergeom}. Below we describe how to express the spinorial Lie derivative (\ref{spinorial_Lie_final}) of a mode solution $\psi_{M \ell \sigma}^{(s,\tilde{s})}$ as a linear combination of other solutions with the same $M$.

\noindent \textbf{Case 1:} $\bm{N}$ \textbf{even} $\bm{(>2)}$. Using Eq.~(\ref{even_gammas}) one finds
\begin{equation}\label{N-1_0_generators}
  \gamma^{N-1} \gamma^{0} = \begin{pmatrix}  
   -\widetilde{ \gamma}^{N-1} & 0 \\
0 & \widetilde{ \gamma}^{N-1}
    \end{pmatrix} .
\end{equation}
Let $\psi$ be the eigenspinor $ \psi^{(\pm,\tilde{s})}_{M \ell \, \ell_{N-2} \, \tilde{\sigma}}$, where $\tilde{\sigma}$ stands for quantum numbers other than $\ell, \ell_{N-2}$. Since the partial derivatives in Eq.~(\ref{spinorial_Lie_final}) refer only to the coordinates $\{t,\theta_{N-1}\}$ we want to extract the $t$ and 
 $\theta_{N-1}$ dependence from our analytically continued eigenspinors. By combining Eqs.~(\ref{negative_spin_even_dS}), (\ref{positive_spin_even_dS}) and (\ref{odd_eigenspinor}) we can express the spinors $\psi^{(\pm,\tilde{s})}_{M \ell \, \ell_{N-2} \, \tilde{\sigma}} (t,\Omega_{N-1})$ in terms of eigenspinors on $S^{N-2}$ ( $\hat{\tilde{\chi}}^{\tilde{(s)}}_{\pm \ell_{N-2} \, \tilde{\sigma}}(\Omega_{{N-2}})$) as follows:
\begin{align}\label{negative_spin_dS_in_terms_S_(N-2)}
 \psi^{(-,\tilde{s})}_{M \ell \, \ell_{N-2} \, \tilde{\sigma}}(t,\Omega_{N-1})=&\frac{c_{N}(M \ell)}{\sqrt{2}} \frac{c_{N-1}( \ell \, \ell_{N-2})}{\sqrt{2}}  \nonumber \\ &\times \begin{pmatrix}U^{(\tilde{s})}_{-M \ell \, \ell_{N-2} \, \tilde{\sigma}}(t,\theta_{N-1},\Omega_{N-2}) \\ D^{(\tilde{s})}_{-M \ell \, \ell_{N-2} \, \tilde{\sigma}}(t,\theta_{N-1},\Omega_{N-2}) \end{pmatrix}, 
\end{align}
\begin{align}\label{positive_spin_dS_in_terms_S_(N-2)}
   \psi^{(+,\tilde{s})}_{M \ell \, \ell_{N-2} \, \tilde{\sigma}}(t,\Omega_{N-1})=&\frac{c_{N}(M \ell)}{\sqrt{2}} \frac{c_{N-1}( \ell \, \ell_{N-2})}{\sqrt{2}}  \nonumber \\ &\times \begin{pmatrix}D^{(\tilde{s})}_{+M \ell \, \ell_{N-2} \, \tilde{\sigma}}(t,\theta_{N-1},\Omega_{N-2}) \\ U^{(\tilde{s})}_{+M \ell \, \ell_{N-2} \, \tilde{\sigma}}(t,\theta_{N-1},\Omega_{N-2}) \end{pmatrix},
\end{align}
where
\begin{align}\label{negative_spin_dS_in_terms_S_(N-2)_upper_comp}
  U^{(\tilde{s})}_{\mp M \ell \, \ell_{N-2} \, \tilde{\sigma}}&(t,\theta_{N-1},\Omega_{N-2}) \nonumber \\ =&\phi_{M \ell}(t) \Big( \tilde{\phi}_{\ell \, \ell_{N-2}}(\theta_{N-1})    \hat{\tilde{\chi}}^{(\tilde{s})}_{- \ell_{N-2} \, \tilde{\sigma}}(\Omega_{N-2})   \nonumber \\  &  \mp i  \tilde{\psi}_{\ell \, \ell_{N-2}}(\theta_{N-1})\hat{\tilde{\chi}}^{(\tilde{s})}_{+\ell_{N-2} \, \tilde{\sigma}}(\Omega_{{N-2}}) \Big)
\end{align}
and $ D^{(\tilde{s})}_{\mp M \ell \, \ell_{N-2} \, \tilde{\sigma}}$ is given by an analogous expression with $\phi_{M \ell}(t) \rightarrow i \psi_{M \ell}(t)$. 
By substituting Eqs.~(\ref{negative_spin_dS_in_terms_S_(N-2)})-(\ref{negative_spin_dS_in_terms_S_(N-2)_upper_comp}) into the expression for the spinorial Lie derivative~(\ref{spinorial_Lie_final}) and making use of Eqs.~(\ref{raising_phi_psi})-(\ref{tilde_kappa_minus}) we find after a lengthy calculation
 \begin{align}\label{spinor_Lie_N=even,final_result_neg_spin}
  \mathcal{L}^{s}_{\xi} \psi^{(\mp,\tilde{s})}_{M\ell  \sigma }=& R^{(N)}_{M \ell \, \ell_{N-2}} \psi^{(\mp,\tilde{s})}_{M\,\ell+1 \,\sigma } + L^{(N)}_ {M\ell \, \ell_{N-2}} \psi^{(\mp,\tilde{s})}_{M\, \ell-1 \, \sigma}\nonumber \\&+C^{(N)}_{M\ell \, \ell_{N-2} }     \psi^{(\pm,\tilde{s})}_{M\ell  \sigma},
\end{align}
where the coefficients on the right-hand side are given by the following expressions:
\begin{align}\label{Spin_Lie_coef_R}
    R^{(N)}_{M \ell \, \ell_{N-2}}& \nonumber \\
    =&\frac{c_{N}(M \ell) \, c_{N-1}( \ell \, \ell_{N-2})}{c_{N}(M \,\ell+1) \, c_{N-1}(\ell+1,\ell_{N-2}) } \frac{ k^{(+)} \tilde{k}^{(+)}}{2 (\ell + \frac{N-1}{2})}  \\ =&\frac{-i}{2} \frac{\sqrt{(\frac{N}{2}+\ell)^{2}+M^{2}}}{\frac{N}{2}+\ell} \nonumber \\ &\times \sqrt{(\ell-\ell_{N-2}+1)(\ell+\ell_{N-2}+N-1)}, 
\end{align}
\begin{align}\label{Spin_Lie_coef_L}
     L^{(N)}_{M \ell \, \ell_{N-2}}& \nonumber \\
     =&\frac{(-1)\times c_{N}(M \ell) \, c_{N-1}( \ell \, \ell_{N-2})}{c_{N}(M ,\ell-1) \, c_{N-1}(\ell-1, \ell_{N -2}) } \frac{ k^{(-)} \tilde{k}^{(-)}}{2 (\ell + \frac{N-1}{2})} \\=& -\big( R^{(N)}_{M,\, \ell-1, \ell_{N-2}} \big)^{*},
\end{align}
and
\begin{align}\label{Spin_Lie_coef_C}
      C^{(N)}_{M\ell \, \ell_{N-2} }=-i \frac{M(\ell_{N-2}+\frac{N-2}{2})       }{ 2(\ell+\frac{N-2}{2}) (\ell+\frac{N}{2})}.
\end{align}
Notice that in the last term of the linear combination in Eq.~(\ref{spinor_Lie_N=even,final_result_neg_spin}) the spin projection sign is flipped. We have checked the validity of the above results by using the de~Sitter invariance of the inner product (\ref{inner_prod}). More specifically, we have verified that $(\mathcal{L}^{s}_{\xi}\psi_{M\ell}, \psi_{M\,\ell\pm1})+(\psi_{M\ell}, \mathcal{L}^{s}_{\xi}\psi_{M\,\ell\pm1})=0$.
(Some details regarding the derivation of Eq.~(\ref{spinor_Lie_N=even,final_result_neg_spin}) can be found in Appendix~\ref{Appendix_Spinor_Lie} along with the $N=2$ case.)
It is clear from Eq.~(\ref{spinor_Lie_N=even,final_result_neg_spin}) that our positive frequency solutions transform to other positive frequency solutions with the same $M$ under the transformation generated by $\xi$. Based on this observation we can conclude that the vacuum corresponding to these positive frequency modes is de~Sitter invariant (see, e.g. Refs.~\cite{Waldbook} and \cite{Higuchi_1991}).

\noindent \textbf{Case 2:} $\bm{N}$ \textbf{odd}. Using Eq.~(\ref{odd_gammas}) we find
\begin{equation}\label{N-1_0_generators_odd}
  \gamma^{N-1} \gamma^{0} = i\begin{pmatrix} 
  0 & -\bm{1} \\
\bm{1} & 0
    \end{pmatrix} .
\end{equation}
As in the case with $N$ even, it is convenient to express the analytically continued eigenspinors $\psi_{M \ell \, \ell_{N-2} \,\tilde{\sigma}}^{(s ,\tilde{s})}(t, \Omega_{N-1})$ (Eq.~(\ref{odd_eigenspinor_dS})) in terms of eigenspinors on $S^{N-2}$ ($\tilde{\chi}^{(\tilde{s})}_{\pm \ell_{N-2} \, \tilde{\sigma}}(\Omega_{N-2})$). By combining Eqs.~(\ref{chi_hat_in_terms_of_chi}), (\ref{psi_minus_even_S_N}) and (\ref{psi_plus_even_S_N}), we can rewrite Eq.~(\ref{odd_eigenspinor_dS}) as
\begin{align}\label{odd_psi_minus_dS_in_terms_S_(N-2)}
    \psi^{(-,\tilde{s})}_{M \ell \, \ell_{N-2} \, \tilde{\sigma}}(t &,\theta_{N-1},\Omega_{N-2})=\frac{c_{N}(M \ell)}{\sqrt{2}} \frac{c_{N-1}( \ell  \, \ell_{N-2})}{\sqrt{2}} \frac{1}{\sqrt{2}}\nonumber \\&\times
    \begin{pmatrix}  (1+i)\tilde{\phi}_{\ell \, \ell_{N-2}} [\phi_{M\ell}+ i \psi_{M\ell}] \tilde{\chi}^{(\tilde{s})}_{-\ell_{N-2} \tilde{\sigma}}  \\  (-1+i)i\tilde{\psi}_{\ell \, \ell_{N-2}}[\phi_{M\ell}- i \psi_{M\ell}] \tilde{\chi}^{(\tilde{s})}_{-\ell_{N-2} \tilde{\sigma}} \end{pmatrix}
\end{align}
and
\begin{align}\label{odd_psi_plus_dS_in_terms_S_(N-2)}
     \psi^{(+,\tilde{s})}_{M \ell \, \ell_{N-2} \, \tilde{\sigma}}(t &,\theta_{N-1},\Omega_{N-2})=\frac{c_{N}(M \ell)}{\sqrt{2}} \frac{c_{N-1}( \ell  \, \ell_{N-2})}{\sqrt{2}} \frac{1}{\sqrt{2}}\nonumber \\&\times
    \begin{pmatrix}  (1+i)i\tilde{\psi}_{\ell \, \ell_{N-2}} [\phi_{M\ell}+ i \psi_{M\ell}] \tilde{\chi}^{(\tilde{s})}_{+\ell_{N-2} \tilde{\sigma}}  \\  (-1+i)\tilde{\phi}_{\ell \, \ell_{N-2}}[\phi_{M\ell}- i \psi_{M\ell}] \tilde{\chi}^{(\tilde{s})}_{+\ell_{N-2} \tilde{\sigma}} \end{pmatrix}.
\end{align}
Working as in the case with $N$ even, we find after a lengthy calculation
\begin{align}\label{spinor_Lie_N=odd,final_result_neg_spin}
  \mathcal{L}^{s}_{\xi} \psi^{(\mp,\tilde{s})}_{M\ell  \sigma }=& R^{(N)}_{M \ell \, \ell_{N-2}} \psi^{(\mp,\tilde{s})}_{M\,\ell+1 \,\sigma } + L^{(N)}_ {M\ell \, \ell_{N-2}} \psi^{(\mp,\tilde{s})}_{M\, \ell-1 \, \sigma}\nonumber \\&+C^{(N)}_{M\ell \, \ell_{N-2} }     \psi^{(\mp,\tilde{s})}_{M\ell  \sigma}.
\end{align}
Notice that, unlike the even-dimensional case, the two spin projections do not mix with each other. As in the case with $N$ even, we conclude that the vacuum is de~Sitter invariant.
%Though, contrary to the even-dimensional case, for $N=$ odd the %two different spin projections do not mix under the infinitesimal %transformation generated by $\xi$.

%%%%%%%%%%%%%%%%%%%%%%%%%%%%%%%%%%%%%%%%%%%%%%%%%%%%%%%%%%%%%%%%%%%%%%%%%%%%%%%%%%%%%%%%%%%%%%%%%%%%%%%%%%%%%%%%%%%%%%%%%%%%%%%%%%%%%%%5
\section{Canonical quantization}
In this section we follow the canonical quantization procedure and give the mode expansion for the free quantum Dirac field on $N$-dimensional de~Sitter space-time with ($N-1$)-sphere spatial sections using the analytically continued spinor modes of $\slashed{\nabla}|_{S^{N}}$. As mentioned earlier, our analytically continued eigenspinors can be used as the analogs of the flat space-time positive frequency modes. However, the latter are not the only solutions of the Dirac equation~(\ref{Dirac_equation}) on $dS_{N}$. New solutions (i.e. the negative frequency modes) can be obtained by separating variables. Below we present the negative frequency solutions before proceeding to the canonical quantization. (Note that the negative frequency solutions can also be obtained using charge conjugation as demonstrated in Appendix~\ref{appendix_charge_conjugation}.) 
\subsection{Negative frequency solutions}
\noindent \textbf{Case 1:} $\bm{N}$ \textbf{even}. 
By making the replacements~(\ref{replacements}) in the expression for the iterated Dirac operator on $S^{N}$~(\ref{itterated_Dirac_op_on_S_N_even}) one finds
 \begin{align}\label{itterated_Dirac_op_on_dS_even}
  \Big[ \big( \frac{\partial}{\partial t}&+ \frac{N-1}{2}\tanh{t} \big)^{2}-\frac{1}{\cosh^{2}{t}}\tilde{\slashed{\nabla}}^{2}\pm \frac{\sinh{t}}{\cosh^{2}{t}}\tilde{\slashed{\nabla}}\Big]\varphi_{\pm} \nonumber \\
  &=-M^{2}\varphi_{\pm}.
 \end{align}
Then by separating variables (as in Ref.~\cite{Camporesi}) one finds the negative frequency solutions
\begin{equation}\label{neg_freq_dS_even_neg_spin}
     V^{(-,\tilde{s})}_{M\ell \sigma }(t,\Omega_{N-1})=\frac{c_{N}(M \ell)}{\sqrt{2}}\begin{pmatrix} \phi^{*}_{M\ell}(t)\chi^{(\tilde{s})}_{+\ell \sigma }(\Omega_{{N-1}})
 \\  i \psi^{*}_{M\ell} (t)\chi^{(\tilde{s})}_{+\ell \sigma }(\Omega_{{N-1}})\end{pmatrix},
    \end{equation}
    \begin{equation}\label{neg_freq_dS_even_positive_spin}
      V^{(+,\tilde{s})}_{M\ell \sigma }(t,\Omega_{N-1})=\frac{c_{N}(M \ell)}{\sqrt{2}}\begin{pmatrix} i\psi^{*}_{M\ell}(t)\chi^{(\tilde{s})}_{-\ell \sigma }(\Omega_{N-1})
 \\   \phi^{*}_{M\ell} (t)\chi^{(\tilde{s})}_{-\ell \sigma }(\Omega_{N-1})\end{pmatrix}.
    \end{equation}
 These are normalized using the inner product (\ref{inner_prod}) as
    \begin{align}\label{neg_freq_normalization}
       & ( V^{(s,\tilde{s})}_{M\ell \sigma  },V^{(s',\tilde{s}')}_{M\ell' \sigma'  })=\delta_{ss'}\delta_{\tilde{s} \tilde{s}'} \delta_{\ell \ell'} \delta_{\sigma \sigma'} 
    \end{align}
and they are orthogonal to the positive frequency solutions, i.e.
    \begin{align}\label{neg_freq_pos_freq_inner_prod}
      & ( \psi^{(s,\tilde{s})}_{M\ell \sigma  },V^{(s', \tilde{s}')}_{M\ell' \sigma'  })=0 .
    \end{align}
    As we can see, the negative frequency modes are given by the positive frequency solutions (\ref{negative_spin_even_dS}) and (\ref{positive_spin_even_dS}) by replacing the functions $\phi_{M \ell}(t),\psi_{M \ell}(t)$ with their complex conjugate functions and by exchanging $\chi_{\pm  }(\Omega_{{N-1}})$ and $\chi_{\mp  }(\Omega_{{N-1}})$. 
    The time derivatives of the spinors (\ref{neg_freq_dS_even_neg_spin})-(\ref{neg_freq_dS_even_positive_spin}) reproduce the flat space-time behaviour in the large $\ell$ limit, i.e. the complex conjugate of Eqs.~(\ref{pos_freq_flat_spacetime_time_der_one}) and (\ref{pos_freq_flat_spacetime_time_der_two}). 
    
    \noindent \textbf{Case 2:} $\bm{N}$ \textbf{odd}. Working as in the even-dimensional case the negative frequency modes are found to be
     \begin{align} \label{negative_freq_odd_N_eigenspinor_dS}
      V^{(s,\tilde{s})}_{ M \ell \sigma }&(t,\Omega_{{N-1}})= \frac{c_{N}(M \ell)}{\sqrt{2}} (  \phi^{*}_{M\ell}(t) \nonumber \\ &\times \hat{\chi}^{(s,\tilde{s})}_{+\ell \sigma }(\Omega_{N-1})
      + i  \psi^{*}_{M\ell}(t)\hat{\chi}^{(s,\tilde{s})}_{-\ell \sigma}(\Omega_{{N-1}}) )
 \end{align}
     and they satisfy the conditions (\ref{neg_freq_normalization}) and (\ref{neg_freq_pos_freq_inner_prod}). 
 %%%%%%%%%%%%%%%%%%%%%%%%%%%%%%%%%%%%%%%%%%%%%%%%%%%%%%%%%%%%%%%% 
  \subsection{Canonical Quantization} 
  The Lagrangian density for a free spinor field $\Psi$ is
   \begin{align}
       \pazocal{L}&=\sqrt{-g}\hspace{1mm}\overline{\Psi} \Big(\gamma^{\mu}\nabla_{\mu}-M \Big) \Psi \label{spinor_lagrangian_1}\\
       &=\sqrt{-g}\hspace{1mm}i{\Psi}_{A}^{\dagger} (\gamma^{0})^{A}_{\hspace{3mm}B} \Big((\gamma^{\mu})^{B}_{\hspace{3mm}C}(\nabla_{\mu}\Psi)^{C}-M \Psi^{B}\Big) \label{spinor_lagrngian_2},
   \end{align}
   where we have written out the spinor indices explicitly in the second line ($A,B,C=1,...,2^{[N/2]}$). The corresponding equation of motion for $\Psi$ is the Dirac equation~(\ref{Dirac_equation}). By the standard canonical quantization procedure we find
     \begin{align}
  & \{ \Psi(t, \bm{\theta})^{A} , \Psi^{\dagger}(t, \bm{\theta}')_{B} \} = \frac{1}{\sqrt{-g(t,\bm{\theta})}} \delta^{(N-1)}{(\bm{\theta}- \bm{\theta}')} \delta^{A}_{\hspace{3mm}B},\label{anticommutation_relations_psi_psidagger} \\
  & \{ \Psi(t, \bm{\theta})^{A} , \Psi(t, \bm{\theta}')^{B} \} = \{ \Psi^{\dagger}(t, \bm{\theta})_{A} , \Psi^{\dagger}(t, \bm{\theta}')_{B} \}=0\label{anticommutation_relations_psi_psi}.
   \end{align}
 The mode expansion for the free Dirac field is 
   \begin{equation}\label{mode_expansion}
   {\Psi}(t,\bm{\theta})=\sum_{\ell,{\sigma}}\sum_{s,\tilde{s}} \Big( {a}_{M\ell {\sigma}}^{(s,\tilde{s})}\psi^{(s, \tilde{s})}_{M \ell  {\sigma}}(t,\bm{\theta}) +{b}_{M\ell  {\sigma}}^{(s,\tilde{s})\dagger}V^{(s, \tilde{s})}_{M \ell  {\sigma}}(t,\bm{\theta}) \Big) ,
\end{equation}
where we are summing over all angular momentum quantum numbers and over all the possible spin projections. (There are 
${[N/2]}$ spin projection indices in total.) Using the normalization conditions (\ref{dS_normalization}), (\ref{neg_freq_normalization}) and the orthogonality condition (\ref{neg_freq_pos_freq_inner_prod}) we may express the annihilation operators, $a^{(s,\tilde{s})}_{M\ell \sigma}$ and $b^{(s,\tilde{s})}_{M\ell \sigma}$, as
\begin{align}
    a_{M\ell  \sigma}^{(s,\tilde{s})}&=(\psi^{(s,\tilde{s})}_{M \ell \sigma}{(t,\bm{\theta})}, \Psi{(t, \bm{\theta)}}  )\nonumber \\
    &=\int d\bm{\theta} \sqrt{-g}\, \psi^{(s,\tilde{s})}_{M \ell \sigma }{(t,\bm{\theta})^{\dagger}}  \hspace{1mm}\Psi{(t, \bm{\theta)}} \label{particle_annihilation}
\end{align}
and
\begin{align}
 b_{M\ell \sigma}^{(s,\tilde{s})} =\int d\bm{\theta} \sqrt{-g}   \Psi^{\dagger}{(t, \bm{\theta)}}\hspace{1mm}V^{(s,\tilde{s})}_{M \ell \sigma }{(t,\bm{\theta})} \label{antiparticle_annihilation}.
\end{align}
By combining Eqs.~(\ref{particle_annihilation})-(\ref{antiparticle_annihilation}) with the anti-commutation relations~(\ref{anticommutation_relations_psi_psidagger}) and (\ref{anticommutation_relations_psi_psi}) we obtain
\begin{align}\label{anticommutation_relations_for_cr_ann_operators}
   & \{a^{(s,\tilde{s})}_{M\ell \sigma} , a^{(s' ,\tilde{s}')\dagger}_{M\ell' \sigma'} \}=\delta_{ss'} \delta_{\tilde{s} \tilde{s}'} \delta_{\ell \ell '} \delta_{\sigma \sigma'} ,\\
    & \{b^{(s,\tilde{s})}_{M\ell \sigma} , b^{(s',\tilde{s}')\dagger}_{M\ell' \sigma'} \}=\delta_{ss'}\delta_{\tilde{s} \tilde{s}'} \delta_{\ell \ell '} \delta_{\sigma \sigma'} ,
\end{align}
while all the other anti-commutators are zero. The de~Sitter invariant vacuum is defined by 
\begin{equation}
  a^{(s,\tilde{s})}_{M\ell \sigma} \ket{0}=  b^{(s,\tilde{s})}_{M\ell \sigma} \ket{0}=0,
\end{equation}
for all $\ell, \sigma, (s, \tilde{s})$.
Using the mode expansion of the Dirac field (\ref{mode_expansion}) we can obtain the mode-sum form for the Wightman two-point function
\begin{align}
    W \big( (t,\bm{\theta}), (t',\bm{\theta}') \big) &\equiv \bra{0} \Psi(t,\bm{\theta})\overline{\Psi}(t',\bm{\theta}') \ket{0} \label{Wightman_definition}\\
    &= \sum_{\ell, \sigma} \sum_{s,\tilde{s}} \psi^{(s,\tilde{s})}_{M \ell \sigma}(t,\bm{\theta}) \overline{\psi}^{(s,\tilde{s})}_{M \ell \sigma}(t',\bm{\theta}').\label{Wightman_mode_sum}
\end{align}
The high frequency behaviour of our mode solutions (\ref{pos_freq_flat_spacetime_time_der_one})-(\ref{pos_freq_flat_spacetime_time_der_two}) implies that we should adopt the $-i \epsilon$ prescription (i.e. the time variable $t$ should be understood to have an infinitesimal negative imaginary part: $t \rightarrow t - i \epsilon$, $ \epsilon >0$).

%%%%%%%%%%%%%%%%%%%%%%%%%%%%%%%%%%%%%%%%%%%%%%%%%%%%%%%%%%%%%%%%%%%%%%%%%%%%%%%%%%%%%%%%%%%%%%%%%%%%%%%%%%%%%%%%%%%%%%%%%%%%%%%%%%%%%%%%%%%%%%%%%%%%%%%%%%%%%%%%%%%%%%%%%%%%%%%%%%%%%%%%%%%%%%%%%%%%%%%%
\section{The Wightman two-point function}\label{Section_2_point_function}
In this section we first review the basics about the construction of Dirac spinor Green's functions on $dS_{N}$ using intrinsic geometric objects following the work of M\"{u}ck~\cite{Muck:1999mh}. (M\"{u}ck gave the coordinate independent construction of the  spinor Green's function in terms of intrinsic geometric objects on maximally symmetric spaces of arbitrary dimensions using Dirac spinors. An analogous construction on 4-dimensional maximally symmetric spaces using two-component spinors was first presented in Ref.~\cite{Allen_spinors}.)
Then using the mode-sum method (\ref{Wightman_mode_sum}) we obtain a closed-form expression for the massless spinor Wightman two-point function on $dS_{N}$ that agrees with the construction presented in Ref.~\cite{Muck:1999mh}. Using this massless two-point function we infer the analytic expression for the spinor parallel propagator and then obtain the massive spinor Wightman two-point function in a closed form.

\subsection{The spinor parallel propagator on \texorpdfstring{$dS_{N}$}{dSN} }
Let $\ket{\psi}$ be a state invariant under the action of the de~Sitter group. Then two-point functions (such as $\bra{\psi} \Psi(x) \overline{\Psi}(x') \ket{\psi}$) define maximally symmetric bispinors~\cite{allen1986}. These bispinors can be expressed in terms of the following ``preferred geometric objects'': the  geodesic distance (\ref{spacelike_sep}), the unit tangent vectors (\ref{tangent_vectors}) to the geodesic with endpoints $x,x'$ and the bispinor of parallel transport $\Lambda(x,x')$, also known as the spinor parallel propagator~\cite{Muck:1999mh, Clutton_Brock_1975}.  
The spinor parallel propagator parallel transports a spinor $\psi(x')$ from $x'$ to $x$ along the (shortest) geodesic joining these points, i.e.  
\begin{align}\label{parallel_transport_of_spinors}
    \psi_{||}(x)^{A} = \Lambda(x,x')^{A}_{\hspace{3mm} A'}\, \psi(x')^{A'},
\end{align}
where $ \psi_{||}(x)$ is the parallelly transported spinor. The following relations can be used as the defining properties of the spinor parallel propagator for arbitrary space-time dimension~\cite{Muck:1999mh}
\begin{align} 
      &\Lambda(x',x)=  [\Lambda(x,x')]^{-1},\label{defining_properties_1} \\  &\gamma^{\nu'}(x')=\Lambda(x',x) \gamma^{\mu}(x) g^{\nu'}_{\hspace{3mm}\mu}(x',x) \Lambda(x,x'), \label{defining_properties_2}\\
       &n^{\mu} \nabla_{\mu}\Lambda(x,x')=0, \label{defining_properties_3}
   \end{align} 
where the parallel transport equation (\ref{defining_properties_3}) holds along the geodesic connecting $x$ and $x'$. Equation~(\ref{defining_properties_2}) describes the parallel transport of gamma matrices. In Appendix~\ref{appendix_defining_test} we show that our result for the spinor parallel propagator (given by Eq.~(\ref{spinor_parallel_propagator_N_even})) is consistent with the defining properties~(\ref{defining_properties_1})-(\ref{defining_properties_3}). 
 On $dS_{N}$ the covariant derivatives of $\Lambda(x.x')$ can be expressed as~\cite{Muck:1999mh}
\begin{align}
    &\nabla_{\mu} \Lambda (x,x')=-\frac{1}{2}\tan(\frac{\mu}{2}) ( \gamma_{\mu} \slashed{n} - n_{\mu})\Lambda(x,x'), \\
   & \nabla_{\mu'} \Lambda (x,x')=\frac{1}{2}\tan(\frac{\mu}{2}) \Lambda(x,x')( \gamma_{\mu'} \slashed{n}' - n_{\mu'}),
\end{align}
where $\slashed{n} \equiv \gamma^{\mu}(x) n_{\mu}(x,x')$ and $ \slashed{n}' \equiv \gamma^{\mu'}(x') n_{\mu'}(x,x')$. Note that $\slashed{n}^{2}=\bm{1}$ and $(\slashed{n}')^{2}=\bm{1}'$, where $\bm{1}, \bm{1}'$ are the identity spinor matrices at $x$ and $x'$, respectively. 
%%%%%%%%%%%%%%%%%%%%%%%%%%%%%%%%%%%%%%%%%%%%%%%%%%%%%%%%%%%%%%%%%%%%%%%%%%%%%%%%%%%%%%%%%%%%%%%%%%%%%%%%%%%%%%%%%%%%%%%%%%%%%%%%%%%
\subsection{Constructing spinor Green's function on \texorpdfstring{$dS_{N}$}{dSN} using intrinsic geometric objects}
\noindent \textbf{The massive case}.
The massive spinor Green's function $S_{M}(x,x')$ on $dS_{N}$ satisfies the inhomogeneous Dirac equation
\begin{equation}\label{inhomogeneous_Dirac_equation}
  [ \big( \slashed{\nabla} -M \big)  S_{M}(x,x') ]^{A}_{\hspace{3mm}A'}=\frac{\delta^{(N)}(x-x')}{\sqrt{-g(x)}} \delta^{A}_{\hspace{3mm}A'}.
\end{equation}
The Green's function $S_{M}(x,x')$ can be expressed in terms of intrinsic geometric objects as follows~\cite{Muck:1999mh}:
\begin{equation}\label{Ansatz}
    S_{M}(x,x')=(\alpha_{M}(\mu) + \beta_{M}(\mu) \slashed{n}) \Lambda(x,x'),
\end{equation}
where $\alpha_{M}(\mu) , \beta_{M}(\mu)$ are scalar functions of the geodesic distance. By requiring that $S_{M}(x,x')$ in Eq.~(\ref{Ansatz}) satisfies Eq.~(\ref{inhomogeneous_Dirac_equation}) we find the following system of ordinary differential equations for $\alpha_{M}(\mu) ,\beta_{M}(\mu) $:
\begin{align}
&\frac{d\alpha_{M}}{d\mu}  -\frac{N-1}{2}\tan\frac{\mu}{2}\,\alpha_{M} -M \beta_{M} =0,\label{system_of_dif_equations_alpha} \\
&\frac{d \beta_{M}}{d\mu}  +\frac{N-1}{2}\cot\frac{\mu}{2}\,\beta_{M} -M \alpha_{M}=\frac{\delta(x-x')}{\sqrt{-g(x)}}.\label{system_of_dif_equations_beta}
\end{align}
Using the variable $z=\cos^{2}{(\mu/2)}$ (see Eq.~(\ref{globally_defined_geodesic_distance})) this system of equations is solved by~\cite{Muck:1999mh}
\begin{align}\label{alpha_massive}
    \alpha_{M}(z)=-M &\frac{| \Gamma(\frac{N}{2}+iM) |^{2}}{\Gamma(\frac{N}{2}+1)(2\pi)^{N/2}2^{N/2}} \sqrt{z}\nonumber \\
   & \times F(\frac{N}{2}-iM,\frac{N}{2}+iM ;\frac{N}{2}+1;z)
\end{align}
and
\begin{align}\label{beta_massive_in_terms_of_alpha}
    \beta_{M}(z)= -\frac{\sqrt{1-z}}{M}\, [\sqrt{z}\, \frac{d}{dz}+\frac{N-1}{2 \sqrt{z}} \,] \alpha_{M}(z) .
\end{align}
 Using Eqs.~(\ref{alpha_massive}) and (\ref{hyper_lower_c}) we can rewrite Eq.~(\ref{beta_massive_in_terms_of_alpha}) as
\begin{align}\label{beta_massive}
  \beta_{M}(z)= & \frac{| \Gamma(\frac{N}{2}+iM) |^{2}}{\Gamma(\frac{N}{2}+1)(2\pi)^{N/2}\,2^{N/2}} \nonumber \\ 
  &\times  \sqrt{1-z} \,\frac{N}{2} F(\frac{N}{2}-iM, \frac{N}{2}+iM; \frac{N}{2};z).
\end{align}
(Note that there is a misprint in the corresponding equation for $\beta_{M}(z)$ - equation (29) - in Ref.~\cite{Muck:1999mh}. Equation~(\ref{beta_massive}) of the present paper and equation~(29) of Ref.~\cite{Muck:1999mh} agree with each other after inserting a missing prefactor.)
The proportionality constant for $\alpha_{M}(\mu)$ (and hence for $\beta_{M}(\mu)$) has been determined by requiring that the singularity in Eq.~(\ref{alpha_massive}) for $\mu \rightarrow 0$ has the same strength as the singularity of the flat space-time Green's function~\cite{Muck:1999mh}. This ensures that the spinor Green's function (\ref{Ansatz}) has the desired short-distance behaviour. (Note that since $\slashed{n},\,\alpha_{M}$ and $\beta_{M}$ are known, the only remaining step for obtaining an explicit expression for the two-point function (\ref{Ansatz}) is to derive an analytic expression for the spinor parallel propagator.)

\noindent \textbf{The massless case}. Letting $M=0$ in Eqs.~(\ref{alpha_massive}) and (\ref{beta_massive}) we find
\begin{align}
    &\alpha_{0}(z)=0,\label{alpha_massless} \\
    & \beta_{0}(z)=\frac{\Gamma(N/2)}{2^{ N/2}(2\pi)^{{N}/{2}}} \frac{1}{(1-z)^{(N-1)/2}},\label{beta_massless}
\end{align}
($z=\cos^{2}(\mu/2)$) where we used Eq.~(\ref{Fabb}). These are just the solutions (with the appropriate singularity strength) of the decoupled system
\begin{align}
&\frac{d\alpha_{0}}{d\mu}  -\frac{N-1}{2}\tan\frac{\mu}{2}\,\alpha_{0} =0, \label{decoupled_system_of_dif_equations_alpha}\\
&\frac{d \beta_{0}}{d\mu}  +\frac{N-1}{2}\cot\frac{\mu}{2}\,\beta_{0} =\frac{\delta(x-x')}{\sqrt{-g(x)}}.\label{decoupled_system_of_dif_equations_beta}
\end{align}
The massless Green's function is then given as follows:
\begin{align}
    S_{0}(x,x')&= \beta_{0}(z) \slashed{n} \Lambda(x,x')\label{Ansatz_massless} \\
    &=\frac{\Gamma(N/2)}{2^{ N/2}(2\pi)^{{N}/{2}}} {(1-z)^{-(N-1)/2}}\slashed{n} \Lambda(x,x').\label{Ansatz_massless_with_prop_const}
\end{align}
We find that the defining properties of $\Lambda(x,x')$ (Eqs.~(\ref{defining_properties_1})-(\ref{defining_properties_3})) translate to the following properties for the massless Green's function:
\begin{align}
&[S_{0}(x,x')]^{-1} \slashed{n}=\frac{1}{\beta_{0}^{2}}\slashed{n}' S_{0}(x',x), \label{defining_property_Green_1} \\
  & [S_{0}(x,x') ]^{-1}=-\frac{1}{\beta^{2}_{0}} S_{0}(x',x), \label{defining_property_Green_2}\\
&(n^{\mu} \nabla_{\mu} + \frac{N-1}{2} \cot\frac{\mu}{2} \,)S_{0}(x,x')=0. \label{defining_property_Green_3}
  \end{align}
Note that by combining Eqs.~(\ref{defining_property_Green_1}) and (\ref{defining_property_Green_2}) one obtains 
\begin{align}\label{defining_property_Green_new}
    \slashed{n}S_{0}(x,x')=-S_{0}(x,x')\slashed{n}'.
\end{align}
or equivalently 
\begin{align}\label{parallel_transport_of_slashed_n}
  [ \Lambda(x,x')]^{-1} \,\slashed{n}\, \Lambda(x,x')=- \slashed{n}'.   
\end{align}
This equation conveniently describes the parallel transport property of $\slashed{n}$. 
%%%%%%%%%%%%%%%%%%%%%%%%%%%%%%%%%%%%%%%%%%%%%%%%%%%%%%%%%%%%%%%%%%%%%%%%%%%%%%%%%%%%%%%%%%%%%%%%%%%%%%%%%%%%%%%%%%%%%%%%%%%%%%%%%%%%%
\subsection{Analytic expressions for the massless and massive Wightman two-point function and the spinor parallel propagator}\label{subsection_analytic_expressions_Wightman_and_propagator}
In the massive case the mode-sum approach for the Wightman function~(\ref{Wightman_mode_sum}) leads to complicated series involving products of hypergeometric functions and it seems that their corresponding closed-form expressions do not exist in the literature. Fortunately, the situation is simpler in the massless case and we can obtain a closed-form expression for the Wightman two-point function. This directly results in the knowledge of the spinor parallel propagator $\Lambda(x,x')$ due to Eq.~(\ref{Ansatz_massless}). The spinor parallel propagator $\Lambda(x,x')$ in turn can be used to obtain an analytic expression for the massive spinor Wightman two-point function via Eq.~(\ref{Ansatz}).

Below we present the closed-form expression we have obtained by the mode-sum method for the massless Wightman two-point function in agreement with Eq.~(\ref{Ansatz_massless}). We present the details of the lengthy calculation in Appendix~\ref{appendix_2pt_function_calculations} (as well as the result for the $N=2$ case).

\noindent \textbf{Case 1:} $\bm{N}$ \textbf{even} $\bm{(N>2)}$.
By letting $M=0$ in Eqs.~(\ref{negative_spin_even_dS})-(\ref{positive_spin_even_dS}) we obtain the massless positive frequency modes 
\begin{equation}\label{massless_negative_spin_even_dS}
     \psi^{(-,\tilde{s})}_{ 0\ell \sigma}(t,\Omega_{N-1})=\frac{2 ^{(2-N)/2}} {\sqrt{2}} \phi_{0\ell}(t)\begin{pmatrix} \chi^{(\tilde{s})}_{-\ell \sigma}(\Omega_{{N-1}})
 \\ 0\end{pmatrix},
 \end{equation}
\begin{equation}\label{massless_positive_spin_even_dS}
     \psi^{(+,\tilde{s})}_{ 0\ell \sigma}(t,\Omega_{{N-1}})=\frac{2 ^{(2-N)/2}} {\sqrt{2}}  \phi_{0\ell}(t)\begin{pmatrix}0
 \\ \chi^{(\tilde{s})}_{+\ell \sigma}(\Omega_{{N-1}})\end{pmatrix}.
 \end{equation}
Now the function describing the time dependence has the following form:
\begin{align}\label{scalar_massless}
     \phi_{0 \ell}(t) =\frac{(\tan{\frac{x}{2}})^{\ell}}{(\cos\frac{x}{2})^{N-1}},
\end{align}
where $\cos(x/2)$ is given in Eq.~(\ref{cosx/2}) and
\begin{align}
    \tan{\frac{x}{2}}=\frac{1-i \sinh{t}}{\cosh{t}}.
\end{align}
Exploiting the rotational symmetry of $S^{N-1}$ we may let $\theta_{N-1}'=\theta_{N-2}'=...=\theta_{2}'=\theta_{1}'=0$ in the mode-sum~(\ref{Wightman_mode_sum}). After a long calculation we obtain the following $2^{N/2}$-dimensional bispinorial matrix:
\begin{align}\label{massless_wightman_2point_function_mode_sum_N_even}
    W_{0}[ &(t,\bm{\theta}),(t',\bm{0}) ] \nonumber \\
    & =(\beta_{0}{(\mu)} \slashed{n} )|_{\bm{\theta}'=\bm{0}}\,\, \exp{[\,\frac{\lambda(t,\theta_{N-1},t')}{2} \gamma^{0}\gamma^{N-1}\,] } \nonumber \\
    & \times \prod_{j=2}^{N-1} \exp{[\,\frac{\theta_{N-j}}{2} \gamma^{N-j+1}\gamma^{N-j}\,] },
\end{align}
where 
\begin{align}\label{slashed_n_theta_i'=0}
   \slashed{n}&|_{\bm{\theta}'=0} \nonumber \\
  & =  \gamma^{0}n_{0}{[(t,\bm{\theta}),(t',\bm{0})]} +  \gamma^{N-1}\,n_{N-1}{[(t,\bm{\theta}),(t',\bm{0})]}
\end{align}
(see Eqs.~(\ref{orthonormal_basis_components_tangent_vectors_n0})-(\ref{orthonormal_basis_components_tangent_vectors_n_N-1})). By comparing Eq.~(\ref{massless_wightman_2point_function_mode_sum_N_even}) with Eq.~(\ref{Ansatz_massless}) we find 
\begin{align}\label{spinor_parallel_propagator_N_even}
    \Lambda\Big( (t,\bm{\theta}),(t',\bm{0})\Big)&= \exp{[\,\frac{\lambda(t,\theta_{N-1},t')}{2} \gamma^{0}\gamma^{N-1}\,] } \nonumber \\
    & \times \prod_{j=2}^{N-1} \exp{[\,\frac{\theta_{N-j}}{2} \gamma^{N-j+1}\gamma^{N-j}\,] }.
    \end{align}
 The biscalar $\lambda(t,\theta_{N-1},t')$ is defined by the following relations:
    \begin{align}\label{lambda_biscalar_parameter_parallel_transport}
     & \cosh{ \frac{\lambda}{2}}=\frac{w_{+}n_{+}+w_{-}n_{-}}{2i\sin{(\mu /2)}}=\frac{w_{1}n_{0}+w_{2}n_{N-1}}{\sin{(\mu/2)}}, \nonumber \\
     & \sinh{ \frac{\lambda}{2}}=\frac{w_{+}n_{+}-w_{-}n_{-}}{2i\sin{(\mu /2)}}=\frac{w_{1}n_{N-1}+w_{2}n_{0}}{\sin{(\mu/2)}},
    \end{align}
    where $n_{0} \equiv  n_{0}[(t,\bm{\theta}),(t',\bm{0})], n_{N-1} \equiv n_{N-1}[(t,\bm{\theta}),(t',\bm{0})]$ and
\begin{align}
   & w_{1}(t,\theta_{N-1},t')=\sinh{\frac{t-t'}{2} } \,\cos{\frac{\theta_{N-1}}{2}}, \label{2-point-func-scalars_w1}\\
    & w_{2}(t,\theta_{N-1},t')=\cosh{\frac{t+t'}{2} } \,\sin{\frac{\theta_{N-1}}{2}},\label{2-point-func-scalars_w2} \\
 & w_{\pm} (t,\theta_{N-1},t')\equiv i [  w_{1}(t,t',\theta_{N-1}) \pm w_{2}(t,\theta_{N-1},t') ],\label{2-point-func-scalars_wpm} \\
 &n_{\pm}\equiv n_{0} \pm n_{N-1}\label{2-point-func-scalars_n_pm}.
\end{align}
(This definition of $\lambda$ is motivated naturally in the mode-sum construction of the massless Wightman function given in Appendix~\ref{appendix_2pt_function_calculations}.) It is worth mentioning that the biscalar functions $w_{+}$ and $w_{-}$ satisfy $w_{+} w_{-}=\sin^{2}{(\mu/2)}$, i.e. $\beta_{0}{(\mu)} \propto (w_{+}w_{-})^{-(N-1)/2}$ (see Eqs.~(\ref{geodesic_distance_dS}) and (\ref{beta_massless})). We have verified that Eqs.~(\ref{lambda_biscalar_parameter_parallel_transport}) are consistent with the relation $ \cosh^{2}{ ({\lambda}/{2})} - \sinh^{2}{ ({\lambda}/{2})}=1$. 
    
    It is natural that the spinor parallel propagator (\ref{spinor_parallel_propagator_N_even}) is given by a product of $N-1$  matrices $\in$ Spin($N-1,1$); these correspond to one boost and $N-2$ rotations (see Appendix~\ref{appendix_2pt_function_calculations}). 
    
    As mentioned earlier, we do not follow the mode-sum method for the construction of the massive Wightman function. A closed-form expression for the latter can be found using our result for the spinor parallel propagator (\ref{spinor_parallel_propagator_N_even}). To be specific, by substituting Eq.~(\ref{spinor_parallel_propagator_N_even}) into Eq.~(\ref{Ansatz}) one can straightforwardly obtain an analytic expression for the massive Wightman function (with $x=(t,\bm{\theta})$ and $x'=(t',\bm{0})$) in terms of intrinsic geometric objects. In Appendix~\ref{appendix_conjecture} we compare the mode-sum form of the massive Wightman function with timelike separated points, $x=(t,\bm{0})$ and $x'=(t',\bm{0})$, with the expression coming from Eq.~(\ref{Ansatz}) with $\mu=i(t-t')$. Based on this comparison we make a conjecture for the closed-form expression of a series containing the Gauss hypergeometric function.

   \noindent \textbf{Case 2:} $\bm{N}$ \textbf{odd}. The massless positive frequency solutions (\ref{odd_eigenspinor_dS}) are given by
    \begin{align}\label{massless_positive_freq_N=odd}
        \psi^{(s,\tilde{s})}_{ 0 \ell \sigma}(t,\Omega_{{N-1}})= \frac{2 ^{(2-N)/2}} {\sqrt{2}} \,  \phi_{0\ell}(t)\, \hat{\chi}^{(s,\tilde{s})}_{-\ell \sigma}(\Omega_{N-1}).
    \end{align}
    Working as in the even-dimensional case we obtain again Eqs.~(\ref{massless_wightman_2point_function_mode_sum_N_even}) and (\ref{spinor_parallel_propagator_N_even}) (where $\gamma^{0}$ is given by Eq.~(\ref{odd_gammas})) and then we can construct the massive two-point function using Eq.~(\ref{Ansatz}).
    
%%%%%%%%%%%%%%%%%%%%%%%%%%%%%%%%%%%%%%%%%%%%%%%%%%%%%%%%%%%%%%%%%%%
%%%%%%%%%%%%%%%%%%%%%%%%%%%%%%%%%%%%%%%%%%%%%%%%%%%%%%%%%%%%%%%%%%%%%%%%%%%%%%%%%%%%%%%%%%%%%%%%%%%%%%%%%%%%%%%%%%%%%%%%%%%%%%%%%%%%%%%%%%%%%%%%%%%%%%%%%%%%%%%%%%%%%%%%%%%%%%%%%%%%%%%%%%%%%%%%%%%%%%%%%%%%%%%%%%%%%%%%%%%%%%%%%%%%%%%%%%%%%%%%%%%%%%%%%%%%%%%%%%%%%%%%%%%%%%%%%%%%%%%%%%%%%%%%%%%%%%%%
\section{Summary and conclusions}
\label{summary and conclusions}
In this paper we analytically continued the eigenspinors of the Dirac operator on $S^{N}$ to obtain solutions to the Dirac equation on $dS_{N}$ that serve as analogs of the positive frequency modes of flat space-time. Our generalised positive frequency solutions were used to define a vacuum for the free Dirac field. The negative frequency solutions were also constructed. The de~Sitter invariance of the vacuum was demonstrated by showing that the positive frequency solutions transform among themselves under infinitesimal Spin($N$,1) transformations. 

In order to check the validity of our mode functions, the Wightman function for massless spinors was calculated using the mode-sum method and it was expressed in a form that is in agreement with the  construction in terms of intrinsic geometric objects ($\mu, \slashed{n},\Lambda$) given in Ref.~\cite{Muck:1999mh}. An analytic expression for the spinor parallel propagator was found. This expression was tested using the defining properties of the spinor parallel propagator as presented in Ref.~\cite{Muck:1999mh} (see Appendix~\ref{appendix_defining_test}). Note that it has been checked that the spinor Green's functions expressed in terms of $\mu,\slashed{n}, \Lambda$ have Minkowskian singularity strength in the limit $\mu \rightarrow 0$~\cite{Muck:1999mh}.  Thus, the conditions for the unique vacuum~\cite{KAY199149} are satisfied by the vacuum for the free massless Dirac field defined in this paper.

Although we did not obtain a closed-form expression for the massive spinor Wightman function by the mode-sum method using our analytically continued eigenspinors, we constructed it in terms of intrinsic geometric objects. Since the short-distance behaviour has been checked in Ref.~\cite{Muck:1999mh}, the requirements for a preferred vacuum are again satisfied. The mode-sum method and the geometric construction of Ref.~\cite{Muck:1999mh} should give the same result for the massive Wightman function. This observation leads to the series conjecture of Appendix~\ref{appendix_conjecture}.

    %%%%%%%%%%%%%%%%%%%%%%%%%%%%%%%%%%%%%%%%%%%%%%%%%%%%%%%%%%%%%%%%%%%%%%%%%%%%%%%%%%%%%%%%%%%%%%%%%%%%%%%%%%%%%%%%%%%%%%%%%%%%%
    \acknowledgements 
    The author is grateful to Atsushi Higuchi for guidance, encouragement and useful discussions. He also thanks Wolfgang M\"{u}ck for communications. Subsection~\ref{Unitarity_of_rep_and_norm_factors} was part of the author's MSc thesis at Imperial College London. This work was supported by a studentship from the Department of Mathematics, University of York.
    %%%%%%%%%%%%%%%%%%%%%%%%%%%%%%%%%%%%%%%%%%%%%%%%%%%%%%%%%%%%%%%%%%%%%%%%%%%%%%%%%%%%%%%%%%%%%%%%%%%%%%%%%%%%%%%%%%%%%%%%%%%%%%%%%%%%%%%%%%%%%%%%%%%%%%%%%%%%%%%%%%%%%%%%%%%%%%%%%%%%%%%%%%
%%%%%%%%%%%%%%%%%%%%%%%%%%%%%%%%%%%%%%%%%%%%%%%%%%%%%%%%%%%%%%%%%%%
\appendix

\section{Charge conjugation and negative frequency solutions} \label{appendix_charge_conjugation}
In this Appendix we demonstrate how the negative frequency solutions given by Eqs.~(\ref{neg_freq_dS_even_neg_spin})-(\ref{neg_freq_dS_even_positive_spin}) and (\ref{negative_freq_odd_N_eigenspinor_dS}) are constructed by charge conjugating our analytically continued eigenspinors. First, let us review charge conjugation for Dirac spinors on $dS_{N}$ and on spheres following Ref.~\cite{Tanii}. For convenience, our discussion will be based on the unitary matrices $B_{\pm}$ that relate the gamma matrices to their complex conjugate matrices by similarity transformations, i.e.
\begin{equation} \label{Beta_eqn}
     (\gamma^{a})^{*}= B_{+} \gamma^{a} B_{+}^{-1},\hspace{5mm}   -(\gamma^{a})^{*}= B_{-}\gamma^{a} B_{-}^{-1},
 \end{equation}
 and not in terms of the conventional charge conjugation matrices $C_{\pm}$ that relate $\gamma^{a}$ to $(\gamma^{a})^{T}$. These two ways of defining charge conjugation are equivalent~\cite{Tanii}. From this point we will refer to the matrices $B_{\pm}$ as the charge conjugation matrices. (We should note that the representation we use for the gamma matrices~(\ref{even_gammas}), (\ref{odd_gammas}) is different from the one used in Ref.~\cite{Tanii}. Also, note that charge conjugation matrices are defined up to a phase factor and that $\gamma^{N} \equiv -i \gamma^{0}$.)

\subsection{Charge conjugation on \texorpdfstring{$N$}{N}-dimensional de Sitter space-time and on spheres}
For convenience, let us work in $d=\tau+s$ dimensions, with $\tau \in \set{0,1}$ being the number of timelike dimensions and $s$ being the number of spacelike dimensions.

For $d$ even dimensions there are both $B_{+}$ and $B_{-}$. For $d$ odd dimensions we can use one of the matrices from the $(d-1)$-dimensional case~\cite{Tanii}. (As it will be clear in the next subsections, one needs to modify the charge conjugation matrix on $dS_{d-1}$ before using it on $dS_{d}$. This is not the case in Ref.~\cite{Tanii}, because a different representation for $\gamma^{a}$'s is used.) More specifically, on odd-dimensional spaces with Lorentzian (Euclidean) metric signature there is only $B_{+}$ ($B_{-}$) for $[d/2]$ odd and only $B_{-}$ ($B_{+}$) for $[d/2]$ even. (See Refs.~\cite{Tanii} and \cite{Freedman} for more details.)
 
 Let $\Psi$ be a $2^{[d/2]}$-dimensional Dirac spinor transforming under Spin($s,\tau$).
Its charge conjugated spinor is defined with either one of the following two ways:
\begin{equation}\label{definition_of_charge_conj_spinors_dS}
    \Psi^{C_{+}} := B_{+}^{-1} \Psi^{*} \hspace{4mm}\text{or}\hspace{4mm} \Psi^{C_{-}} := B_{-}^{-1} \Psi^{*}.
\end{equation}
Suppose now that $\Psi_{\pm}$ is an eigenspinor of the Dirac operator with eigenvalue $\kappa^{\pm}_{(\tau,s)}$, i.e.
\begin{align}\label{Dirac_eqn_for_charge_conjugation}
  \slashed{\nabla}_{(\tau,s)} \Psi_{\pm}= \kappa^{\pm}_{(\tau,s)}\Psi_{\pm} ,
\end{align}
 where $\slashed{\nabla}_{(1,N-1)} \equiv \slashed{\nabla}|_{dS_{N}}$ is the Dirac operator on $dS_{N}$ with $\kappa^{\pm}_{(1,N-1)}\equiv \pm M$ and $\slashed{\nabla}_{(0,N-1)} \equiv \tilde{\slashed{\nabla}}$ is the Dirac operator on $S^{N-1}$ with $\kappa^{\pm}_{(0,N-1)}\equiv  \pm i(\ell+(N-1)/2)$. The charge conjugated counterparts of the eigenspinors of the Dirac operator are also eigenspinors. This can be understood as follows: taking the complex conjugate of Eq.~(\ref{Dirac_eqn_for_charge_conjugation}) and using Eqs.~(\ref{Beta_eqn}) and (\ref{definition_of_charge_conj_spinors_dS}) we find
 \begin{align}
   &\slashed{\nabla}_{(\tau,s)}  \Psi_{\pm}^{C_{+}}=+(\kappa^{\pm}_{(\tau,s)})^{*}\,\Psi_{\pm}^{C_{+}},\label{charge_conjuageted_Dirac_eqn_+}\\
   &\slashed{\nabla}_{(\tau,s)}  \Psi_{\pm}^{C_{-}}=-(\kappa^{\pm}_{(\tau,s)})^{*}\,\Psi_{\pm}^{C_{-}},\label{charge_conjuageted_Dirac_eqn-}
 \end{align}
where we also used $(\Sigma^{ab})^{*}=B_{\pm} \Sigma^{ab}B_{\pm}^{-1}$. It is clear from Eqs.~(\ref{charge_conjuageted_Dirac_eqn_+})-(\ref{charge_conjuageted_Dirac_eqn-}) that performing charge conjugation with $B_{-}$ changes the sign of the mass term on $dS_{N}$. Also, Eqs.~(\ref{charge_conjuageted_Dirac_eqn_+})-(\ref{charge_conjuageted_Dirac_eqn-}) imply the following relations for the eigenspinors of the Dirac operator on $S^{n}$ (with $\Psi_{\pm}=\chi^{(\tilde{s})}_{\pm \ell_{n} \sigma}$ and $\kappa^{\pm}_{(0,n)}=\pm i(\ell_{n} +n/2)$):
\begin{equation}\label{charge_conjugated_spinors_sphere}
    (\chi^{(\tilde{s})}_{\pm \ell_{n} \sigma})^{C_{-}}\propto \chi^{(\tilde{s}')}_{\pm \ell_{n} \sigma},\hspace{4mm}(\chi^{(\tilde{s})}_{\pm \ell_{n} \sigma})^{C_{+}}\propto \chi^{(\tilde{s}')}_{\mp \ell_{n} \sigma},
\end{equation}
where $n$ is arbitrary, $\sigma$ stands for angular momentum quantum numbers other than $\ell_{n}$ and $\tilde{s}$ represents the $[n/2]$ spin projection indices that correspond to this eigenspinor. The label $\tilde{s}'$ is no necessarily equal to $\tilde{s}$.

 Below we use the tilde notation for quantities defined on $S^{N-1}$.
  
%%%%%%%%%%%%%%%%%%%%%%%%%%%%%%%%%%%%%%%%%%%%%%%%%%%%%%%%%%%%%%%%%%%%%%%%%%%%%%%%%%%%%%%%%%%%%%%%%%%%%%%%%%%%%%%%%%%%%%%%%%%%%%%%%%%%%%%%%%%%%%%%%%%%
\subsection{Negative frequency solutions for \texorpdfstring{$N$}{N} even}
\noindent \textbf{Case 1:} $\bm{N/2}$ \textbf{even}.
The charge conjugation matrices $B_{\pm}$, satisfying Eq.~(\ref{Beta_eqn}) on $dS_{N}$, are given by the following products of gamma matrices:
\begin{align}
  & B_{+}= {\gamma}^{1}\prod_{ r=1 }^{(N-4)/4}{\gamma}^{4r}{\gamma}^{4r+1}, \label{B+_N=even-N/2=even} \\
 & B_{-}= {\gamma}^{0}\prod_{ r=1 }^{N/4}{\gamma}^{4r-2}{\gamma}^{4r-1}\label{B-_N=even-N/2=even}.
\end{align}
On the odd-dimensional spatial part $S^{N-1}$ there is only $\tilde{B}_{-}$ since $[(N-1)/2]$ is odd.
This is given by
\begin{align}
   \tilde{B}_{-}= \tilde{\gamma}^{1}\prod_{ r=1 }^{(N-4)/4}\tilde{\gamma}^{4r}\tilde{\gamma}^{4r+1}.
\end{align}
For convenience, we choose to define charge conjugation using $B_{+}$, which preserves the sign of the mass term in the Dirac equation. Using the representation~(\ref{even_gammas}) for the gamma matrices we can express $B_{+}$ as follows:
\begin{align}\label{even_dS_N/2_even_Bplus_in_terms_of_Bminus(N-2)}
B_{+}=\begin{pmatrix} 0 && i\tilde{B}_{-}\\ -i\tilde{B}_{-}&& 0 \end{pmatrix}.
\end{align}
 The charge conjugated counterparts of the positive frequency solutions $\psi^{(-,\tilde{s})}_{M \ell \sigma}$ (Eq.~(\ref{negative_spin_even_dS})) can be constructed using Eqs.~(\ref{charge_conjugated_spinors_sphere}) and (\ref{even_dS_N/2_even_Bplus_in_terms_of_Bminus(N-2)}). Then we have (omitting the normalization factors)
    \begin{align}\label{construction_of_neg_freq_N_even_N/2_even}
(\psi^{(-,\tilde{s})}_{M\ell \sigma}(t,\Omega_{N-1}))^{C_{+}}=&(-i)\begin{pmatrix} i\psi^{*}_{M \ell}(t) (\chi^{(\tilde{s})}_{- \ell \sigma}(\Omega_{N-1}))^{\tilde{C}_{-}} \\ \phi^{*}_{M \ell}(t)
(\chi^{(\tilde{s})}_{- \ell\sigma}(\Omega_{N-1}))^{\tilde{C}_{-}}\end{pmatrix} \nonumber  \\
& \propto
\begin{pmatrix} i\psi^{*}_{M \ell}(t) \chi^{(\tilde{s}')}_{- \ell \sigma}(\Omega_{N-1}) \\ \phi^{*}_{M \ell}(t) \chi^{(\tilde{s}')}_{- \ell\sigma}(\Omega_{N-1})\end{pmatrix}.
\end{align}
After normalizing these modes we find the negative frequency solutions~(\ref{neg_freq_dS_even_positive_spin}). Similarly, starting from the positive frequency solutions $\psi^{(+,\tilde{s})}_{M \ell \sigma}$ (Eq.~(\ref{positive_spin_even_dS})) we find the negative frequency modes~(\ref{neg_freq_dS_even_neg_spin}).
%%%%%%%%%%%%%%%%%%%%%%%%%%%%%%%%%%%%%%%%%%%%
%%%%%%%%%%%%%%%%%%%%%%%%%%%%%%%%%%%%%%%%%%%%

\noindent \textbf{Case 2:} $\bm{N/2}$ \textbf{odd}.
The charge conjugation matrices on $dS_{N}$ are given by
\begin{align}
  & B_{+}= {\gamma}^{0}{\gamma}^{1}\prod_{r=1 }^{(N-2)/4}{\gamma}^{4r}{\gamma}^{4r+1}, \label{B+_N=even-N/2=odd}  \\
 & B_{-}=\bm{1}\times \prod_{ r=1 }^{(N-2)/4}{\gamma}^{4r-2}{\gamma}^{4r-1}.\label{B-_N=even-N/2=odd} 
\end{align}
Since $[(N-1)/2]$ is even, the only charge conjugation matrix on $S^{N-1}$ is $\tilde{B}_{+}$. The matrices $B_{-}$ and $\tilde{B}_{+}$ are related to each other as follows:
\begin{align}
  B_{-}=&\prod_{ r=1 }^{(N-2)/4}\begin{pmatrix}
   {\tilde{\gamma}}^{4r-2}\tilde{{\gamma}}^{4r-1} && 0\\
  0 && {\tilde{\gamma}}^{4r-2} \tilde{{\gamma}}^{4r-1}
  \end{pmatrix}\\
  =&\begin{pmatrix}\tilde{B}_{+} && 0 \\ 0 && \tilde{B}_{+} \end{pmatrix}  .
\end{align}
In order to construct the negative frequency solutions it is convenient to use the charge conjugation matrix $B_{-}$ that flips the sign of the mass term in the Dirac equation and the ``negative mass'' spinors $\psi^{(s,\tilde{s})}_{-M \ell \sigma}$ (Eqs.~(\ref{neg_mass_negative_spin_even_dS})-(\ref{neg_mass_positive_spin_even_dS})). 
Then, by working as in the case with $N/2$ even, we obtain the negative frequency solutions~(\ref{neg_freq_dS_even_neg_spin})-(\ref{neg_freq_dS_even_positive_spin}) (with $ V^{(-,\tilde{s}')}_{M\ell \sigma} \equiv (\psi^{(-,\tilde{s})}_{-M\ell \sigma})^{C_{-}}$ and $V^{(+,\tilde{s}')}_{M\ell \sigma}  \equiv(\psi^{(+,\tilde{s})}_{-M\ell \sigma})^{C_{-}} $).
%%%%%%%%%%%%%%%%%%%%%%%%%%%%%%%%%%%%%%%%%%%%%%%%%%%%%%%%%%%%%%%%%%%
%%%%%%%%%%%%%%%%%%%%%%%%%%%%%%%%%%%%%%%%%%%%%%%%%%%%%%%%%%%%%%%%%%%%%%%%%%%%%%%%%%%%%%%%%%%%%%%%%%%%%%%%%%%%%%%%%%%%%%%%%%%%%%%%%%%%%%
\subsection{Negative frequency solutions for \texorpdfstring{$N$}{N} odd}
\noindent \textbf{Case 1:} $\bm{[N/2]}$ \textbf{even}.
The only charge conjugation matrix on $dS_{N}$ is $B_{-}$, which changes the sign of the mass term of the Dirac equation. It is given by 
\begin{align}
 & B_{-}= {\gamma}^{0}\prod_{ r=1 }^{(N-1)/4}{\gamma}^{4r-2}{\gamma}^{4r-1}\label{B-_N=odd-[N/2]=even}.
\end{align}
 Note that this is the matrix~(\ref{B-_N=even-N/2=even}) with $N \rightarrow N-1$, where now $\gamma^{0}$ is given by Eq.~(\ref{odd_gammas}). Then Eq.~(\ref{B-_N=odd-[N/2]=even}) may be expressed in terms of the charge conjugation matrix on $S^{N-1}$ as
\begin{align}
 & B_{-}= i{\gamma}^{N} \tilde{B}_{+}=\tilde{B}_{+} i\gamma^{N} .
\end{align}
By performing charge conjugation for the spinors $\psi^{(\tilde{s}_{N-1})}_{-M \ell \sigma}$ (Eq.~(\ref{neg_mass_odd_N_eigenspinor_dS})) we find
         \begin{align}\label{charge_conjugated_negative_mass_spinor_N=odd_[N/2]=even}
             (\psi^{(\tilde{s}_{N-1})}_{- M \ell \sigma }&(t,\Omega_{N-1}) )^{C_{-}} \nonumber\\
             =&-i \Big[ \phi^{*}_{M\ell}(t) \gamma^{N}\left( \hat{\chi}^{(\tilde{s}_{N-1})}_{-\ell \sigma}(\Omega_{N-1})\right)^{\tilde{C}_{+}}
     \nonumber\\  &+ i\psi^{*}_{M\ell}(t)\gamma^{N} \left(\hat{\chi}^{(\tilde{s}_{N-1})}_{+\ell \sigma }(\Omega_{N-1})\right)^{\tilde{C}_{+}} \Big],
         \end{align}
         where $\tilde{s}_{N-1}$ represents the spin projection indices $s_{N-1},s_{N-3},...,s_{4},s_{2}$ on the lower-dimensional spheres and the charge conjugated counterparts of the ``hatted" spinors can be found using Eqs.~(\ref{chi_hat_in_terms_of_chi})-(\ref{gamma_changes_sign_for_chi_hat}) and Eq.~(\ref{charge_conjugated_spinors_sphere}). More specifically, by introducing the proportionality constant $c$, such that $(\chi^{(\tilde{s}_{N-1})}_{-\ell \sigma })^{\tilde{C}_{+}}=c \chi^{(\tilde{s}_{N-1}')}_{+\ell \sigma } $ we find
         \begin{align}
             (\hat{\chi}^{(\tilde{s}_{N-1})}_{\pm \ell \sigma })^{\tilde{C}_{+}}= -i\,c\hat{\chi}^{(\tilde{s}_{N-1}')}_{\pm \ell \sigma }.
         \end{align}
By substituting this equation into Eq.~(\ref{charge_conjugated_negative_mass_spinor_N=odd_[N/2]=even}) we obtain the negative frequency solution~(\ref{negative_freq_odd_N_eigenspinor_dS}).
         
         \noindent \textbf{Case 2:} $\bm{[N/2]}$ \textbf{odd}. The only charge conjugation matrix on $dS_{N}$ is $B_{+}$. This is given by
         \begin{align}
               B_{+}=& {\gamma}^{0}{\gamma}^{1}\prod_{r=1 }^{(N-3)/4}{\gamma}^{4r}{\gamma}^{4r+1}, \label{B+_N=odd-[N/2]=odd}\\
              =&{\gamma}^{0}\tilde{B}_{-}=-   \tilde{B}_{-}   {\gamma}^{0}.\label{charge_conj_matrix_N=odd_B+_in_terms_of_Btilde_minus_[N/2]=odd}
         \end{align}
         As in the case with $[N/2]$ even, we introduce the proportionality constant $m$, such that $(\chi^{(\tilde{s}_{N-1})}_{-\ell \sigma })^{\tilde{C}_{-}}=m \chi^{(\tilde{s}_{N-1}')}_{-\ell \sigma } $, and we find
         \begin{align}
             (\hat{\chi}^{(\tilde{s}_{N-1})}_{\pm \ell \sigma })^{\tilde{C}_{-}}= \mp m \hat{\chi}^{(\tilde{s}_{N-1}')}_{\pm \ell \sigma }.
         \end{align}
         Then we use the matrix~(\ref{charge_conj_matrix_N=odd_B+_in_terms_of_Btilde_minus_[N/2]=odd}) in order to find the charge conjugate of the spinors $\psi^{(\tilde{s}_{N-1})}_{M \ell \sigma}$ (Eq.~(\ref{odd_eigenspinor_dS})) and working as in the previous case we obtain the negative frequency solution~(\ref{negative_freq_odd_N_eigenspinor_dS}).

%%%%%%%%%%%%%%%%%%%%%%%%%%%%%%%%%%%%%%%%%%%%%%%%%%%%%%%%%%%%%%%%%%%%%%%%%%%%%%%%%%%%%%%%%%%%%%%%%%%%%%%%%%%%%%%%%%%%%%%%%%%%%%%%%%%%%%

\section{Some raising and lowering operators for the parameters of the Gauss hypergeometric function}\label{appendix_raising_lowering_hypergeom}
The Gauss hypergeometric function $F(a,b;c;z)$ satisfies~\cite{NIST:DLMF}
\begin{align}
&\frac{d}{dz}F(a,b;c;z)= \frac{ab}{c}F(a+1,b+1;c+1;z), \label{hyper_raise_all}\\
&(z \frac{d}{dz}+c-1 ) F(a,b;c;z)=(c-1) F(a,b;c-1;z), \label{hyper_lower_c}\\
&(z \frac{d}{dz}+a ) F(a,b;c;z)= a F(a+1,b;c;z). \label{hyper_raise_a}
\end{align}
By combining Eq.~(\ref{hyper_raise_a}) with the following relation~\cite{hyper}:
\begin{align}
   (c-b)& F(a+1,b-1;c;z)+  (b-a-1)(1-z) \nonumber \\& \times F(a+1,b;c;z)
   = (c-a-1) F(a,b;c;z),
\end{align}
we find
\begin{align}
    \Big( a(b-c)&+a(-b+a+1)z-(-b+a+1)z(1-z)\frac{d}{dz}\Big)\nonumber \\ &\times   F(a,b;c;z)= a (b-c) F(a+1,b-1;c;z) \label{hyper_raise_a_lower_b}.
\end{align}
Using Eqs.~(\ref{hyper_raise_all}) and (\ref{hyper_lower_c}) we can show the ladder relations (\ref{raising_phi_psi}) and (\ref{lowering_phi_psi}), while using Eq.~(\ref{hyper_raise_a_lower_b}) we can show the ladder relations (\ref{raising_tilde}) and (\ref{lowering_tilde}).
%%%%%%%%%%%%%%%%%%%%%%%%%%%%%%%%%%%%%%%%%%%%%%%%%%%%%%%%%%%%%%%%%%%%%%%%%%%%%%%%%%%%%%%%%%%%%%%%%%%%%%%%%%%%%%%%%%%%%%%%%%%%%%%%%%%%%%

\begin{widetext}
\section{Transformation properties of the positive frequency solutions under Spin(\texorpdfstring{$N$}{N},1)} \label{Appendix_Spinor_Lie}
\subsection{Transformation properties for \texorpdfstring{$N>2$}{N>2}; some details for the derivation of Eq.~(\ref{spinor_Lie_N=even,final_result_neg_spin})}
Here, we present some details for the derivation of Eq.~(\ref{spinor_Lie_N=even,final_result_neg_spin}) that expresses the spinorial Lie derivative~(\ref{spinorial_Lie_final}) of the analytically continued eigenspinors~(\ref{negative_spin_even_dS})-(\ref{positive_spin_even_dS}) as a linear combination of solutions of the Dirac equation. The case with $N$ odd (i.e. Eq.~(\ref{spinor_Lie_N=odd,final_result_neg_spin})) can be proved similarly and its derivation is not presented.

In order to obtain Eq.~(\ref{spinor_Lie_N=even,final_result_neg_spin}) it is useful to introduce the following relations (where $\theta \equiv \theta_{N-1}$):
\begin{align} \label{useful_ladder_relation_phiM_tildephi}
    \xi^{\mu}\partial_{\mu}( \phi_{M\ell}{(t)} \tilde{\phi}_{\ell \, \ell_{N-2}}{(\theta)}) +i \frac{\phi_{M\ell}(t)}{2 \cosh{t}}\sin{\theta}\, \tilde{\psi}_{\ell \, \ell_{N-2}}{(\theta)}=&\frac{1}{2(\ell + \frac{N-1}{2})}  \Big(  T^{(+)}_{\phi} \times \tilde{T}^{(+)}_{\tilde{\phi}} -  T^{(-)}_{\phi} \times  \tilde{T}^{(-)}_{\tilde{\phi}}      \Big) \phi_{M\ell}{(t)} \tilde{\phi}_{\ell \, \ell_{N-2}}({\theta})\nonumber \\ 
   & +\frac{M(\ell_{N-2}+\frac{N-2}{2})}{2( \ell + \frac{N}{2}) (  \ell +\frac{ N-2}{2})} \psi_{M\ell}{(t)} \tilde{\phi}_{\ell \, \ell_{N-2} }{(\theta)},
\end{align}
%%%%%%%%%%%%%%%%%%%%%%%%%%%%%%%%%%%%%%%%%%%%%%%%%%%%%%%%%%%%%%%%%%
\begin{align} \label{useful_ladder_relation_phiM_tildepsi}
     \xi^{\mu}\partial_{\mu}( \phi_{M\ell}{(t)} \tilde{\psi}_{\ell \, \ell_{N-2}}{(\theta)}) -i \frac{\phi_{M\ell}{(\theta)}}{2 \cosh{t}}\sin{\theta}\, \tilde{\phi}_{\ell \, \ell_{N-2}}{(\theta)} = &\frac{1}{2(\ell + \frac{N-1}{2})}  \Big(  T^{(+)}_{\phi} \times \tilde{T}^{(+)}_{\tilde{\psi}} -  T^{(-)}_{\phi} \times  \tilde{T}^{(-)}_{\tilde{\psi}}      \Big) \phi_{M\ell}{(t)} \tilde{\psi}_{\ell \, \ell_{N-2}}{(\theta)}\nonumber \\ 
  &-\frac{M(\ell_{N-2}+\frac{N-2}{2})}{2( \ell + \frac{N}{2}) (  \ell +\frac{ N-2}{2})} \psi_{M\ell}{(t)} \tilde{\psi}_{\ell \, \ell_{N-2}}{(\theta)},
\end{align}
%%%%%%%%%%%%%%%%%%%%%%%%%%%%%%%%%%%%%%%%%%%%%%%%%%%%%%%%%%%%%%%%
\begin{align} \label{useful_ladder_relation_psiM_tildepsi}
     \xi^{\mu}\partial_{\mu}( \psi_{M\ell}{(t)} \tilde{\psi}_{\ell \, \ell_{N-2}}{(\theta)}) +i \frac{\psi_{M\ell}{(t)}}{2 \cosh{t}}\sin{\theta}\, \tilde{\phi}_{\ell \, \ell_{N-2}}{(\theta)} = &\frac{1}{2(\ell + \frac{N-1}{2})}  \Big(  T^{(+)}_{\psi} \times \tilde{T}^{(+)}_{\tilde{\psi}} -  T^{(-)}_{\psi} \times  \tilde{T}^{(-)}_{\tilde{\psi}}      \Big) \psi_{M\ell}{(t)} \tilde{\psi}_{\ell \, \ell_{N-2}}{(\theta)}\nonumber \\ 
  &+\frac{M(\ell_{N-2}+\frac{N-2}{2})}{2( \ell + \frac{N}{2}) (  \ell +\frac{ N-2}{2})} \phi_{M\ell}{(t)} \tilde{\psi}_{\ell \, \ell_{N-2}}{(\theta)},
\end{align}
%%%%%%%%%%%%%%%%%%%%%%%%%%%%%%%%%%%%%%%%%%%%%%%%%%%%%%%%%%%%%%
\begin{align} \label{useful_ladder_relation_psiM_tildephi}
    \xi^{\mu}\partial_{\mu}( \psi_{M\ell}{(t)} \tilde{\phi}_{\ell \, \ell_{N-2}}{(\theta)}) -i \frac{\psi_{M\ell}{(t)}}{2 \cosh{t}}\sin{\theta} \,\tilde{\psi}_{\ell \, \ell_{N-2}}{(\theta)}=&\frac{1}{2(\ell + \frac{N-1}{2})}  \Big(  T^{(+)}_{\psi} \times \tilde{T}^{(+)}_{\tilde{\phi}} -  T^{(-)}_{\psi} \times  \tilde{T}^{(-)}_{\tilde{\phi}}      \Big) \psi_{M\ell}{(t)} \tilde{\phi}_{\ell \, \ell_{N-2}}{(\theta)}\nonumber \\ 
   & -\frac{M(\ell_{N-2}+\frac{N-2}{2})}{2( \ell + \frac{N}{2}) (  \ell +\frac{ N-2}{2})} \phi_{M\ell}{(t)} \tilde{\phi}_{\ell \, \ell_{N-2} }{(\theta)}.
\end{align}
We can prove relation (\ref{useful_ladder_relation_phiM_tildephi}) as follows: We express $\tilde{\psi}_{\ell \, \ell_{N-2}}$ on the left-hand side in terms of $\tilde{\phi}_{\ell \, \ell_{N-2}},d\tilde{\phi}_{\ell \, \ell_{N-2}}/d\theta_{N-1} $ using Eq.~(\ref{phi->psi}). As for the right-hand side, we expand $T^{(\pm)}_{\phi}, \tilde{T}^{(\pm)}_{\tilde{\phi}}$ using Eqs.~(\ref{raising_phi_psi})-(\ref{lowering_tilde}) and then we express $\psi_{M\ell}$ in terms of  ${\phi}_{M\ell},d{\phi}_{M\ell  }/dt $ using Eq.~(\ref{psiM_in_terms_of_phiM_and_derivs}). Then it is straightforward to show that the two sides are equal. Relations (\ref{useful_ladder_relation_phiM_tildepsi}), (\ref{useful_ladder_relation_psiM_tildepsi}) and (\ref{useful_ladder_relation_psiM_tildephi}) can be proved in the same way. 

Let us now derive Eq.~(\ref{spinor_Lie_N=even,final_result_neg_spin}) for the negative spin projection solution (the positive spin projection case can be treated in the same way). Substituting Eqs.~(\ref{N-1_0_generators}) and (\ref{negative_spin_dS_in_terms_S_(N-2)}) into Eq.~(\ref{spinorial_Lie_final}) we find 
\begin{align}
    \mathcal{L}^{s}_{\xi} \psi^{(-,\tilde{s})}_{M\ell\, \ell_{N-2} \, \tilde{\sigma} }= C_{1}C_{2}   \begin{pmatrix}\xi^{\mu}\partial_{\mu} U^{(\tilde{s})}_{M \ell \, \ell_{N-2} \, \tilde{\sigma}}-\frac{\sin{\theta}}{2 \cosh{t}}\tilde{\gamma}^{N-1}  U^{(\tilde{s})}_{M \ell \, \ell_{N-2} \, \tilde{\sigma}}\\ \\ \xi^{\mu}\partial_{\mu}D^{(\tilde{s})}_{M \ell \, \ell_{N-2} \, \tilde{\sigma}}+\frac{\sin{\theta}}{2 \cosh{t}}\tilde{\gamma}^{N-1}  D^{(\tilde{s})}_{M \ell \, \ell_{N-2} \, \tilde{\sigma}} \end{pmatrix},
\end{align}
where $C_{1}\equiv {c_{N}(M \ell)}/{\sqrt{2}}$ and $C_{2}\equiv{c_{N-1}( \ell \, \ell_{N-2})}/{\sqrt{2}}$. 
Then, using 
$$\tilde{\gamma}^{N-1}\hat{\tilde{\chi}}^{(\tilde{s})}_{\pm\ell_{N-2} \, \tilde{\sigma}}(\Omega_{{N-2}})=\hat{\tilde{\chi}}^{(\tilde{s})}_{\mp\ell_{N-2} \, \tilde{\sigma}}(\Omega_{{N-2}})$$
(see Eq.~(\ref{gamma_changes_sign_for_chi_hat})) and Eq.~(\ref{negative_spin_dS_in_terms_S_(N-2)_upper_comp}), it is straightforward to find
\begin{align}
 \mathcal{L}^{s}_{\xi}& \psi^{(-,\tilde{s})}_{M\ell\, \ell_{N-2} \, \tilde{\sigma} }\nonumber \\
 & =C_{1}C_{2} \nonumber\\
 &\times\begin{pmatrix}
    \hat{\tilde{\chi}}^{(\tilde{s})}_{-\ell_{N-2} \, \tilde{\sigma}}\big(\xi^{\mu} \partial_{\mu}[\phi_{M\ell}\tilde{\phi}_{\ell \,\ell_{N-2}}]+i\frac{\sin{\theta}}{2 \cosh{t}}\phi_{M\ell}\tilde{\psi}_{\ell \,\ell_{N-2}}\big)-i\hat{\tilde{\chi}}^{(\tilde{s})}_{+\ell_{N-2} \, \tilde{\sigma}} \big(\xi^{\mu} \partial_{\mu}[\phi_{M\ell}\tilde{\psi}_{\ell \,\ell_{N-2}}]-i\frac{\sin{\theta}}{2 \cosh{t}}\phi_{M\ell}\tilde{\phi}_{\ell \,\ell_{N-2}} \big) \\ \\
    i \hat{\tilde{\chi}}^{(\tilde{s})}_{-\ell_{N-2} \, \tilde{\sigma}} \big(\xi^{\mu} \partial_{\mu}[\psi_{M\ell}\tilde{\phi}_{\ell \,\ell_{N-2}}]-i\frac{\sin{\theta}}{2 \cosh{t}}\psi_{M\ell}\tilde{\psi}_{\ell \,\ell_{N-2}} \big)+\hat{\tilde{\chi}}^{(\tilde{s})}_{+\ell_{N-2} \, \tilde{\sigma}} \big(\xi^{\mu} \partial_{\mu}[\psi_{M\ell}\tilde{\psi}_{\ell \,\ell_{N-2}}]+i\frac{\sin{\theta}}{2 \cosh{t}}\psi_{M\ell}\tilde{\phi}_{\ell \,\ell_{N-2}}\big)
    \end{pmatrix}.
\end{align}
At this point we can use relations (\ref{useful_ladder_relation_phiM_tildephi})-(\ref{useful_ladder_relation_psiM_tildephi}) to find 
\begin{align}
\mathcal{L}^{s}_{\xi}& \psi^{(-,\tilde{s})}_{M\ell\, \ell_{N-2} \, \tilde{\sigma} }\nonumber \\
& =C_{1}C_{2} \nonumber \\
 &\times \Bigg( \frac{1}{2(\ell + \frac{N-1}{2})}\, \begin{pmatrix}
   T^{(+)}_{\phi} \phi_{M \ell}\\ 
    iT^{(+)}_{\psi} \psi_{M \ell}
   \end{pmatrix}  \Big[\hat{\tilde{\chi}}^{(\tilde{s})}_{-\ell_{N-2} \, \tilde{\sigma}}\,\tilde{T}^{(+)}_{\tilde{\phi}} \tilde{\phi}_{\ell \, \ell_{N-2}}- i  \hat{\tilde{\chi}}^{(\tilde{s})}_{+\ell_{N-2} \, \tilde{\sigma}}\, \tilde{T}^{(+)}_{\tilde{\psi}} \tilde{\psi}_{\ell \, \ell_{N-2}} \Big] \nonumber \\
  &-\frac{1}{2(\ell + \frac{N-1}{2})}\, \begin{pmatrix}
   T^{(-)}_{\phi} \phi_{M \ell}\\
    iT^{(-)}_{\psi} \psi_{M \ell}
   \end{pmatrix}  \Big[\hat{\tilde{\chi}}^{(\tilde{s})}_{-\ell_{N-2} \, \tilde{\sigma}}\,\tilde{T}^{(-)}_{\tilde{\phi}} \tilde{\phi}_{\ell \, \ell_{N-2}}- i  \hat{\tilde{\chi}}^{(\tilde{s})}_{+\ell_{N-2} \, \tilde{\sigma}}\, \tilde{T}^{(-)}_{\tilde{\psi}} \tilde{\psi}_{\ell \, \ell_{N-2}} \Big] \nonumber \\
  & -i\frac{M(\ell_{N-2}+\frac{N-2}{2})}{2( \ell + \frac{N}{2}) (  \ell +\frac{ N-2}{2})}\begin{pmatrix}
  i \psi_{M \ell}\\
     \phi_{M \ell}
   \end{pmatrix}  \Big[\hat{\tilde{\chi}}^{(\tilde{s})}_{-\ell_{N-2} \, \tilde{\sigma}}\, \tilde{\phi}_{\ell \, \ell_{N-2}}+ i  \hat{\tilde{\chi}}^{(\tilde{s})}_{+\ell_{N-2} \, \tilde{\sigma}}\,  \tilde{\psi}_{\ell \, \ell_{N-2}} \Big] \Bigg).
    \end{align}
Then using Eqs.~(\ref{negative_spin_dS_in_terms_S_(N-2)}) and (\ref{negative_spin_dS_in_terms_S_(N-2)_upper_comp}) as well as the ladder relations~(\ref{raising_phi_psi})-(\ref{lowering_tilde}) we obtain Eq.~(\ref{spinor_Lie_N=even,final_result_neg_spin}). 

\subsection{Transformation properties for \texorpdfstring{$N=2$}{N=2}.}
The massive positive frequency solutions~(\ref{negative_spin_even_dS})-(\ref{positive_spin_even_dS}) for $N=2$ are given by
\begin{align}
     &\psi^{(-)}_{ M\ell }(t,{\varphi})=\frac{c_{2}(M \ell)}{2\sqrt{ \pi}} \begin{pmatrix} \phi_{M\ell}(t)
 \\  i\psi_{M\ell}(t)\end{pmatrix} {e^{-i(\ell+1/2)\varphi}}, \\
    & \psi^{(+)}_{ M\ell }(t,{\varphi})=\frac{c_{2}(M \ell)}{2\sqrt{ \pi}} \begin{pmatrix} i \psi_{M\ell}(t) \\ \phi_{M\ell}(t) \\
 \end{pmatrix} {e^{+i(\ell+1/2)\varphi}},
 \end{align}
where $0\leq \varphi \equiv \theta_{1}< 2 \pi$ and $\ell = 0,1,...\,$. By calculating the spinorial Lie derivative with respect to the boost Killing vector (\ref{Killing_vector}) we arrive again at Eq.~(\ref{spinorial_Lie_final}), where $\partial \psi^{(\pm)}_{ M\ell } / \partial \varphi = \pm i(\ell + \frac{1}{2}) \psi^{(\pm)}_{ M\ell }$. By expressing $\cos{\varphi}$ and $\sin{\varphi}$ in terms of $\exp{(\pm i \varphi)}$ and using the ladder operators (\ref{raising_phi_psi}), (\ref{lowering_phi_psi}) with $N=2$ it is straightforward to find
\begin{align}
  \mathcal{L}^{s}_{\xi} \psi^{(\pm)}_{M\ell  }=&\frac{k^{(+)}}{2}\frac{c_{2}(M\ell)}{c_{2}(M,\,\ell+1)} \psi^{(\pm)}_{M\,\ell+1 } + \frac{k^{(-)}}{2}\frac{c_{2}(M\ell)}{c_{2}(M,\,\ell-1)} \psi^{(\pm)}_{M\,\ell-1 } \\
  =&-\frac{i}{2}(\ell+1-iM)\psi^{(\pm)}_{M\,\ell+1 } - \frac{i}{2}(\ell+iM) \psi^{(\pm)}_{M\,\ell-1 }.\label{spinor_Lie_N=2,final_result}
\end{align}
By using Eq.~(\ref{spinor_Lie_N=2,final_result}) we have verified that $(\mathcal{L}^{s}_{\xi}\psi_{M \ell}, \psi_{M \,\ell\pm1})+(\psi_{M \ell}, \mathcal{L}^{s}_{\xi}\psi_{M \,\ell\pm1})  =0$, in agreement with the de~Sitter invariance of the inner product (\ref{inner_prod}).
%%%%%%%%%%%%%%%%%%%%%%%%%%%%%%%%%%%%%%%%%%%%%%%%%%%%%%%%%%%%%%%%%%%%%%%%%%%%%%%%%%%%%%%%%%%%%%%%%%%%%%%%%%%%%%%%%%%%%%%%%%%%%%%%%%%%%%%%%%%%%%%%%%%%%%%%%%%%%%%%%%%%%%%%%%%%%%%%%%%%%%%%%%%%%%%%%%%%%%%%%%%%
\section{Derivation of the massless Wightman two-point function using the mode-sum method}\label{appendix_2pt_function_calculations}
In this Appendix we present the derivation of the massless Wightman two-point function using the mode-sum method (\ref{Wightman_mode_sum}) for $N$ even. The derivation of the two-point function for $N$ odd has many similarities with the even-dimensional case and therefore is just briefly discussed.
The case with $N=2$ is presented separately at the end.

Let us first introduce the notation and some useful relations used in the calculations. The functions (\ref{phi_nl})-(\ref{psi_nl}) used in the recursive construction of the eigenspinors of the Dirac operator on $S^{N-r}$ ($N-r=1,2,...,N-2$) are denoted as \begin{align}
   \tilde{\phi}_{\ell_{N-r}\, \ell_{N-r-1}}(\theta_{N-r})\equiv\tilde{\phi}^{(N-r)}_{\ell_{N-r}\, \ell_{N-r-1}}, \hspace{4mm} \tilde{\psi}_{\ell_{N-r}\, \ell_{N-r-1}}(\theta_{N-r})\equiv\tilde{\psi}^{(N-r)}_{\ell_{N-i}\, \ell_{N-r-1}}, 
\end{align} 
with $\tilde{\phi}^{(N-r)}_{0\,0}=\cos{(\theta_{N-r}/2)}$ and $\tilde{\psi}^{(N-r)}_{0\,0}=\sin{(\theta_{N-r}/2)}$ (see Eqs.~(\ref{phi_N-1_ell_0}) and (\ref{psi_N-1_ell_0}) below). We let $\bm{\theta}_{N-r}=(\theta_{N-r}, \theta_{N-r-1},..., \theta_{1})$. The dimension of the Spin($N-1$,1) representation is denoted as $D \equiv 2 ^{N/2}$. Also, let $\tilde{s}_{N-2}$ represent the spin projection indices $(s_{N-2}, s_{N-4},...,s_{4},s_{2})$, $\tilde{s}_{N-4}$ represent $( s_{N-4},...,s_{4},s_{2})$ and so forth. Similarly, $\sigma_{N-r}$ represents the angular momentum quantum numbers $(\ell_{N-r}, \ell_{N-r-1},...,\ell_{2}, \ell_{1})$ etc. Note that for $\theta_{N-1}' = \theta_{N-2}'=...= \theta_{1}'=0$ we have
\begin{align}\label{geodesic_distance_for_theta_prime=0}
    \cos{\mu}|_{\bm{\theta}'= \bm{0}}= -\sinh{t} \sinh{t'} + \cosh{t} \cosh{t'} \cos{\theta_{N-1}},
\end{align}
(see Eq.~(\ref{geodesic_distance_dS})) while the only non-zero (vielbein basis) components of the tangent vector $n_{a}|_{\bm{\theta}'=0}$ (see Eqs.~(\ref{orthonormal_basis_components_tangent_vectors_n0})$-$(\ref{orthonormal_basis_components_tangent_vectors_spatial_comps})) are given by
\begin{align}
& n_{0}|_{\bm{\theta}'=\bm{0}}=\frac{1}{\sin{\mu}}(  \cosh{t} \sinh{t'} - \sinh{t} \cosh{t'} \cos{\theta_{N-1}}   ) ,\label{tangent_vec_non_zero_componetns_orth_basis}\\
  &    n_{N-1}|_{\bm{\theta}'=\bm{0}}=\frac{\cosh{t'}}{\sin{\mu}}\sin{\theta_{N-1}}=\frac{1}{\cosh{t}} n_{\theta_{N-1}}|_{\bm{\theta}'=\bm{0}}.\label{tangent_vec_non_zero_componetns_orth_basis_theta_N-1}
\end{align}
(For brevity we will denote $n_{0}|_{\bm{\theta}'=\bm{0}},\,n_{N-1}|_{\bm{\theta}'=\bm{0}}, \, $ and $\slashed{n}|_{\bm{\theta}'=\bm{0}}$ by $n_{0}, n_{N-1}$ and $\slashed{n}$ respectively.) 
Also, notice that Spin$(N-1,1)$ transformation matrices can be expressed as
\begin{align}
   & \exp{({a} \,\Sigma^{0 j})}=\exp{(\frac{a}{2} \,\gamma^{0}\gamma^{j})}=\,\bm{1}\, \cosh{\frac{a}{2}}+ \,\gamma^{0}\gamma^{j}\,\sinh{\frac{a}{2}}, \label{Spin(N-1,1)_boost} \\
    & \exp{({b} \,\Sigma^{k j})}=\exp{(\frac{b}{2} \,\gamma^{k}\gamma^{j})}=\,\bm{1}\, \cos{\frac{b}{2}}+ \,\gamma^{k}\gamma^{j}\,\sin{\frac{b}{2}}\label{Spin(N-1,1)_rotation},
\end{align}
(with $k \neq j$ and $k,j=1,2,...,N-1$) where $a,b$ are the transformation parameters. The corresponding generators are given by Eq.~(\ref{Spin(N-1,1)_generators}). Also, many of the following calculations involve the variables $x= \pi/2 - it, \,x'= \pi/2 - it' $ (see Eq.~(\ref{analytic_con})).
 
We can now start deriving the massless Wightman two-point function for $N$ even. By expanding the summation over the spin projections ($s= \pm$) Eq.~(\ref{Wightman_mode_sum}) becomes
\begin{align}
    W_{0} \big( (t,\bm{\theta}_{N-1}), (t',\bm{0}) \big) = \sum_{\ell=0}^{\infty} \sum_{\sigma_{N-2}} \sum_{\tilde{s}_{N-2}} \Big[ \psi^{(+,\tilde{s}_{N-2})}_{0 \ell \sigma_{N-2}}(t,\bm{\theta}_{N-1}) \overline{\psi}^{(+,\tilde{s}_{N-2})}_{0 \ell \sigma_{N-2}}(t',\bm{0})+ \psi^{(-,\tilde{s}_{N-2})}_{0\ell \sigma_{N-2}}(t,\bm{\theta}_{N-1}) \overline{\psi}^{(-,\tilde{s}_{N-2})}_{0 \ell \sigma_{N-2}}(t',\bm{0}) \Big].
\end{align}
Then using Eqs.~(\ref{massless_negative_spin_even_dS})-(\ref{massless_positive_spin_even_dS}) we find
\begin{align}\label{mode_sum_appendix_before_breaking_down_S_N-1}
  W_{0} \big( (t,\bm{\theta}_{N-1}), (t',\bm{0}) \big) \nonumber  =&-\Big|\frac{c_{N}(M=0)} { \sqrt{2} } \Big|^{2} \sum_{\ell=0}^{\infty} \sum_{\sigma_{N-2}}     \phi_{0\ell}(t)\phi^{*}_{0\ell}(t')  \nonumber \\
  &\times  \sum_{\tilde{s}_{N-4}} \sum_{{s}_{N-2}} \begin{pmatrix} {0} &&  \chi^{(s_{N-2},\tilde{s}_{N-4})}_{-\ell \sigma_{N-2}}(\bm{\theta}_{N-1})  \chi^{(s_{N-2},\tilde{s}_{N-4})}_{-\ell \sigma_{N-2}}(\bm{0})^{\dagger} \\ \chi^{(s_{N-2},\tilde{s}_{N-4})}_{+\ell\sigma_{N-2}}(\bm{\theta}_{N-1})  \chi^{(s_{N-2},\tilde{s}_{N-4})}_{+\ell \sigma_{N-2}}(\bm{0})^{\dagger} &&  {0}           \end{pmatrix},
\end{align}
where $\chi^{(s_{N-2},\tilde{s}_{N-4})}_{\pm \ell\sigma_{N-2}} $ are the eigenspinors on $S^{N-1}$. In order to proceed we need to express the eigenspinors on $S^{N-r}$, with $N-r$ odd, in terms of eigenspinors on $S^{N-r-2}$. Therefore, using Eqs.~(\ref{odd_eigenspinor}), (\ref{psi_minus_even_S_N}) and (\ref{psi_plus_even_S_N}) we derive the following two recursive relations:
\begin{align}\label{chi_minus_S_N-r_in_terms_of_N-r-2}
          \chi^{(-,\tilde{s}_{ N -r-3})}_{\pm \ell_{N-r} \,{\sigma_{N-r-1}} }(\bm{\theta}_{N-r} )=&\,\frac{c_{N-r}(\ell_{N-r}\, \ell_{N-r-1})}{\sqrt{2}} \frac{c_{N-r-1}(\ell_{N-r-1}\, \ell_{N-r-2})}{\sqrt{2}} \frac{1}{\sqrt{2}} \nonumber \\ 
        & \times \begin{pmatrix} (1+i){\tilde{\phi}}^{(N-r-1)}_{\ell_{N-r-1}\, \ell_{N-r-2}} [ \tilde{\phi}_{\ell_{N-r}\, \ell_{N-r-1}} ^{(N-r)}\pm i \tilde{\psi}_{\ell_{N-r}\, \ell_{N-r-1}} ^{(N-r)} ]
 \\ \\-(1+i){\tilde{\psi}}_{\ell_{N-r-1}\, \ell_{N-r-2}} ^{(N-r-1)}[ \tilde{\phi}_{\ell_{N-r}\, \ell_{N-r-1}}^{(N-r)} \mp i \tilde{\psi}_{\ell_{N-r}\, \ell_{N-r-1}}^{(N-r)}  ] \end{pmatrix}     \chi^{(\tilde{s}_{N-r-3})}_{-\ell_{N-r-2}, \,{\sigma_{N-r-3}} }(\bm{\theta}_{N-r-2} ),
\end{align}
%%%%%%%%%%%%%%%%%%%%%%%%%%%%%%%%%%%%%%%%%%%%%%%%%%%%%%%%%%%%%%%%%%
\begin{align}\label{chi_plus_S_N-r_in_terms_of_N-r-2}
          \chi^{(+,\tilde{s}_{ N -r-3})}_{\pm \ell_{N-r} \,{\sigma_{N-r-1}} }(\bm{\theta}_{N-r} )=&\,\frac{c_{N-r}(\ell_{N-r}\, \ell_{N-r-1})}{\sqrt{2}} \frac{c_{N-r-1}(\ell_{N-r-1}\, \ell_{N-r-2})}{\sqrt{2}}\frac{1}{\sqrt{2}} \nonumber \\ 
        & \times \begin{pmatrix} (-1+i){\tilde{\psi}}^{(N-r-1)}_{\ell_{N-r-1}\, \ell_{N-r-2}} [ \tilde{\phi}_{\ell_{N-r}\, \ell_{N-r-1}} ^{(N-r)}\pm i \tilde{\psi}_{\ell_{N-r}\, \ell_{N-r-1}} ^{(N-r)} ]
 \\ \\(-1+i){\tilde{\phi}}_{\ell_{N-r-1}\, \ell_{N-r-2}} ^{(N-r-1)}[ \tilde{\phi}_{\ell_{N-r}\, \ell_{N-r-1}}^{(N-r)} \mp i \tilde{\psi}_{\ell_{N-r}\, \ell_{N-r-1}}^{(N-r)}  ] \end{pmatrix}    \chi^{(\tilde{s}_{N-r-3})}_{+\ell_{N-r-2}, \,{\sigma_{N-r-3}} }(\bm{\theta}_{N-r-2} )
\end{align}
(for $r$ odd and $N-3\geq r \geq 1$).
Since $\tilde{\psi}_{\ell_{N-r}\, \ell_{N-r-1}}^{(N-r)}(0)=0$ and $\tilde{\phi}_{\ell_{N-r}\, \ell_{N-r-1}}^{(N-r)}(0)$ is non-zero only for $\ell_{N-r-1}=0$, it is clear from the recursive relations~(\ref{chi_minus_S_N-r_in_terms_of_N-r-2})-(\ref{chi_plus_S_N-r_in_terms_of_N-r-2}) that the only non-vanishing terms in Eq.~(\ref{mode_sum_appendix_before_breaking_down_S_N-1}) are the ones with $\ell_{N-2}= \ell_{N-3}=...=\ell_{2}=\ell_{1}=0$. Thus, only the summation over $\ell_{N-1} \equiv \ell$ survives in the mode-sum. Substituting Eqs.~(\ref{chi_minus_S_N-r_in_terms_of_N-r-2}) and (\ref{chi_plus_S_N-r_in_terms_of_N-r-2}) (with $r=1$) into Eq.~(\ref{mode_sum_appendix_before_breaking_down_S_N-1}) one obtains (after some calculations)
\begin{align}\label{mode_sum_appendix_before_separating_steps}
  W_{0} \big( (t,\bm{\theta}_{N-1}), (t',\bm{0}) \big) \nonumber  =&\,\Big|\frac{c_{N}(M=0)} { \sqrt{2} } \Big|^{2}   \sum_{\ell=0}^{\infty}    \phi_{0\ell}(t)\phi^{*}_{0\ell}(t')\, \Big| \frac{c_{N-1}(\ell0)}{\sqrt{2}} \Big|^{2} \, \Big|\frac{c_{N-2}(00)}{\sqrt{2}} \Big|^{2} \nonumber \\
 & \times \tilde{\phi}^{(N-1)}_{\ell 0}(0)^{*}\, \Big[i \tilde{\phi}_{\ell 0}^{(N-1)}(\theta_{N-1})\, \gamma^{0} + \tilde{\psi}_{\ell 0}^{(N-1)}(\theta_{N-1}) \,\gamma^{N-1}\Big]\,\left[\mathbb{I}_{2} \otimes \begin{pmatrix}
 \tilde{\phi}_{0 0}^{(N-2)} &  \tilde{\psi}_{0 0}^{(N-2)} \\
  -\tilde{\psi}_{0 0}^{(N-2)} &\tilde{\phi}_{0 0}^{(N-2)}
 \end{pmatrix}  \otimes \mathbb{I}_{D/4}\right] \nonumber \\
&\times \left[ \mathbb{I}_{2} \otimes \sum_{\tilde{s}_{N-4}} \begin{pmatrix}\chi^{(\tilde{s}_{N-4})}_{-00}(\bm{\theta}_{N-3})  \chi^{(\tilde{s}_{N-4})}_{-00}(\bm{0})^{\dagger}  & 0 \\
 0 & \chi^{(\tilde{s}_{N-4})}_{+00}(\bm{\theta}_{N-3})  \chi^{(\tilde{s}_{N-4})}_{+00}(\bm{0})^{\dagger}
 \end{pmatrix}\right],
  \end{align}
  where $\mathbb{I}_{d}$ is the identity matrix of dimension $d$.
  Also, we are going to use the following results:
  \begin{align}
 &|c_{N-1}(\ell 0)|^{2}=
 \frac{\Gamma(\ell +1)\Gamma(\ell +N-1)}{2^{N-3}|\Gamma(\ell +\frac{N-1}{2})|^{2}} \label{c_N-1_ell_0},\\
    &  \tilde{\phi}^{(N-1)}_{\ell 0}(\theta_{N-1})=\kappa_{\phi}^{(N-1)}{(\ell 0)}\, \cos{\frac{\theta_{N-1}}{2}} F(\ell+N-1,-\ell;\frac{N-1}{2};\sin^{2}{\frac{\theta_{N-1}}{2}}), \label{phi_N-1_ell_0}\\
      & \tilde{\psi}^{(N-1)}_{\ell 0}(\theta_{N-1})=\,\kappa_{\phi}^{(N-1)}{(\ell 0)} \frac{ (\ell +(N-1)/2)}{(N-1)/2}\, \sin{\frac{\theta_{N-1}}{2}} F(\ell+N-1,-\ell;\frac{N+1}{2};\sin^{2}{\frac{\theta_{N-1}}{2}}),\label{psi_N-1_ell_0}
  \end{align}
  where
  \begin{align}\label{kappa_normalzn_ell_0}
      \kappa_{\phi}^{(N-1)}{(\ell 0)}= \frac{\Gamma(\ell + \frac{N-1}{2})}{\ell! \,\Gamma(\frac{N-1}{2})}
  \end{align}
  (see Eqs.~(\ref{S_N_normlzn_fac}) and (\ref{phi_nl})$-$(\ref{scalar_functions_normalization_factors})).
  
We complete the derivation of the massless two-point function in three steps: 1) we calculate the proportionality constant (and we show that it agrees with the proportionality constant in Eq.~(\ref{Ansatz_massless_with_prop_const}); 2) we obtain a closed-form result for the infinite sum over $\ell$ and we determine the dependence on $\set{t,\theta_{N-1},t'}$; 3) we obtain analytic expressions for the terms of the two-point function that depend only on the angular variables $\theta_{N-2}, \theta_{N-3},...,\theta_{1}$. We call the latter the ``angular part'' of the two-point function and we denote it as follows:
\begin{align}\label{appendix_angular_part_of_2pt_function}
    \widetilde{W}_{0}(\Omega_{N-2})\equiv &\,\left(\prod_{j=3}^{N-1}\Big|\frac{c_{N-j}(00)} { \sqrt{2} } \Big|^{-2} \right) \times \left[\mathbb{I}_{2} \otimes \begin{pmatrix}
 \tilde{\phi}_{0 0}^{(N-2)} &  \tilde{\psi}_{0 0}^{(N-2)} \\
  -\tilde{\psi}_{0 0}^{(N-2)} &\tilde{\phi}_{0 0}^{(N-2)}
 \end{pmatrix}  \otimes \mathbb{I}_{D/4} \right] \nonumber \\
&\times \left[ \mathbb{I}_{2} \otimes \sum_{\tilde{s}_{N-4}} \begin{pmatrix}\chi^{(\tilde{s}_{N-4})}_{-00}(\bm{\theta}_{N-3})  \chi^{(\tilde{s}_{N-4})}_{-00}(\bm{0})^{\dagger}  & 0 \\
 0 & \chi^{(\tilde{s}_{N-4})}_{+00}(\bm{\theta}_{N-3})  \chi^{(\tilde{s}_{N-4})}_{+00}(\bm{0})^{\dagger}
 \end{pmatrix} \right].
\end{align}
After completing these steps it will be clear that the obtained two-point function is of the form~(\ref{Ansatz_massless}) (i.e. it agrees with the construction presented in Ref.~\cite{Muck:1999mh}). 
%%%%%%%%%%%%%%%%%%%%%%%%%%%%%%%%%%%%%%%%%%%%%%%%%%%%%%%%%%%%%%%%%%%%%%%%%%%%%%%%%%%%%%%%%%%%%%%%%%%%%%%%%%%%%%%%%%%%%%%%%%%%%%%%%%%%%%%%%%%%%%%%%%%%%%%%%%%%%%%%%%%%%%%%%%%%%%%%%%%%%%%%%%%%%%%%%%%%%%%%%
\subsection{The proportionality constant}
The proportionality constant for the massless two-point function arises from the normalization factors in Eq.~(\ref{mode_sum_appendix_before_separating_steps}). (Note that apart from $c_{N}(M=0)$ and $c_{N-1}(\ell0)$ there are $N-2$ additional normalization factors; one for each lower-dimensional sphere.) The overall contribution from the normalization factors is given by the following product:
\begin{align}\label{prod_of_normlzn_facs}
 &  \Big|\frac{c_{N}(M=0)} { \sqrt{2} } \Big|^{2}   \,\Big|\frac{c_{N-1}(\ell 0)} { \sqrt{2} } \kappa^{(N-1)}_{\phi}( \ell 0)\Big|^{2} \prod_{j=2}^{N-1}\Big|\frac{c_{N-j}(00)} { \sqrt{2} } \Big|^{2}  \\
 =&\frac{1}{2 \pi}   \Big|\frac{c_{N}(M=0)} { \sqrt{2} } \Big|^{2}   \,\Big|\frac{c_{N-1}(\ell 0)} { \sqrt{2} } \kappa^{(N-1)}_{\phi}( \ell 0)\Big|^{2} \prod_{j=2}^{N-2}\Big|\frac{c_{N-j}(00)} { \sqrt{2} } \Big|^{2}
\end{align}
where $c_{1}(00)\equiv 1/\sqrt{ \pi}$ is the normalization factor for eigenspinors on $S^{1}$, while the normalization factors for eigenspinors on higher-dimensional spheres are given by Eq.~(\ref{S_N_normlzn_fac}). Using Eqs.~(\ref{c_N-1_ell_0}) and (\ref{kappa_normalzn_ell_0}) we observe that
\begin{align}\label{Pochhamer_trick_for_prop_const}
  \Big|{c_{N-1}(\ell 0)}\,  \kappa^{(N-1)}_{\phi}( \ell 0)\Big|^{2}=    \Big|{c_{N-1}(0 0)}\Big|^{2} \, \frac{(N-1)_{\ell}}{\ell!},
\end{align}
where $(N-1)_{\ell}= \Gamma(N-1+\ell)/\Gamma(N-1)$ is the Pochhamer symbol for the rising factorial. Using Eq.~(\ref{Pochhamer_trick_for_prop_const}) we may rewrite Eq.~(\ref{prod_of_normlzn_facs}) as
\begin{align}\label{appendix_proportionality_const_final_expression}
   & \frac{1}{\pi 2^{N} 2^{N-2}} \prod_{j=1}^{N-2} \frac{\Gamma(N-j)}{|\Gamma(\frac{N-j}{2}) |^{2} 2^{N-j-2}} \times \frac{(N-1)_{\ell}}{\ell!} \nonumber\\
   &= \frac{\Gamma(N/2)}{2^{N/2}\,(2 \pi)^{N/2}}\times \frac{(N-1)_{\ell}}{\ell!}, 
\end{align}
where we also used Eqs.~(\ref{S_N_normlzn_fac}), (\ref{dS_normalization_fac}) and the Legendre duplication formula (\ref{Legendre_duplication_formula}). Equation~(\ref{appendix_proportionality_const_final_expression}) clearly gives the desired form for the proportionality constant (see Eq.~(\ref{Ansatz_massless_with_prop_const})). The $\ell$-dependence in Eq.~(\ref{appendix_proportionality_const_final_expression}) will be discussed later (it will be used in the summation over $\ell$).
%%%%%%%%%%%%%%%%%%%%%%%%%%%%%%%%%%%%%%%%%%%%%%%%%%%%%%%%%%%%%%%%%%%%%%%%%%%%%%%%%%%%%%%%%%%%%%%%%%%%%%%%%%%%%%%%%%%%%%%%%%%%%%%%%%%%%%
\subsection{Obtaining a closed-form expression for the series}
Using Eqs.~(\ref{mode_sum_appendix_before_separating_steps}), (\ref{phi_N-1_ell_0}), (\ref{psi_N-1_ell_0}) and (\ref{appendix_proportionality_const_final_expression}) we collect all the $\ell$-dependent terms of the two-point function. Then the mode-sum expression (\ref{mode_sum_appendix_before_separating_steps}) can be written as
\begin{align}\label{Appendix_2pt_function_with_A_and_B}
      W_{0} \big( (t,\bm{\theta}_{N-1}), (t',\bm{0}) \big)   =\frac{\Gamma(N/2)}{2^{N/2}\,(2 \pi)^{N/2}} [iA \gamma^{0} + B \gamma^{N-1}]\, \widetilde{W}_{0}(\Omega_{N-2}),
\end{align}
where
\begin{align}\label{appendix_series_one}
&A=\cos{\frac{\theta_{N-1}}{2}}\sum_{\ell=0}^{\infty}  \frac{{(N-1)_{\ell}}}{\ell!}  \phi_{0\ell}(t)\phi^{*}_{0\ell}(t') \, F{( {\ell+N-1,-\ell; \frac{N-1}{2};} \sin^{2}{\frac{\theta_{N-1}}{2}} )}  , \\
&B=\frac{2\sin{\frac{\theta_{N-1}}{2}}}{N-1}\sum_{\ell=0}^{\infty}  \frac{{(N-1)_{\ell}}}{\ell!} (\ell+\frac{N-1}{2}) \phi_{0\ell}(t)\phi^{*}_{0\ell}(t') \, F{( {\ell+N-1,-\ell; \frac{N+1}{2};} \sin^{2}{\frac{\theta_{N-1}}{2}} )}.
\end{align}
 Using Eq.~(\ref{scalar_massless}) for $\phi_{0\ell}(t), \, \phi^{*}_{0\ell}(t')$ we find
\begin{align}
    &A=\,\frac{\cos{({\theta_{N-1}}/{2})}}{(\cos(x/2)\cos^{*}(x'/2))^{N-1} }\sum_{\ell=0}^{\infty}  \frac{{(N-1)_{\ell}}}{\ell!}  \big(\rho(t,t')\big)^{\ell} \, F{( {\ell+N-1,-\ell; \frac{N-1}{2};} \sin^{2}{\frac{\theta_{N-1}}{2}} )}, \label{Appendix_A_initial}\\
     & B=\frac{2}{N-1}\frac{\sin{({\theta_{N-1}}/{2})}}{(\cos(x/2)\cos^{*}(x'/2))^{N-1} }\sum_{\ell=0}^{\infty}  \frac{{(N-1)_{\ell}}}{\ell!} (\ell+\frac{N-1}{2}) \big(\rho(t,t')\big)^{\ell} \, F{( {\ell+N-1,-\ell; \frac{N+1}{2};} \sin^{2}{\frac{\theta_{N-1}}{2}} )}.\label{Appendix_B_initial}
\end{align}
where 
\begin{align}\label{definition_of_rho}
   \rho(t,t')\equiv \tan{\frac{x}{2}}\left[\tan{\frac{x'}{2}}\right]^{*}=\frac{(1-i\sinh{t})(1+i \sinh{t'})}{\cosh{t}\, \cosh{t'}}.
\end{align}
Let us first find the infinite sum in $A$. By using the formula~\cite{Prudnikov}
\begin{equation}\label{Prudnikovs_series}
    \sum_{k=0}^{\infty} \frac{(a)_{k}}{k!} t^{k} F(-k,a+k; c ;z)=(1-t)^{-a} F(\frac{a}{2}, \frac{a+1}{2}; c; \frac{-4t}{(1-t)^{2}}z  ),\hspace{2mm} |t|< 1
\end{equation}
Eq.~(\ref{Appendix_A_initial}) can be written as
\begin{align}
   A= &\,\frac{\cos{{(\theta_{N-1}}/{2})} } {(\cos{({x}/{2})}\left[\cos{({x'}/{2})}\right]^{*} )^{N-1} }   (1-\rho(t,t'))^{-N+1}F({\frac{N-1}{2},\frac{N}{2};\frac{N-1}{2}} ;\frac{-4\rho(t,t')}{(1-\rho(t,t'))^{2}} \sin^{2}\frac{\theta_{N-1}}{2} )\\
   =& \,\frac{\cos{{(\theta_{N-1}}/{2})} } {(\cos{({x}/{2})} \left[\cos{({x'}/{2})}\right]^{*} )^{N-1} }  \frac{(1-\rho(t,t'))^{-N+1} \,((1-\rho(t,t'))^{2})^{N/2} }{  [ (1-\rho(t,t'))^{2} +4\rho(t,t') \sin^{2}(\frac{\theta_{N-1}}{2} )]^{N/2}},\label{appendix_A_just_before_w1}
\end{align}
where we also used 
\begin{align}\label{Fabb}
    F(a,b;b;z)=\frac{1}{(1-z)^{a}}.
\end{align}
(Note that since $|\rho(t,t')|=1$ the series in Eq.~(\ref{Appendix_A_initial}) diverges. Therefore, we make the replacement $t\rightarrow t-i \epsilon$ with $\epsilon >0$ before applying (\ref{Prudnikovs_series}) and then we let $\epsilon \rightarrow 0$.) By expressing $x,x'$ and $\rho(t,t')$ in terms of $t$ and $t'$ we can write Eq.~(\ref{appendix_A_just_before_w1}) as
\begin{align}
   A=&i \,{\sinh{\frac{t-t'}{2}} \cos{\frac{\theta_{N-1}}{2}}}(\sin^{2}\frac{\mu}{2})^{-N/2}\\
   &= i w_{1}(t,\theta_{N-1},t')\, (\sin^{2}\frac{\mu}{2})^{-N/2}\label{appendix_A_final}.
\end{align}
(The biscalar function $w_{1}$ is given in Eq.~(\ref{2-point-func-scalars_w1}), while $\sin^{2}{(\mu/2)}$ can be found by Eq.~(\ref{geodesic_distance_for_theta_prime=0})).

Let us now find the infinite sum in $B$. We can rewrite Eq.~(\ref{Appendix_B_initial}) as
\begin{align}
   B=&\frac{2}{N-1}\frac{\sin{({\theta_{N-1}}/{2})}}{(\cos(x/2)\left[\cos(x'/2)\right]^{*})^{N-1} } \nonumber \\
   &\times \Big( \rho \frac{\partial}{\partial \rho}+ \frac{N-1}{2}\, \Big) \nonumber \\
   &\times \sum_{\ell=0}^{\infty} \frac{{(N-1)_{\ell}}}{\ell!} \, \rho^{\ell} \, F{( {\ell+N-1,-\ell; \frac{N+1}{2};} \sin^{2}{\frac{\theta_{N-1}}{2}} )} \Big),
\end{align}
where $t$ should be understood as $t- i \epsilon$ ($\epsilon >0$) in order to achieve convergence in this series. (We take the limit $\epsilon \rightarrow 0$ at the end of the calculation.)
At this point we use again the formula (\ref{Prudnikovs_series}) and then we introduce the variable 
\begin{align}
    X\equiv \frac{-4\rho}{(1-\rho)^{2}} \sin^{2}\frac{\theta_{N-1}}{2}.
\end{align} After some calculations we can rewrite 
$B$ as follows:
\begin{align}
    B=&\frac{2}{N-1}\frac{\sin{({\theta_{N-1}}/{2})}}{(\cos(x/2)\left[\cos(x'/2)\right]^{*})^{N-1} } \frac{1+\rho}{(1-\rho)^{N}} \nonumber \\
   & \times [X\frac{\partial}{\partial X}+ \frac{N-1}{2} ]F(\frac{N-1}{2},\frac{N}{2};\frac{N+1}{2};X),
\end{align}
where we notice the appearance of the raising operator for the first parameter of the hypergeometric function (\ref{hyper_raise_a}). Then using Eq.~(\ref{Fabb}) we obtain
\begin{align}
 B=\,\frac{\sin{{(\theta_{N-1}}/{2})} } {(\cos{({x}/{2})}\left[\cos{(x'/2)}\right]^{*} )^{N-1} }  \frac{1+\rho(t,t') }{  [ (1-\rho(t,t'))^{2} +4\rho(t,t') \sin^{2}(\frac{\theta_{N-1}}{2} )]^{N/2}}\frac{((1-\rho(t,t'))^{2})^{N/2}}{(1-\rho(t,t'))^{N}}.\label{appendix_B_just_before_w1}   
\end{align}
After a straightforward calculation Eq.~(\ref{appendix_B_just_before_w1}) can be written as
\begin{align}
   B=&\,{\cosh{\frac{t+t'}{2}} \sin{\frac{\theta_{N-1}}{2}}}\,(\sin^{2}\frac{\mu}{2})^{-N/2}\\
   &=  w_{2}(t,\theta_{N-1},t') (\sin^{2}\frac{\mu}{2})^{-N/2}\label{appendix_B_final}.
\end{align}
(The biscalar function $w_{2}$ is given in Eq.~(\ref{2-point-func-scalars_w2}).)

By combining Eqs.~(\ref{appendix_A_final}) and (\ref{appendix_B_final}), the two-point function (\ref{Appendix_2pt_function_with_A_and_B}) can be written in the following form:
\begin{align}
     W_{0} \big( (t,\bm{\theta}_{N-1}), (t',\bm{0}) \big)   =\frac{\beta_{0}(\mu)}{\sin{(\mu/2)}} [-w_{1} \gamma^{0} + w_{2} \gamma^{N-1}]\, \widetilde{W}_{0}(\Omega_{N-2}),\label{appendix_2-pt-func-do-the-slashed-n-squared=1-trick}
\end{align}
where $\beta_{0}(\mu)$ is given by Eq.~(\ref{beta_massless}). We can simplify this expression by using $\slashed{n}^{2}=\bm{1}$ and observing that
\begin{align}
    \slashed{n}[-w_{1} \gamma^{0} + w_{2} \gamma^{N-1}]&=(w_{1}\,n_{0} + w_{2}\,n_{N-1})\bm{1}+
(w_{2}\,n_{0} + w_{1}\,n_{N-1})\gamma^{0} \gamma^{N-1} \\
&=\sin{\frac{\mu}{2}}\,(\bm{1}\, \cosh{\frac{\lambda}{2}}+ \,\gamma^{0}\gamma^{N-1}\,\sinh{\frac{\lambda}{2}})\\
&=\sin{\frac{\mu}{2}} \exp{(\frac{\lambda}{2}\gamma^{0}\gamma^{N-1})},\label{appendix_slashed_n_times_w_t_theta_N-1_t'}
\end{align}
where in the last line we used Eq.~(\ref{Spin(N-1,1)_boost}). (For the definition of $\lambda$ see Eq.~(\ref{lambda_biscalar_parameter_parallel_transport}).) Substituting Eq.~(\ref{appendix_slashed_n_times_w_t_theta_N-1_t'}) into the expression~(\ref{appendix_2-pt-func-do-the-slashed-n-squared=1-trick}) of the two-point function we find
\begin{align}
     W_{0} \big( (t,\bm{\theta}_{N-1}), (t',\bm{0}) \big) \nonumber  ={\beta_{0}(\mu)}\slashed{n}\, \exp{(\frac{\lambda}{2}\gamma^{0}\gamma^{N-1})}\, \widetilde{W}_{0}(\Omega_{N-2}).
\end{align}
The bispinor $\exp{(\frac{\lambda}{2}\gamma^{0}\gamma^{N-1})}\, \widetilde{W}_{0}(\Omega_{N-2})$ is the spinor parallel propagator (see Eq.~(\ref{spinor_parallel_propagator_N_even})). 
%%%%%%%%%%%%%%%%%%%%%%%%%%%%%%%%%%%%%%%%%%%%%%%%%%%%%%%%%%%%%%%%%%%%%%%%%%%%%%%%%%%%%%%%%%%%%%%%%%%%%%%%%%%%%%%%%%%%%%%%%%%%%%%%%%%%%%%%%%%%%%%%%%%%%%%%%%%%%%%%%%%%%%%%%%%%%%%%%%%%%%%%%%%%%%%%%%%%%%%%%
\subsection{Determining the ``angular part'' of the two-point function}
In this section of the Appendix we show that the ``angular part'' $\widetilde{W}_{0}(\Omega_{N-2})$ (which is defined in Eq.~(\ref{appendix_angular_part_of_2pt_function})) can be written as a product of $N-2$ rotation matrices $\in$ Spin$(N-1,1)$ (see Eq.~(\ref{Spin(N-1,1)_rotation})). As is well known, these rotation matrices can be constructed by exponentiating the generators (\ref{Spin(N-1,1)_generators}).

It is convenient to express the $2^{N/2}$-dimensional gamma matrices $(\ref{even_gammas})$ using the tensor-product notation as follows:
\begin{align}\label{gammas_as_products_of_Pauli}
&\gamma^{0}=i\sigma^{2} \otimes \mathbb{I}_{2^{N/2-1}}, \nonumber \\
&\gamma^{N-r}=[\bigotimes_{i=1}^{[r/2]+1}\sigma^{1}] \otimes \sigma^{3} \otimes \mathbb{I}_{2^{(N-3-r)/2}},\nonumber \\
&\gamma^{N-r-1}=[\bigotimes_{i=1}^{[r/2]+1} \sigma^{1}]\otimes \sigma^{2} \otimes \mathbb{I}_{2^{(N-3-r)/2}} , \hspace{4mm}r\,\,\text{odd},\,\, 1\leq r\leq N-3,\nonumber \\
&\gamma^{1}=\bigotimes_{i=1}^{N/2} \sigma^{1},
\end{align}
where the Pauli matrices are given by
\begin{align}\label{Pauli_matrices}
 &  \sigma^{1}= \begin{pmatrix}
   0 & i \\
   -i & 0
\end{pmatrix}, \hspace{2mm} \sigma^{2}= \begin{pmatrix}
   0 & 1 \\
   1 & 0
\end{pmatrix}, \hspace{2mm} \sigma^{3}= \begin{pmatrix}
   1 & 0 \\
   0 & -1
\end{pmatrix}.
\end{align}
Note that they satisfy $\sigma^{i} \sigma^{j}=\delta^{ij}+i\sum_{k}\epsilon^{ijk}\sigma^{k}$, where $\epsilon^{ijk}$ is the totally antisymmetric tensor (the latter equals $+1$ if $(i,j,k)$ is an even permutation of $(1,2,3)$ and $-1$ if it is an odd permutation).
The form of the Pauli matrices we use here is related to their conventional form as follows: $\sigma_{1}=\sigma^{2},\sigma_{2}=-\sigma^{1},\sigma_{3}=\sigma^{3}$, where lower indices are used to label the conventional Pauli matrices.
For later convenience, consider the rotation generators $\gamma^{N-r+1}\gamma^{N-r}/2$ and $\gamma^{N-r}\gamma^{N-r-1}/2$ ($r$ odd) of Spin$(N-1,1)$ (see Eq.~(\ref{Spin(N-1,1)_generators})). Using Eqs.~(\ref{gammas_as_products_of_Pauli}) for the gamma matrices the generators can be written as
\begin{align}
    &\frac{1}{2}\gamma^{N-r+1}\gamma^{N-r}=\mathbb{I}_{2^{[r/2]}}\otimes (-\frac{i}{2} \sigma^{3}) \otimes \sigma^{3} \otimes  \mathbb{I}_{2^{(N-3-r)/2}}, \hspace{5mm} r\,\,\text{odd},\,\,N-3\geq r \geq 3, \\
    &\frac{1}{2}\gamma^{N-r}\gamma^{N-r-1}=\mathbb{I}_{2^{[r/2]}}\otimes \mathbb{I}_{2} \otimes (-\frac{i}{2} \sigma^{1}) \otimes  \mathbb{I}_{2^{(N-3-r)/2}}, \hspace{5mm} r\,\,\text{odd},\,\,N-3\geq r \geq 1 
\end{align}
and the corresponding rotation matrices with parameters $\theta_{N-r},\, \theta_{N-r-1}$ are respectively found to be
\begin{align}\label{rotation_matrx_odd_pol_angle}
 \exp{[\frac{\theta_{N-r}}{2}\gamma^{N-r+1}\gamma^{N-r}]} &=\mathbb{I}_{2^{[r/2]}}\otimes \exp[{-\frac{i \theta_{N-r}}{2} \sigma^{3} \otimes \sigma^{3}}] \otimes  \mathbb{I}_{2^{(N-3-r)/2}},\hspace{3mm}r\,\,\text{odd},\,\,N-3\geq r \geq 3
\end{align}
and
\begin{align}\label{rotation_matrx_even_pol_angle}
\exp{[\frac{\theta_{N-r-1}}{2}\gamma^{N-r}\gamma^{N-r-1}]}=&\mathbb{I}_{2^{[r/2]}}\otimes \exp{[ \mathbb{I}_{2} \otimes (-\frac{i \theta_{N-r-1}}{2} \sigma^{1}) }]\otimes  \mathbb{I}_{2^{(N-3-r)/2}},\hspace{3mm}r\,\,\text{odd},\,\,N-3\geq r \geq 1 .
\end{align}
Similarly, one can show that
\begin{align}\label{rotation_gamma_2_gamma_1}
    \exp[\frac{\theta_{1}}{2} \gamma^{2} \gamma^{1}]=& \mathbb{I}_{2^{(N-2)/2}}\otimes \exp{(-i\frac{\theta_{1}}{2}  \sigma^{3})} .
\end{align}

Our goal is to  express the ``angular part" of the two-point function (\ref{appendix_angular_part_of_2pt_function}) as a product consisting of rotation matrices such as (\ref{rotation_matrx_odd_pol_angle}), (\ref{rotation_matrx_even_pol_angle}) and (\ref{rotation_gamma_2_gamma_1}). By using Eq.~(\ref{rotation_matrx_even_pol_angle}) we can write the ``angular part''~(\ref{appendix_angular_part_of_2pt_function}) of the two point function as follows:
\begin{align}\label{appendix_angular_part_of_2pt_function_define_X}
    \widetilde{W}_{0}(\Omega_{N-2})= &\, \exp{[\frac{\theta_{N-2}}{2} \gamma^{N-1} \gamma^{N-2}]} \left[\mathbb{I}_{2} \otimes\begin{pmatrix}X^{(N-3)}_{-}  & 0 \\
 0 & X^{(N-3)}_{+}
 \end{pmatrix} \right],
\end{align}
where we also used $\tilde{\phi}^{(N-2)}_{0\,0}=\cos{(\theta_{N-2}/2)},\, \tilde{\psi}^{(N-2)}_{0\,0}=\sin{(\theta_{N-2}/2)}$ (see Eqs.~(\ref{phi_N-1_ell_0})-(\ref{psi_N-1_ell_0})) and we defined
\begin{align}\label{definition_of_X_plus_minus}
   X_{\pm}^{(N-r)}\equiv& \prod_{j=r}^{N-1}\Big|\frac{c_{N-j}(00)} { \sqrt{2} } \Big|^{-2}  \nonumber \\
   &\times  \left[ \sum_{\tilde{s}_{N-r-3}} \sum_{{s}_{N-r-1}}   \chi^{({s}_{N-r-1}, \tilde{s}_{N-r-3})}_{\pm00}(\bm{\theta}_{N-r})  \chi^{({s}_{N-r-1}, \tilde{s}_{N-r-3})}_{\pm00}(\bm{0})^{\dagger} \right],\hspace{4mm}r\,\,\text{odd},\,\, N-3\geq r\geq 3,
\end{align}
with $X^{(1)}_{\pm} \equiv \exp{[\pm i \theta_{1}/2]}$.
In order to proceed we use the recursive relations (\ref{chi_minus_S_N-r_in_terms_of_N-r-2})-(\ref{chi_plus_S_N-r_in_terms_of_N-r-2}) to find the following recursive relation:
\begin{align}
X_{\pm}^{(N-r)}=& \begin{pmatrix}
(\tilde{\phi}^{(N-r)}_{00 } \pm i \tilde{\psi}^{(N-r)}_{00})\bm{1} & {0} \\ \\
{0} &(\tilde{\phi}^{(N-r)}_{00} \mp i \tilde{\psi}^{(N-r)}_{00})\bm{1}
\end{pmatrix}\begin{pmatrix}
 \tilde{\phi}_{0 0}^{(N-r-1)}\bm{1} &  \tilde{\psi}_{0 0}^{(N-r-1)} \bm{1}\\ \\
  -\tilde{\psi}_{0 0}^{(N-r-1)}\bm{1} &\tilde{\phi}_{0 0}^{(N-r-1)} \bm{1}
 \end{pmatrix} \nonumber\\
  & \times \begin{pmatrix}
  X_{-}^{(N-r-2)} & {0} \\ \\
 {0} &  X_{+}^{(N-r-2)}
  \end{pmatrix} \nonumber \\
= &\left[ \begin{pmatrix}
\exp{(\pm i \frac{\theta_{N-r}}{2})} & {0} \\ \\
{0} &\exp{(\mp i \frac{\theta_{N-r}}{2})}
\end{pmatrix}\begin{pmatrix}
 \cos{\frac{\theta_{N-r-1}}{2}} & \sin{\frac{\theta_{N-r-1}}{2}} \\ \\
  -\sin{\frac{\theta_{N-r-1}}{2}} &\cos{\frac{\theta_{N-r-1}}{2}} 
 \end{pmatrix}  \otimes \mathbb{I}_{2^{(N-3-r)/2}} \right]\nonumber\\
  & \times \begin{pmatrix}
  X_{-}^{(N-r-2)} & {0} \\ \\
 {0} &  X_{+}^{(N-r-2)}
  \end{pmatrix},\hspace{8mm}r\,\,\text{odd},\,\, N-3\geq r\geq 3,\label{final_recursive_relation_angular_part}
\end{align}
 where we expanded the summation over the spin projection index $s_{N-r-1}$ in Eq.~(\ref{definition_of_X_plus_minus}) and we used $\tilde{\phi}^{(n)}_{0\,0}=\cos{(\theta_{n}/2)},\, \tilde{\psi}^{(n)}_{0\,0}=\sin{(\theta_{n}/2)}$.
Then, by combining Eqs.~(\ref{rotation_matrx_odd_pol_angle}), (\ref{rotation_matrx_even_pol_angle}) and (\ref{final_recursive_relation_angular_part}),
we can show that
\begin{align}\label{angular_part_final_recursive_relation_HONESTLY}
    \mathbb{I}_{2^{[r/2]}}\otimes  \begin{pmatrix}
  X_{-}^{(N-r)} & {0} \\ \\
 {0} &  X_{+}^{(N-r)}
  \end{pmatrix}=& \exp{[\frac{\theta_{N-r}}{2}\gamma^{N-r+1}\gamma^{N-r}   ]}\exp{[\frac{\theta_{N-r-1}}{2}\gamma^{N-r}\gamma^{N-r-1}   ]} \nonumber\\
  &\times \left[ \mathbb{I}_{2^{[(r+2)/2]}}\otimes  \begin{pmatrix}
  X_{-}^{(N-(r+2))} & {0} \\ \\
 {0} &  X_{+}^{(N-(r+2))} \end{pmatrix}  \right],\hspace{4mm}r\,\,\text{odd},\,\, N-3\geq r\geq 3.
\end{align}
We can now sequentially apply the recursive relation~(\ref{angular_part_final_recursive_relation_HONESTLY}) for $r=3,5,...,N-3$ in the expression for the ``angular part'' of the two-point function~(\ref{appendix_angular_part_of_2pt_function_define_X}). It is straightforward to find
 \begin{align}\label{angular_as_product_of_rorotations}
      \widetilde{W}_{0}(\Omega_{N-2})=& \exp{[\frac{\theta_{N-2}}{2} \gamma^{N-1} \gamma^{N-2}]}\, \exp{[\frac{\theta_{N-3}}{2} \gamma^{N-2} \gamma^{N-3}]}\, ...\,\exp{[\frac{\theta_{2}}{2} \gamma^{3} \gamma^{2}]} \,\exp{[\frac{\theta_{1}}{2} \gamma^{2} \gamma^{1}]} \\
      =&\prod_{j=2}^{N-1} \exp{[\,\frac{\theta_{N-j}}{2} \gamma^{N-j+1}\gamma^{N-j}\,] }.
 \end{align}
 %%%%%%%%%%%%%%%%%%%%%%%%%%%%%%%%%%%%%%%%%%%%%%%%%%%%%%%%%%%%%%%%%%%%%%%%%%%%%%%%%%%%%%%%%%%%%%%%%%%%%%%%%%%%%%%%%%%%%%%%%%%%%%%%%%%%
 \subsection{Massless Wightman two-point function for \texorpdfstring{$N$}{N} odd}
 The derivation for $N$ odd shares many similarities with the case with $N$ even. Therefore, we just outline the steps involved in the calculation.
 
Substituting the massless positive frequency modes~(\ref{massless_positive_freq_N=odd}) into the mode-sum expression~(\ref{Wightman_mode_sum}) and working as in the case with $N$ even it is straightforward to derive Eq.~(\ref{Appendix_2pt_function_with_A_and_B}) (where $A$, $B$ and the proportionality constant are calculated in the same way as for $N$ even). The ``angular part'' of the two-point function is given by
\begin{align}\label{appendix_angular_part_of_2pt_function_ODDD}
    \widetilde{W}_{0}(\Omega_{N-2})\equiv \left(\prod_{j=2}^{N-1}\Big|\frac{c_{N-j}(00)} { \sqrt{2} } \Big|^{-2} \right)  \times \sum_{\tilde{s}_{N-3}} \begin{pmatrix}\chi^{(\tilde{s}_{N-3})}_{-00}(\bm{\theta}_{N-2})  \chi^{(\tilde{s}_{N-3})}_{-00}(\bm{0})^{\dagger}  & 0 \\
 0 & \chi^{(\tilde{s}_{N-3})}_{+00}(\bm{\theta}_{N-2})  \chi^{(\tilde{s}_{N-3})}_{+00}(\bm{0})^{\dagger}
 \end{pmatrix}.\end{align}
 The gamma matrices (\ref{odd_gammas}) have dimension $D=2^{[N/2]}$ and can be expressed in terms of Pauli matrices as follows:
 \begin{align}\label{gammas_as_products_of_Pauli_ODDD}
 &\gamma^{0}=i\sigma^{3} \otimes \mathbb{I}_{2^{[N/2]-1}},\nonumber \\
&\gamma^{N-1}=\sigma^{2} \otimes \mathbb{I}_{2^{[N/2]-1}},\nonumber \\
&\gamma^{N-r-1}=[\bigotimes_{i=1}^{[r/2]+1}\sigma^{1}] \otimes \sigma^{3} \otimes \mathbb{I}_{2^{(N-4-r)/2}},\nonumber \\
&\gamma^{N-r-2}=[\bigotimes_{i=1}^{[r/2]+1} \sigma^{1}]\otimes \sigma^{2} \otimes \mathbb{I}_{2^{(N-4-r)/2}} , \hspace{4mm}r=\text{odd},\,\, 1\leq r\leq N-4,\\
&\gamma^{1}=\bigotimes_{i=1}^{[N/2]} \sigma^{1}.
\end{align}
Then we can show Eqs.~(\ref{rotation_matrx_odd_pol_angle}), (\ref{rotation_matrx_even_pol_angle}) and (\ref{angular_part_final_recursive_relation_HONESTLY}) with $N \rightarrow N-1$ and $r\,\,\text{odd},\,\, 1\leq r\leq N-4$. As in the even-dimensional case, Eq.~(\ref{angular_part_final_recursive_relation_HONESTLY}) (with $N \rightarrow N-1$) can be sequentially applied for $r=1,3,...,N-4$ in Eq.~(\ref{appendix_angular_part_of_2pt_function_ODDD}). Then one obtains the final expression (\ref{angular_as_product_of_rorotations}) for the ``angular part'' .
%%%%%%%%%%%%%%%%%%%%%%%%%%%%%%%%%%%%%%%%%%%%%%%%%%%%%%%%%%%%%%%%%%%%%%%%%%%%%%%%%%%%%%%%%%%%%%%%%%%%%%%%%%%%%%%%%%%%%%%%%%%%%%%%%%%%%%
\subsection{Massless Wightman two-point function on \texorpdfstring{$dS_{2}$}{dS2}}
In this subsection we derive the massless spinor Wightman two-point function on $dS_{2}$ using the mode-sum method~(\ref{Wightman_mode_sum}) and we show that it agrees with Eq.~(\ref{Ansatz_massless}). The derivation is slightly different but simpler than the case with $N>2$. Note that in this subsection we use the same (bi)scalar functions that we introduced for the case with $N>2$, with $\theta_{N-1} \rightarrow \varphi - \varphi'$ and $0\leq \varphi , \varphi '<2 \pi $. (See Eqs.~(\ref{lambda_biscalar_parameter_parallel_transport})-(\ref{2-point-func-scalars_n_pm}).) The geodesic distance and the tangent vector components are 
\begin{align}
    &\cos{\mu}= -\sinh{t} \sinh{t'} + \cosh{t} \cosh{t'} \cos(\varphi - \varphi'),\\
     &n_{0}=\frac{1}{\sin{\mu}}(  \cosh{t} \sinh{t'} -
    \sinh{t} \cosh{t'} \cos{(\varphi - \varphi')}   ), \\
    &n_{1}=\frac{1}{\sin{\mu}} \cosh{t'} \sin{(\varphi - \varphi')},
\end{align}
where $n_{1}= n_{\varphi}/\cosh{t}$.
(Note that by letting $\varphi-\varphi'\rightarrow \theta_{N-1}$ in these expressions we obtain Eqs.~(\ref{tangent_vec_non_zero_componetns_orth_basis}) and (\ref{tangent_vec_non_zero_componetns_orth_basis_theta_N-1}). This makes many steps of the calculation the same as in the case with $N>2$.)
The massless positive frequency solutions (\ref{massless_negative_spin_even_dS})-(\ref{massless_positive_spin_even_dS}) for $N=2$ are given by
\begin{align}
     &\psi^{(-)}_{ 0\ell }(t,{\varphi})=\frac{1}{2\sqrt{ \pi}} \begin{pmatrix} \phi_{0\ell}(t)
 \\  0 \end{pmatrix} {e^{-i(\ell+1/2)\varphi}},\label{negative_spin_even_dS2} \\
    & \psi^{(+)}_{ 0\ell }(t,{\varphi})=\frac{1}{2\sqrt{ \pi}} \begin{pmatrix}  0 \\ \phi_{0\ell}(t) \\
 \end{pmatrix} {e^{+i(\ell+1/2)\varphi}},\label{positive_spin_even_dS2}
 \end{align}
where $\phi_{0 \ell}(t)$ is given as functions of $x=\pi/2-it$ by Eq.~(\ref{scalar_massless}) ($\ell = 0 , 1 , ...$).
After a straightforward calculation the mode-sum method~(\ref{Wightman_mode_sum}) gives the following expression for the Wightman two-point function:
\begin{align}\label{appendix_N=2_two-point-fun_after_sum}
  W_{0}[(t,\varphi), (t',\varphi') ] = -\frac{1}{4\pi}\frac{1}{\cos{(x/2)} \left[ \cos{(x'/2)} \right]^{*}}
    \begin{pmatrix}
    0 & M_{-} \\
    M_{+}& 0
    \end{pmatrix},
\end{align}
where
\begin{align}
    M_{\pm}& ={e^{\pm i (\varphi-\varphi')/2}} \sum_{\ell=0}^{\infty} \left( \tan{\frac{x}{2}} \left[\tan{\frac{x'}{2}}\right]^{*}\, e^{\pm i (\varphi-\varphi')} \right)^{\ell} \nonumber \\
    &={e^{\pm i (\varphi-\varphi')/2}}\, \left({1- \tan{\frac{x}{2}} \left[\tan{\frac{x'}{2}}\right]^{*}}\, e^{\pm i (\varphi-\varphi')} \right)^{-1}.
\end{align}
Since $\left|\tan{\frac{x}{2}} \left[ \tan{\frac{x'}{2}}\right]^{*}\, e^{\pm i (\varphi-\varphi')}\right|=1$ we let $t \rightarrow t-i\epsilon$ (i.e. $x \rightarrow x - \epsilon$, where $\epsilon >0$) in order for the series to converge and then we let $\epsilon \rightarrow 0$. By expressing $x,x'$ in terms of $t,t'$ we can show the following relations:
\begin{align}
    M_{\pm}&=\frac{1}{w_{\mp}} \\
    &=\frac{w_{\pm}}{\sin^{2}(\mu/2)},\label{appendix_N=2_plus_minus}
\end{align}
where in the second line we used the identity $w_{+} w_{-}=\sin^{2}(\mu/2)$. 
By substituting Eq.~(\ref{appendix_N=2_plus_minus}) into the two-point function (\ref{appendix_N=2_two-point-fun_after_sum}) it is straightforward to find
\begin{align}
     W_{0} [ (t,\varphi), (t',\varphi') ] =\frac{\beta_{0}(\mu)}{\sin{(\mu/2)}} [-w_{1} \gamma^{0} + w_{2} \gamma^{1}].
\end{align}
By repeating the same calculation that resulted in Eq.~(\ref{appendix_slashed_n_times_w_t_theta_N-1_t'}) we find
\begin{align}
     W_{0} [ (t,\varphi), (t',\varphi') ] ={\beta_{0}(\mu)} \slashed{n}\exp{(\frac{\lambda}{2}\gamma^{0}\gamma^{1})},
\end{align}
where the exponential is the spinor parallel propagator (\ref{spinor_parallel_propagator_N_even}) for $dS_{2}$. 
%%%%%%%%%%%%%%%%%%%%%%%%%%%%%%%%%%%%%%%%%%%%%%%%%%%%%%%%%%%%%%%%%%%%%%%%%%%%%%%%%%%%%%%%%%%%%%%%%%%%%%%%%%%%%%%%%%%%%%%%%%%%%%%%%%%%%
\section{Testing our result for the spinor parallel propagator}\label{appendix_defining_test}
In this Appendix we show that our result for the spinor parallel propagator (i.e. Eq.~(\ref{spinor_parallel_propagator_N_even})) satisfies the defining properties (\ref{defining_properties_1})$-$ (\ref{defining_properties_3}), as introduced in Ref.~\cite{Muck:1999mh}.
\subsection{Parallel transport equation}
Our result for the spinor parallel propagator (\ref{spinor_parallel_propagator_N_even}) has to satisfy the parallel transport equation (\ref{defining_properties_3}). Starting from Eq.~(\ref{defining_properties_3}) and expressing the spinor parallel propagator in terms of the massless Wightman function (using Eq.~(\ref{Ansatz_massless})) one can obtain Eq.~(\ref{defining_property_Green_3}). For convenience, we use Eq.~(\ref{defining_property_Green_3}) rather than Eq.~(\ref{defining_properties_3}) in order to test the parallel transport-property of the spinor parallel propagator.
Let us express our two-point function in the form~(\ref{appendix_2-pt-func-do-the-slashed-n-squared=1-trick}).
    Since no derivatives act on the ``angular part'' it is straightforward to write Eq.\ (\ref{defining_property_Green_3}) as follows:
     \begin{align}
    (D({t,\theta_{N-1},t'})+\frac{N-1}{2}\cot{\frac{\mu}{2}})  \, \big[\,\sin^{-N}\frac{\mu}{2}\, (-w_{1} (t,\theta_{N-1},t')\,\gamma^{0} + w_{2}(t,\theta_{N-1},t')\,\gamma^{N-1}) \, \Big]=0, \label{p_t_equation_t,t',theta_N-1}
    \end{align}
    where the differential operator $D({t,\theta_{N-1},t'})$ is defined as
    \begin{align}
        D({t,\theta_{N-1},t'}) &\equiv  [n^{t} \partial_{t}+n^{\theta_{N-1}} \partial_{\theta_{N-1}} -n^{\theta_{N-1}} \, \frac{\sinh{t}}{2} \gamma^{0}\gamma^{N-1}    \, ]_{\bm{\theta}'=\bm{0}} .
    \end{align}
    (The tangent vectors for $\bm{\theta}'=\bm{0}$ are given by Eqs.~(\ref{tangent_vec_non_zero_componetns_orth_basis})-(\ref{tangent_vec_non_zero_componetns_orth_basis_theta_N-1}).)
    Now our initial problem has reduced to a partial differential equation involving only the coordinates $t,\theta_{N-1}$ and $t'$. This is expected because geodesics on $S^{N-1}$ lie along the line $(\theta_{N-2},...,\theta_{2},\theta_{1})=(\theta_{N-2}',...,\theta_{2}',\theta_{1}')=(0,...,0,0)$. In the rest of this Appendix we implicitly let $\bm{\theta}'=\bm{0}$ in all relevant quantities unless otherwise stated (and hence $n^{\mu}\partial_{\mu}$ will stand for $[n^{t} \partial_{t}+n^{\theta_{N-1}} \partial_{\theta_{N-1}}]_{\bm{\theta}'=\bm{0}}$).
   The parallel transport equation (\ref{p_t_equation_t,t',theta_N-1}) gives rise to partial differential equations involving just the biscalars $w_{1}(t,\theta_{N-1},t')$ and $w_{2}(t,\theta_{N-1},t')$ (see Eqs.~(\ref{2-point-func-scalars_w1})-(\ref{2-point-func-scalars_w2})). Below we derive these differential equations. Their validity has been tested using Mathematica 11.2.
    
    \noindent \textbf{Case 1:} $\bm{N}$ \textbf{even}.
Using the expressions for the $\gamma^{a}$'s (\ref{even_gammas}) we find
\begin{align}
\frac{-1}{\sin^{N}({\mu}/{2})}\,(-w_{1} (t,\theta_{N-1},t')\,\gamma^{0} + w_{2}(t,\theta_{N-1},t')\,\gamma^{N-1})&=\frac{1}{\sin^{N}({\mu}/{2})}
    \begin{pmatrix}
    0 & 0 & w_{-} & 0 \\
    0 & 0 & 0 & w_{+} \\
    w_{+} & 0 & 0 & 0\\
    0 & w_{-} & 0 & 0
     \end{pmatrix}\\
     & \equiv \begin{pmatrix}
     0 & W_{1}\\W_{2}&0
     \end{pmatrix},
\end{align}
where $W_{1}$ and $W_{2}$ represent $2^{N/2-1}$-dimensional matrices and their matrix elements can be read from above. Here $0$ stands for the matrix having all entries zero. Then Eq.~(\ref{p_t_equation_t,t',theta_N-1}) can be expanded in matrix-component form as follows:
\begin{align}
  &\begin{pmatrix}
  \bm{1} [ n^{\mu}\partial_{\mu} + \frac{N-1}{2}\cot{(\mu/2)}\,]-\frac{1}{2}n^{\theta_{N-1}}\sinh{t}\, \tilde{\gamma}^{N-1}& 0 \\ \\
    0&  \bm{1}[n^{\mu}\partial_{\mu} + \frac{N-1}{2}\cot{(\mu/2)}\,]+\frac{1}{2}n^{\theta_{N-1}}\sinh{t}\, \tilde{\gamma}^{N-1}
    \end{pmatrix}  \begin{pmatrix}
     0 & W_{1}\\ \\ W_{2}&0
     \end{pmatrix} \nonumber \\
     \nonumber\\
     &=\begin{pmatrix}
     0 & 0 \\ 
     0 & 0
     \end{pmatrix}.
\end{align}
 After a straightforward calculation we obtain the following two equations for the biscalar functions $w_{1}$ and $w_{2}$:
\begin{align}
    &(n^{\mu}\partial_{\mu} + \frac{N-1}{2}\cot{\frac{\mu}{2}}\,)\,\frac{w_{1}}{\sin^{N}(\mu/2)}=-\frac{n^{\theta_{N-1}}}{2}\sinh{t}\frac{w_{2}}{\sin^{N}(\mu/2)},\nonumber \\
     &(n^{\mu}\partial_{\mu} + \frac{N-1}{2}\cot{\frac{\mu}{2}}\,)\,\frac{w_{2}}{\sin^{N}(\mu/2)}=-\frac{n^{\theta_{N-1}}}{2}\sinh{t}\frac{w_{1}}{\sin^{N}(\mu/2)}\label{eqations_for_w1_w2}.
\end{align}
Then we use $\partial_{\alpha} \mu=n_{\alpha}$ in order to simplify Eqs.~(\ref{eqations_for_w1_w2}). Thus, we obtain the following system of differential equations for $w_{1}$ and $w_{2}$:
\begin{align}\label{p_t_eq_final_w1_w2}
   &( n^{t} \partial_{t}+n^{\theta_{N-1}} \partial_{\theta_{N-1}})w_{1}-\frac{1+\cos{\mu}}{2\sin{\mu}}w_{1} = -w_{2}\,\frac{n^{\theta_{N-1}}}{2}\sinh{t},\,\nonumber \\
   & ( n^{t} \partial_{t}+n^{\theta_{N-1}} \partial_{\theta_{N-1}})w_{2}-\frac{1+\cos{\mu}}{2\sin{\mu}}w_{2} = -w_{1}\,\frac{n^{\theta_{N-1}}}{2}\sinh{t},
\end{align}
where we also used $n^{\mu} n_{\mu}=1, \, \cot{(\mu/2)}=(1+ \cos{\mu})/ \sin{\mu}$. 
These equations are expressed in a particularly convenient form since there is a factor of $1/\sin{\mu}$ (which can be cancelled) in  each term (see Eqs.~(\ref{tangent_vectors_general}) and (\ref{tangent_vec_non_zero_componetns_orth_basis})). We have verified that the formulas we have derived for $w_{1}, w_{2}$ (given by Eqs.~(\ref{2-point-func-scalars_w1}) and (\ref{2-point-func-scalars_w2})) satisfy the differential equations (\ref{p_t_eq_final_w1_w2}) using Mathematica 11.2. Thus, our result for the spinor parallel propagator (\ref{spinor_parallel_propagator_N_even}) satisfies the parallel transport equation (\ref{defining_properties_3}).

 \noindent \textbf{Case 2:} $\bm{N}$ \textbf{odd}.
Using the gamma matrices (\ref{odd_gammas}) we find
\begin{align}
-w_{1}\,\gamma^{0} + w_{2}\,\gamma^{N-1}&=
    \begin{pmatrix}
    -i w_{1} \bm{1} && w_{2} \bm{1} \\ \\ 
    w_{2}\bm{1} && i w_{1}\bm{1}
     \end{pmatrix}, \hspace{4mm} \gamma^{0}\gamma^{N-1}= \begin{pmatrix}
     0 && i \bm{1} \\
     -i \bm{1} &&  0
     \end{pmatrix}.
\end{align}
Then, as in the case with $N$ even, we substitute these into Eq.~(\ref{p_t_equation_t,t',theta_N-1}) and we obtain the system~(\ref{p_t_eq_final_w1_w2}). The latter can be solved by our results for $w_{1}, w_{2}$ (Eqs.~(\ref{2-point-func-scalars_w1}) and (\ref{2-point-func-scalars_w2})). Thus, our result for the spinor parallel propagator (\ref{spinor_parallel_propagator_N_even}) satisfies the parallel transport equation, as required.
%%%%%%%%%%%%%%%%%%%%%%%%%%%%%%%%%%%%%%%%%%%%%%%%%%%%%%%%%%%%%%%%%%%%%%%%%%%%%%%%%%%%%%%%%%%%%%%%%%%%%%%%%%%%%%%%%%%%%%%%%%%%%%%%%%%%%%%%%%%%%%%%%%%%%%%%%%%%%%%%%%%%%%%%%%%%%%%%%%%%%%%%%%%%%%%%%%%%%%%%%
\subsection{Parallel-transport property of \texorpdfstring{$\slashed{n}$}{slashedn}}
In this subsection we show that our result for the spinor parallel propagator (\ref{spinor_parallel_propagator_N_even}) satisfies Eq.~(\ref{parallel_transport_of_slashed_n}) describing the parallel-transport property of $\slashed{n}$. Let $L$ and $R$ denote the left- and right-hand sides of Eq.~(\ref{parallel_transport_of_slashed_n}) respectively, i.e.
\begin{align}
    L&\equiv\Big(\Lambda[(t,\bm{\theta}),(t',\bm{0})] \Big)^{-1}\, \gamma^{a}n_{a}|_{\bm{\theta}'=\bm{0}}\,\,\Lambda[(t,\bm{\theta}),(t',\bm{0})], \label{LHS_p_t_slashed_n}\\
     R& \equiv-\gamma^{a'}n_{a'}|_{\bm{\theta}'=\bm{0}},\label{RHS_p_t_slashed_n}
\end{align}
where the inverse of the spinor parallel propagator can be readily found using Eq.~(\ref{spinor_parallel_propagator_N_even}).
 The components of $n_{a}|_{\bm{\theta}'=\bm{0}}$ are given in Eq.~(\ref{tangent_vec_non_zero_componetns_orth_basis}), while the components of $n_{a'}|_{\bm{\theta}'=\bm{0}}$ are found to be (see the paragraph below Eqs.~(\ref{orthonormal_basis_components_tangent_vectors_n0})- (\ref{orthonormal_basis_components_tangent_vectors_spatial_comps}))
\begin{align}\label{tangent_vec__orth_basis_theta_i'=0_at_point_x'}
 n_{0'}|_{\bm{\theta}'=\bm{0}}&=\frac{1}{\sin{\mu}}(  \cosh{t'} \sinh{t} - \sinh{t'} \cosh{t} \cos{\theta_{N-1}}   ) , \nonumber \\
     n_{(N-1)'}|_{\bm{\theta}'=\bm{0}}&=-\frac{\cosh{t}}{\sin{\mu}}\sin{\theta_{N-1}}\cos{\theta_{N-2}}, \nonumber \\
   n_{(N-2)'}|_{\bm{\theta}'=\bm{0}}&=-\frac{\cosh{t}}{\sin{\mu}}\sin{\theta_{N-1}}\sin{\theta_{N-2}}\cos{\theta_{N-3}},\nonumber \\
& \vdots \nonumber \\
 n_{2'}|_{\bm{\theta}'=\bm{0}}&=-\frac{\cosh{t}}{\sin{\mu}}(\prod_{i=1}^{N-2}\sin{\theta_{N-i}})\cos{\theta_{1}},\nonumber \\
 n_{1'}|_{\bm{\theta}'=\bm{0}}&=-\frac{\cosh{t}}{\sin{\mu}}
 \prod_{i=1}^{N-1}\sin{\theta_{N-i}}.
\end{align}
 We will show that the two sides of Eq.~(\ref{parallel_transport_of_slashed_n}) are equal by rearranging the terms in $L$. Substituting Eqs.~(\ref{slashed_n_theta_i'=0}) and (\ref{spinor_parallel_propagator_N_even}) into Eq.~(\ref{LHS_p_t_slashed_n}) we find
\begin{align}\label{LHS_bring_the_gammas_to_the_left}
  L  &=\, e^{-\frac{\theta_{1}}{2} \gamma^{2} \gamma^{1} }...\,e^{-\frac{\theta_{N-2}}{2} \gamma^{N-1} \gamma^{N-2}} e^{-\frac{\lambda}{2} \gamma^{0} \gamma^{N-1}}\, [\gamma^{0}n_{0} +\gamma^{N-1}n_{N-1}    ]_{\bm{\theta}'=\bm{0}}\nonumber\\ &\times e^{\frac{\lambda}{2} \gamma^{0} \gamma^{N-1}}\, e^{\frac{\theta_{N-2}}{2} \gamma^{N-1} \gamma^{N-2}}...\,e^{\frac{\theta_{1}}{2} \gamma^{2} \gamma^{1} }\nonumber \\
  &= \gamma^{0}n_{0}|_{\bm{\theta}'=\bm{0}}\,\big(e^{-\frac{\theta_{1}}{2} \gamma^{2} \gamma^{1} }...\,e^{-\frac{\theta_{N-2}}{2} \gamma^{N-1} \gamma^{N-2}}\big)\,e^{{\lambda} \gamma^{0} \gamma^{N-1}}\big(e^{\frac{\theta_{N-2}}{2} \gamma^{N-1} \gamma^{N-2}}\,...\,e^{\frac{\theta_{1}}{2} \gamma^{2} \gamma^{1}}\big)+\gamma^{N-1}n_{N-1}|_{\bm{\theta}'=\bm{0}}\nonumber\\
  &\times\,\big(e^{-\frac{\theta_{1}}{2} \gamma^{2} \gamma^{1} }...\,e^{-\frac{\theta_{N-3}}{2} \gamma^{N-2} \gamma^{N-3}}\big)e^{+\frac{\theta_{N-2}}{2} \gamma^{N-1} \gamma^{N-2}}\,e^{{\lambda} \gamma^{0} \gamma^{N-1}}\big(e^{\frac{\theta_{N-2}}{2} \gamma^{N-1} \gamma^{N-2}}\,...\,e^{\frac{\theta_{1}}{2} \gamma^{2} \gamma^{1}}\big),
\end{align}
 where we used the fact that if two matrices $A, B$ anti-commute $\exp{(-A)}B=B\exp{(A)}$. Our goal is to express $L$ as a sum of $N$ terms, where each term will be of the form: $\gamma^{a} \times $ (scalar) like Eq.~(\ref{RHS_p_t_slashed_n}). In order to simplify Eq.~(\ref{LHS_bring_the_gammas_to_the_left}) we use $\exp{(\lambda \gamma^{0} \gamma^{N-1})}=\bm{1}\cosh{\lambda}+\gamma^{0}\gamma^{N-1} \sinh{\lambda}$ and find
\begin{align}
    L=&[n_{0} \cosh{\lambda}-n_{N-1}\sinh{\lambda}]_{\bm{\theta}'=\bm{0}}\gamma^{0}+[-n_{0} \sinh{\lambda}+n_{N-1}\cosh{\lambda}]_{\bm{\theta}'=\bm{0}}\gamma^{N-1} \nonumber \\
    & \times \big(e^{-\frac{\theta_{1}}{2} \gamma^{2} \gamma^{1} }...\,e^{-\frac{\theta_{N-3}}{2} \gamma^{N-2} \gamma^{N-3}}\big)e^{{\theta_{N-2}} \gamma^{N-1} \gamma^{N-2} } \big(    e^{\frac{\theta_{N-3}}{2} \gamma^{N-2} \gamma^{N-3} }...e^{\frac{\theta_{1}}{2} \gamma^{2} \gamma^{1} } \big).
\end{align}
Similarly, by expanding $\exp{({\theta_{N-2}} \gamma^{N-1} \gamma^{N-2} )}$ (and then all the other exponentials of the form $\exp{({\theta_{j}} \gamma^{j+1} \gamma^{j} )}$ that will appear, with $j=N-3,...,2,1$) we find 
\begin{align}\label{LHS_one_step_before_end}
   L&=[n_{0} \cosh{\lambda}-n_{N-1}\sinh{\lambda}]_{\bm{\theta}'=\bm{0}}\gamma^{0}-[-n_{0} \sinh{\lambda}+n_{N-1}\cosh{\lambda}]_{\bm{\theta}'=\bm{0}}\, \nonumber \\
   & \times(\frac{\cosh{t}}{\sin{\mu}} \sin{\theta_{N-1}})^{-1} [n_{(N-1)'} \gamma^{N-1}+ n_{(N-2)'} \gamma^{N-2} + ...+  n_{2'}\gamma^{2} +n_{1'} \gamma^{1}]_{\bm{\theta}'=\bm{0}},
\end{align}
where we also used Eq.~(\ref{tangent_vec__orth_basis_theta_i'=0_at_point_x'}). We have verified using Mathematica 11.2 that 
\begin{align}
  [n_{0} \cosh{\lambda}-n_{N-1}\sinh{\lambda}]_{\bm{\theta}'=\bm{0}}&=-n_{0'}, \\
 [-n_{0} \sinh{\lambda}+n_{N-1}\cosh{\lambda}]_{\bm{\theta}'=\bm{0}}&=\frac{\cosh{t}}{\sin{\mu}} \sin{\theta_{N-1}},
\end{align}
where $\cosh{\lambda}$ and $\sinh{\lambda}$ can be found by Eqs.~(\ref{lambda_biscalar_parameter_parallel_transport}). By substituting these formulas into Eq.~(\ref{LHS_one_step_before_end}) we find that $L=R$, i.e. our expression for the spinor parallel propagator~(\ref{spinor_parallel_propagator_N_even}) satisfies Eq.~(\ref{parallel_transport_of_slashed_n}).
%%%%%%%%%%%%%%%%%%%%%%%%%%%%%%%%%%%%%%%%%%%%%%%%%%%%%%%%%%%%%%%%%%%%%%%%%%%%%%%%%%%%%%%%%%%%%%%%%%%%%%%%%%%%%%%%%%%%%%%%%%%%%%%%%%%%%%%%%%%%%%%%%%%%%%%%%%%%%%%%%%%%%%%%%%%%%%%%%%%%%%%%%%%%%%%%%%%%%%%%%
\subsection{The inverse of the spinor parallel propagator}
Finally, we show that our result for the spinor parallel propagator (\ref{spinor_parallel_propagator_N_even}) satisfies the defining property given by Eq.~(\ref{defining_properties_1}).
First, let us derive an expression for the two-point function with interchanged points, i.e. $W_{0} (x',x)= W_{0}[(t',\bm{0}),(t,\bm{\theta})]$. This can be found by the following relation:
\begin{align}
   W_{0} (x',x)=- \gamma^{0} W_{0}(x,x')^{\dagger} \gamma^{0}.
\end{align}
Combining this equation with Eq.~(\ref{massless_wightman_2point_function_mode_sum_N_even}) and using $\slashed{n}^{\dagger} =\gamma^{0} \slashed{n} \gamma^{0}$ and $\gamma^{0} \Lambda(x,x')^{\dagger}\gamma^{0} =-[\Lambda(x,x')]^{-1}$ (this can be verified using Eq.~(\ref{spinor_parallel_propagator_N_even})) we find
\begin{align}
    W_{0}[(t',\bm{0}),(t,\bm{\theta})]&=-\beta_{0}(\mu) \Big(\Lambda[(t,\bm{\theta}),(t',\bm{0})]\Big)^{-1} \slashed{n}|_{\bm{\theta}'=\bm{0}}\label{agrees_with_inverse_of_massless_green}\\ 
    &= \beta_{0}(\mu)\,\slashed{n}'|_{\bm{\theta}'=\bm{0}}\,\Big(\Lambda[(t,\bm{\theta}),(t',\bm{0})]\Big)^{-1},
\end{align}
where in the last line we used Eq.~(\ref{parallel_transport_of_slashed_n}). Equation~(\ref{Ansatz_massless}) implies that the massless spinor Green's function with interchanged points, $x \leftrightarrow x'$, has the following form: $ S_{0}(x',x)= \beta_{0}(\mu) \slashed{n}' \, \Lambda(x',x)$. Thus, we conclude that our expression for the spinor parallel propagator satisfies: $\Big(\Lambda[(t,\bm{\theta}),(t',\bm{0})]\Big)^{-1}= \Lambda[(t',\bm{0}),(t,\bm{\theta}) ]$ in agreement with the defining property~(\ref{defining_properties_1}). 
%%%%%%%%%%%%%%%%%%%%%%%%%%%%%%%%%%%%%%%%%%%%%%%%%%%%%%%%%%%%%%%%%%%%%%%%%%%%%%%%%%%%%%%%%%%%%%%%%%%%%%%%%%%%%%%%%%%%%%%%%%%%%%%%%%%%%%%%%%%%%%%%%%%%%%%%%%%%%%%%%%%%%%%%%%%%%%%%%%%%%%%%%%%%%%%%%%%%%%%%%
\section{A conjecture for the closed-form expression of a series containing the Gauss hypergeometric function} \label{appendix_conjecture}
In Sec.\ \ref{subsection_analytic_expressions_Wightman_and_propagator} and in Appendix \ref{appendix_2pt_function_calculations} we showed that the mode-sum approach (\ref{Wightman_mode_sum}) for the massless Wightman two-point function reproduces the result of Ref.~\cite{Muck:1999mh} (i.e. Eq.~(\ref{Ansatz_massless})).
Motivated by this result, we compare the mode-sum method for the massive Wightman two-point function~(\ref{Wightman_mode_sum}) with Eq.~(\ref{Ansatz}) and we make a conjecture regarding the closed-form expression of a series containing the Gauss hypergeometric function for $N$ even. For simplicity, we specialize to timelike separated points with $\bm{\theta}=\bm{\theta}'= \bm{0}$. For brevity, we represent the Gauss hypergeometric function as follows:
\begin{align}
    F^{(a,b)}_{(c)}(z) \equiv F(a,b;c;z) .
\end{align} 
We first present our conjecture and then we give some details for the reasoning for this conjecture.

The conjecture is
\begin{align}
   F^{(a,b)}_{(c+1)}\big(\cosh^{2}{t}\big)=\sum_{\ell=0}^{\infty} \frac{(a)_{\ell} (b)_{\ell}}{(c)_{\ell} (c+1)_{\ell}}\frac{(N-1)_{\ell}}{\ell!} \Big( \frac{\cosh^{2}{t}}{4}\Big)^{\ell}\,  F^{(a+\ell,b+\ell)}_{(c+\ell)}\big(\frac{1-i\sinh{t}}{2}\big)\,F^{(a+\ell,b+\ell)}_{(c+1+\ell)}\big({\frac{1-i\sinh{t}}{2}}\big), \label{1st_series_conjecture_a}
\end{align}
where
\begin{align}
   & a= \frac{N}{2} + i M ,\hspace{4mm} b\equiv \frac{N}{2} - i M, \hspace{4mm} c=\frac{N}{2}.
\end{align}
 By introducing the variable $w \equiv (1-i\sinh{t})/2$ we may rewrite the conjecture as follows:
\begin{align}
        F^{(a,b)}_{(c+1)}\big(4 w(1-w)\big)=\sum_{\ell=0}^{\infty} \frac{(a)_{\ell} (b)_{\ell}}{(c)_{\ell} (c+1)_{\ell}  }\frac{(N-1)_{\ell}}{\ell!} \,\big (w(1-w) \big)^{\ell}\,  F^{(a+\ell,b+\ell)}_{(c+\ell)}\big(w \big)\,F^{(a+\ell,b+\ell)}_{(c+1+\ell)}\big(w\big),  \label{1st_series_conjecture_b}
\end{align}
where $4w(1-w)=4|w|^{2}=\cosh^{2}{t}$. The time variable should be understood as $t -i \epsilon$ with $\epsilon >0$ (see Eq.~(\ref{Wightman_mode_sum})). This way, the branch cut of the hypergeometric function $F(A,B;C;X)$ along the real axis for $X > 1$ is avoided.

Below we describe the calculations that lead to the conjecture~(\ref{1st_series_conjecture_a}).
For later convenience let $C_{M}$ be the proportionality constant of the two-point function that appears in Eq.~(\ref{Ansatz}), i.e.
\begin{align}
    C_{M}\equiv \frac{|\Gamma(\frac{N}{2}+iM)|^{2}}{\Gamma(\frac{N}{2}+1)(4 \pi)^{N/2}}.
\end{align}
 For $\bm{\theta}=\bm{\theta}'=0$ Eq.~(\ref{Ansatz}) gives the following expression for the two-point function:
\begin{align}\label{appendix_massive_Ansatz_tmelike}
S_{M}[(t,\bm{0}),(t',\bm{0})]=\alpha_{M}(\mu)\bm{1}+\beta_{M}(\mu)i\gamma^{0} ,
\end{align}
where $\mu = i (t - t'), \, \slashed{n}=i \gamma^{0}$ and $\Lambda=\bm{1}$. The first term in Eq.~(\ref{appendix_massive_Ansatz_tmelike}) is diagonal while the second is off-diagonal. 
On the other hand, the mode-sum~(\ref{Wightman_mode_sum}) for massive spinors gives the following expression:
\begin{align}\label{appendix_massive_mode_sum}
    W_{M}[(t,\bm{0}),(t',\bm{0})]
    =\sum_{\ell \,m}& \Big|\frac{c_{N}(M\ell)} { \sqrt{2} } \Big|^{2}\Big[ \begin{pmatrix}
    i \phi_{M \ell}{(t)} \psi^{*}_{M\ell}{(t')}& - \phi_{M \ell}{(t)}\phi^{*}_{M \ell}{(t')} \\ \\- \psi_{M \ell}{(t)}\psi^{*}_{M \ell}{(t')} & -i \psi_{M \ell}{(t)}\phi^{*}_{M \ell}{(t')}
    \end{pmatrix} \otimes (\chi_{-\ell m} (\bm{0})\chi_{-\ell m} (\bm{0})^{\dagger}) \nonumber  \\ \nonumber \\
    &+ \begin{pmatrix}
    -i \psi_{M \ell}{(t)} \phi^{*}_{M\ell}{(t')} &- \psi_{M \ell}{(t)}\psi^{*}_{M \ell}{(t')} \\ \\- \phi_{M \ell}{(t)}\phi^{*}_{M \ell}{(t')} & i \phi_{M \ell}{(t)}\psi^{*}_{M \ell}{(t')}
    \end{pmatrix} \otimes (\chi_{+ \ell m} (\bm{0})\chi_{+\ell m} (\bm{0})^{\dagger}) \Big],
\end{align}
where the functions $\phi_{M \ell}$ and $\psi_{M \ell}$ are given by Eqs.~(\ref{phiM_a}) and (\ref{psiM_a}) and $m$ stands for the angular momentum quantum numbers and spin projection indices on the lower-dimensional spheres. Using relations~(\ref{appendix_proportionality_const_final_expression}) and (\ref{angular_part_final_recursive_relation_HONESTLY}) the two-point function~(\ref{appendix_massive_mode_sum}) can be written as
\begin{align}\label{appendix_massive_2pt-fun-mode-sum-timelike}
    W_{M}[(t,\bm{0}),(t',\bm{0})]
    = & C_{M}{\Gamma(\frac{N}{2})\Gamma(\frac{N}{2}+1)}  \nonumber \\
&\times\sum_{\ell=0}^{\infty}\frac{(N-1)_{\ell}}{\ell!} \Big|\frac{(\frac{N}{2}+iM)_{\ell}}{\Gamma(\frac{N}{2}+\ell)}\Big|^{2} \Big[ M_{\ell}(t,t')\bm{1} +  N_{\ell}(t,t') i \gamma^{0} \Big],
\end{align}
where $M_{\ell}(t,t'),N_{\ell}(t,t') $ are given by
\begin{align}
   M_{\ell}(t,t') & = -i\big(-\phi_{M \ell}{(t)} \psi^{*}_{M\ell}{(t')}+\psi_{M \ell}{(t)}\phi^{*}_{M \ell}{(t')} \big),\\ \nonumber \\
    N_{\ell}(t,t') & = \phi_{M \ell}{(t)} \phi^{*}_{M\ell}{(t')}+\psi_{M \ell}{(t)}\psi^{*}_{M \ell}{(t')}.
\end{align}
By equating Eqs.~(\ref{appendix_massive_Ansatz_tmelike}) and (\ref{appendix_massive_2pt-fun-mode-sum-timelike}) we find the following conjectured equalities:
\begin{align}
 & \sum_{\ell=0}^{\infty}\frac{(N-1)_{\ell}}{\ell!} \Big|\frac{(\frac{N}{2}+iM)_{\ell}}{\Gamma(\frac{N}{2}+\ell)}\Big|^{2}   M_{\ell}(t,t')=\frac{\alpha_{M}(t-t')}{ C_{M}{\Gamma(\frac{N}{2})\Gamma(\frac{N}{2}+1)}}, \label{conjecture_general_1}\\ 
 &\sum_{\ell=0}^{\infty}\frac{(N-1)_{\ell}}{\ell!} \Big|\frac{(\frac{N}{2}+iM)_{\ell}}{\Gamma(\frac{N}{2}+\ell)}\Big|^{2}   N_{\ell}(t,t')=\frac{\beta_{M}(t-t')}{ C_{M}{\Gamma(\frac{N}{2})\Gamma(\frac{N}{2}+1)}},\label{conjecture_general_2}
\end{align}
where the first relation is obtained by comparing the diagonal parts of Eqs.~(\ref{appendix_massive_Ansatz_tmelike}) and (\ref{appendix_massive_2pt-fun-mode-sum-timelike}), while the second is obtained by comparing the off-diagonal parts. Equations~(\ref{conjecture_general_1}) and (\ref{conjecture_general_2}) are the most general series conjectures we can find for the time-like case with $\mu=i (t-t')$. (We have checked that these conjectures are true for $t'=i \pi/2$ with $\phi^{*}_{M\ell}(t'=i\pi/2) = \delta_{\ell 0}$ and $\psi^{*}_{M\ell}(t'=i\pi/2) = 0$.) 
 By substituting Eqs.~(\ref{phiM_a}), (\ref{psiM_a}) and (\ref{alpha_massive}) into Eq.~(\ref{conjecture_general_1}) and letting $t'=- t$ we find our conjecture~(\ref{1st_series_conjecture_b}). We also made use of the following relation: $\cos(x/2)\, \left[\sin(x'/2)\right]^{*}=\frac{1}{2}(\cosh{\frac{t-t'}{2}} + i \sinh{\frac{t+t'}{2}})= \left[ \sin(x/2)\, \left[ \cos(x'/2)\right]^{*} \right]^{*}$ (see Eqs.~(\ref{cosx/2})-(\ref{sinx/2})).

In this Appendix, we made a series conjecture by letting $t'=-t$ in Eq.~(\ref{conjecture_general_1}). One can make additional series conjectures from Eqs.~(\ref{conjecture_general_1}) and (\ref{conjecture_general_2}) by giving various values to $t'$ (or $t$) or by just leaving it arbitrary.

\end{widetext}

\bibliography{apssamp}% Produces the bibliography via BibTeX.

\end{document}